\documentclass{conm-p-l}

\newtheorem{theorem}{Theorem}[section]

\newtheorem{claim}[theorem]{Claim}
\newtheorem{conclusion}[theorem]{Conclusion}

\theoremstyle{definition}
\newtheorem{definition}[theorem]{Definition}

\newtheorem{exercise}{Exercise}[section]

\theoremstyle{remark}

\numberwithin{equation}{section}

\begin{document}

\title[Holomorphic methods]{Holomorphic methods in analysis and mathematical physics}

\author{Brian C. Hall}
\address{Department of Mathematics, University of Notre Dame,
Notre Dame, IN 46556, U.S.A.}
\email{bhall@nd.edu}
\thanks{Supported in part by an NSF Postdoctoral Fellowship.}

\renewcommand{\subjclassname}{\textup{2000} Mathematics Subject Classification}
\date{January, 2000.}

\dedicatory{Dedicated to my ``father'' Leonard Gross, and to the memory of my 
``grandfather'' Irving Segal.}


\maketitle
\tableofcontents


\section{Introduction}

These notes are based on lectures that I gave at the Summer School in
Mathematical Analysis at the Instituto de Matem\'{a}ticas de la Universidad
Nacional Aut\'{o}noma de M\'{e}xico, Unidad Cuernavaca, from June 8 to 18,
1998. I am grateful to Salvador P\'{e}rez Esteva and Carlos Villegas Blas for
organizing the School and for inviting me, and to all the audience members for
their attention and interest. I thank Steve Sontz for corrections to the manuscript.

The notes explain certain parts of the theory of holomorphic function spaces
and the relation of that theory to quantum mechanics. The level is intended
for beginning graduate students. I assume knowledge of the basics of
holomorphic functions of one complex variable, Hilbert spaces, and measure
theory. I do not assume any prior knowledge of holomorphic function spaces or
quantum mechanics. I adopt throughout the physicists' convention that inner
products be linear in the second factor and conjugate-linear in the first factor.

The notes include a number of exercises. Attempting the exercises will greatly
increase the reader's understanding of the material--the best way to learn is
to do. Exercises marked with a star are harder or assume knowledge of more
advanced topics.

Much of the material in these notes has been known for some time, but has not
previously, to my knowledge, been gathered in one place. My aim is to provide
some of the conceptual and mathematical background needed to enter into the
current research in this area. The last two sections give an introduction to
more recent material.

\section{Basics of holomorphic function spaces\label{general.sect}}

This section is slightly more technical than most of the later ones, but
unfortunately we need some preliminary results in order to get started.

Let $U$ be a non-empty open set in $\mathbb{C}^{d}.$ Let $\mathcal{H}\left(
U\right)  $ denote the space of holomorphic (or complex analytic) functions on
$U.$ Recall that a function of several complex variables, $f:U\rightarrow
\mathbb{C},$ is said to be holomorphic if $f$ is continuous and holomorphic in
each variable with the other variables fixed. Let $\alpha$ be a continuous,
strictly positive function on $U.$

\begin{definition}
Let $\mathcal{H}L^{2}\left(  U,\alpha\right)  $ denote the space of $L^{2}$
holomorphic functions with respect to the weight $\alpha,$ that is,
\[
\mathcal{H}L^{2}\left(  U,\alpha\right)  =\left\{  F\in\mathcal{H}\left(
U\right)  \left|  \int_{U}\left|  F\left(  z\right)  \right|  ^{2}%
\alpha\left(  z\right)  \,dz<\infty\right.  \right\}  .
\]
\end{definition}

Here and in what follows $dz$ denotes \textit{not} a line integral, but rather
the $2d$-dimensional Lebesgue measure on $\mathbb{C}^{d}=\mathbb{R}^{2d}.$

\begin{theorem}
\label{first.thm}

\begin{enumerate}
\item  For all $z\in U,$ there exists a constant $c_{z}$ such that
\[
\left|  F\left(  z\right)  \right|  ^{2}\leq c_{z}\left\|  F\right\|
_{L^{2}\left(  U,\alpha\right)  }^{2}
\]
for all $F\in\mathcal{H}L^{2}\left(  U,\alpha\right)  .$

\item $\mathcal{H}L^{2}\left(  U,\alpha\right)  $ is a closed subspace of
$L^{2}\left(  U,\alpha\right)  ,$ and therefore a Hilbert space.
\end{enumerate}
\end{theorem}

Point 1 says that \textit{pointwise evaluation is continuous.} That is, for
each $z\in U,$ the map that takes a function $F\in\mathcal{H}L^{2}\left(
U,\alpha\right)  $ to the number $F\left(  z\right)  $ is a continuous linear
functional on $\mathcal{H}L^{2}\left(  U,\alpha\right)  .$ This is a crucial
property of holomorphic function spaces, which is certainly false for ordinary
(non-holomorphic) $L^{2}$ spaces.

\textit{Proof.} (1) Let $P_{s}\left(  z\right)  $ be the ``polydisk'' of
radius $s,$ centered at $z,$ that is,
\[
P_{s}\left(  z\right)  =\left\{  v\in\mathbb{C}^{d}\left|  \,\left|
v_{k}-z_{k}\right|  <s,\quad k=1,\cdots,d\right.  \right\}  .
\]
Here $z=\left(  z_{1},\cdots,z_{d}\right)  ,$ with each $z_{k}\in\mathbb{C}.$
If $z\in U,$ choose $s$ small enough so that $\overline{P_{s}\left(  z\right)
}\subset U.$ I then claim that
\begin{equation}
F\left(  z\right)  =\left(  \pi s^{2}\right)  ^{-d}\int_{P_{s}\left(
z\right)  }F\left(  v\right)  \,dv.\label{fz}%
\end{equation}
To verify this, consider at first the case $d=1.$ Then we may expand $F$ in a
Taylor series at $v=z$%
\[
F\left(  v\right)  =F\left(  z\right)  +\sum_{n=1}^{\infty}a_{n}\left(
v-z\right)  ^{n}.
\]
This series converges uniformly to $F$ on the compact set $\overline
{P_{s}\left(  z\right)  }\subset U.$ Thus when evaluating the integral on the
RHS of (\ref{fz}) we may interchange the integral with the sum. But now if we
use polar coordinates with the origin at $z,$ then $\left(  v-z\right)
^{n}=r^{n}e^{in\theta}.$ So for $n\geq1,$ the integral over $P_{s}\left(
z\right)  $ (which is just a disk of radius $s$ since $d=1$) give zero. So the
only surviving term is the constant term $F\left(  z\right)  ,$ which gives
$\pi s^{2}\left(  \pi s^{2}\right)  ^{-1}F\left(  z\right)  .$

For the case $d>1,$ we simply do the integral one variable at a time. By the
$d=1$ case, when we do, say, the $v_{1}$ integral, this has the effect of
setting $v_{1}=z_{1}.$ So by the time we have done all $d$ integrals, we get
just $F\left(  z\right)  .$ This establishes (\ref{fz}).

So now rewrite (\ref{fz}) in the form
\begin{align*}
F\left(  z\right)   & =\left(  \pi s^{2}\right)  ^{-d}\int_{U}1_{P_{s}\left(
z\right)  }\left(  v\right)  \frac{1}{\alpha\left(  v\right)  }F\left(
v\right)  \,\alpha\left(  v\right)  \,dv\\
& =\left(  \pi s^{2}\right)  ^{-d}\left\langle 1_{P_{s}\left(  z\right)
}\frac{1}{\alpha},F\right\rangle _{L^{2}\left(  U,\alpha\right)  },
\end{align*}
where $1_{P_{s}\left(  z\right)  }$ is the indicator function of $P_{s}\left(
z\right)  ,$ that is, the function which is one on $P_{s}\left(  z\right)  $
and zero elsewhere. Thus by the Schwarz inequality, we have
\[
\left|  F\left(  z\right)  \right|  ^{2}\leq\left(  \pi s^{2}\right)
^{-2d}\left\|  1_{P_{s}\left(  z\right)  }\frac{1}{\alpha}\right\|
_{L^{2}\left(  U,\alpha\right)  }^{2}\left\|  F\right\|  _{L^{2}\left(
U,\alpha\right)  }^{2}.
\]
Note that because $\overline{P_{s}\left(  z\right)  }\subset U$ and $\alpha$
is positive and continuous, $1/\alpha$ is bounded on $P_{s}\left(  z\right)  ;
$ thus the first $L^{2}$ norm is finite. Note also that we can take any $s$ we
like here, provided only that $\overline{P_{s}\left(  z\right)  }\subset U.$

(2) Looking at the proof of (1), we see that in fact given $z\in U,$ we can
find a neighborhood $V$ of $z$ and a constant $d_{z}$ such that
\[
\left|  F\left(  v\right)  \right|  ^{2}\leq d_{z}\left\|  F\right\|
_{L^{2}\left(  U,\alpha\right)  }^{2}
\]
for all $v\in V$ and all $F\in\mathcal{H}L^{2}\left(  U,\alpha\right)  .$
(That is, the constant in (1) can be taken to be bounded in a neighborhood of
each point.) So now suppose we have a sequence $F_{n}\in\mathcal{H}%
L^{2}\left(  U,\alpha\right)  ,$ and $F\in L^{2}\left(  U,\alpha\right)  $
such that $F_{n}\rightarrow F$ in $L^{2}\left(  U,\alpha\right)  .$ Then
$F_{n}$ is certainly a Cauchy sequence in $L^{2}.$ But then
\[
\sup_{v\in V}\left|  F_{n}\left(  v\right)  -F_{m}\left(  v\right)  \right|
\leq\sqrt{d_{z}}\left\|  F_{n}-F_{m}\right\|  _{L^{2}\left(  U,\alpha\right)
}\rightarrow0\quad\text{as }n,m\rightarrow\infty.
\]
This shows that the sequence $F_{m}$ converges \textit{locally uniformly} to
some limit function, which must be $F.$ (If $L^{2}$ limit and pointwise limit
both exist they must be equal a.e.) But a standard theorem shows that a
locally uniform limit of holomorphic functions is always holomorphic. (Use
Morera's Theorem to show that the limit is still holomorphic in each
variable.) So the limit function $F$ is actually in $\mathcal{H}L^{2}\left(
U,\alpha\right)  ,$ which shows that $\mathcal{H}L^{2}\left(  U,\alpha\right)
$ is closed. \qed

\begin{theorem}
[Reproducing Kernel]\label{rep.thm}Let $\mathcal{H}L^{2}\left(  U,\alpha
\right)  $ be as above. Then there exists a function $K\left(  z,w\right)  ,$
$z,w\in U,$ with the following properties:

\begin{enumerate}
\item $K\left(  z,w\right)  $ is holomorphic in $z$ and anti-holomorphic in
$w,$ and satisfies
\[
K\left(  w,z\right)  =\overline{K\left(  z,w\right)  }.
\]

\item  For each fixed $z\in U,$ $K\left(  z,w\right)  $ is square-integrable
$d\alpha\left(  w\right)  .$ For all $F\in\mathcal{H}L^{2}\left(
U,\alpha\right)  $%
\[
F\left(  z\right)  =\int_{U}K\left(  z,w\right)  F\left(  w\right)
\,\alpha\left(  w\right)  \,dw.
\]

\item  If $F\in L^{2}\left(  U,\alpha\right)  ,$ let $PF$ denote the
orthogonal projection of $F$ onto the closed subspace $\mathcal{H}L^{2}\left(
U,\alpha\right)  .$ Then
\[
PF\left(  z\right)  =\int_{U}K\left(  z,w\right)  F\left(  w\right)
\,\alpha\left(  w\right)  \,dw.
\]

\item  For all $z,u\in U,$%
\[
\int_{U}K\left(  z,w\right)  K\left(  w,u\right)  \,\alpha\left(  w\right)
\,dw=K\left(  z,u\right)  .
\]

\item  For all $z\in U,$%
\[
\left|  F\left(  z\right)  \right|  ^{2}\leq K\left(  z,z\right)  \left\|
F\right\|  ^{2},
\]
and the constant $K\left(  z,z\right)  $ is optimal in the sense that for each
$z\in U$ there exists a non-zero $F_{z}\in\mathcal{H}L^{2}\left(
U,\alpha\right)  $ for which equality holds.

\item  Given any $z\in U,$ if $\phi_{z}\left(  \cdot\right)  \in
\mathcal{H}L^{2}\left(  U,\alpha\right)  $ satisfies
\[
F\left(  z\right)  =\int_{U}\overline{\phi_{z}\left(  w\right)  }F\left(
w\right)  \,\alpha\left(  w\right)  \,dw
\]
for all $F\in\mathcal{H}L^{2}\left(  U,\alpha\right)  ,$ then $\overline
{\phi_{z}\left(  w\right)  }=K\left(  z,w\right)  .$
\end{enumerate}
\end{theorem}

\textit{Proof.} We have already shown that evaluation at a point $z\in U$ is a
continuous linear functional on $\mathcal{H}L^{2}\left(  U,\alpha\right)  .$
Thus by the Riesz Theorem, this linear functional can be represented uniquely
as inner product with some $\phi_{z}\in\mathcal{H}L^{2}\left(  U,\alpha
\right)  ,$ that is,
\begin{align}
F\left(  z\right)   & =\left\langle \phi_{z},F\right\rangle _{L^{2}\left(
U,\alpha\right)  }\label{fz2}\\
& =\int_{U}\overline{\phi_{z}\left(  w\right)  }F\left(  w\right)
\,\alpha\left(  w\right)  \,dw.
\end{align}
So we set $K\left(  z,w\right)  =\overline{\phi_{z}\left(  w\right)  }.$ (I
adopt the convention that the inner product be linear on the right and
conjugate-linear on the left.) By its very construction, $K\left(  z,w\right)
$ satisfies Point 2 of the theorem and is anti-holomorphic in $w.$

Now we apply (\ref{fz2}) to $\phi_{z}$ itself. Thus
\begin{align*}
\phi_{z}\left(  w\right)   & =\left\langle \phi_{w},\phi_{z}\right\rangle
_{L^{2}\left(  U,\alpha\right)  }=\overline{\left\langle \phi_{z},\phi
_{w}\right\rangle }_{L^{2}\left(  U,\alpha\right)  }\\
& =\overline{\phi_{w}\left(  z\right)  }.
\end{align*}
Thus $\overline{K\left(  z,w\right)  }=K\left(  w,z\right)  ,$ and we have
Point 1.

For Point 3, we consider two cases. If $F\in\mathcal{H}L^{2}\left(
U,\alpha\right)  ,$ then Point 3 says the same thing as Point 2. On the other
hand, if $F\in\left[  \mathcal{H}L^{2}\left(  U,\alpha\right)  \right]
^{\bot},$ then the RHS of Point 3 is just $\left\langle \phi_{z}%
,F\right\rangle $ which is zero since $\phi_{z}\in\mathcal{H}L^{2}\left(
U,\alpha\right)  .$ So the RHS of Point 3 is the identity on $\mathcal{H}%
L^{2}\left(  U,\alpha\right)  $ and zero on the orthogonal complement of
$\mathcal{H}L^{2}\left(  U,\alpha\right)  ,$ so it must coincide with $P.$

Point 4 is just Point 2 applied to the square-integrable holomorphic function
$K\left(  w,u\right)  ,$ viewing $w$ as the variable and $u$ as a parameter.

For Point 5 we note evaluation at $z$ is just inner product with an element
$\phi_{z}$ of our Hilbert space. So the norm of this linear functional is just
the norm of $\phi_{z}.$ But
\[
\left\|  \phi_{z}\right\|  ^{2}=\left\langle \phi_{z},\phi_{z}\right\rangle
_{L^{2}\left(  U,\alpha\right)  }=\phi_{z}\left(  z\right)  =K\left(
z,z\right)  .
\]
Saying that we have computed the norm of this linear functional means
precisely that we have found the optimal constant in the inequality $\left|
F\left(  z\right)  \right|  \leq\sqrt{c_{z}}\left\|  F\right\|  .$ The
function $\phi_{z}$ itself is the one that gives equality in Point 5.

For Point 6, note that if $\phi_{z}\left(  \cdot\right)  $ is in
$\mathcal{H}L^{2}\left(  U,\alpha\right)  $ and satisfies $F\left(  z\right)
=\left\langle \phi_{z},F\right\rangle $ for all $F\in\mathcal{H}L^{2}\left(
U,\alpha\right)  ,$ then $\left\langle \phi_{z},F\right\rangle =\left\langle
\overline{K\left(  z,\cdot\right)  },F\right\rangle $ and $\left\langle
\overline{K\left(  z,\cdot\right)  }-\phi_{z},F\right\rangle =0,$ for all
$F\in\mathcal{H}L^{2}\left(  U,\alpha\right)  .$ Since $\overline{K\left(
z,\cdot\right)  }$ and $\phi_{z}$ are both in $\mathcal{H}L^{2}\left(
U,\alpha\right)  ,$ we may take $F=K\left(  z,\cdot\right)  -\phi_{z}\left(
\cdot\right)  ,$ which shows that $K\left(  z,\cdot\right)  -\phi_{z}\left(
\cdot\right)  =0;$ that is, $\phi_{z}\left(  w\right)  =\overline{K\left(
z,w\right)  }.$ \qed

This theorem is really just the continuity of pointwise evaluation, together
with the Riesz Theorem. The reproducing kernel is a useful way of encoding
information about a holomorphic function space. Our next result gives us a way
of calculating the reproducing kernel.

\begin{theorem}
\label{rep.on}Let $\left\{  e_{j}\right\}  $ be any ON basis for
$\mathcal{H}L^{2}\left(  U,\alpha\right)  .$ Then for all $z,w\in U$%
\[
\sum_{j}\left|  e_{j}\left(  z\right)  \overline{e_{j}\left(  w\right)
}\right|  <\infty
\]
and
\[
K\left(  z,w\right)  =\sum_{j}e_{j}\left(  z\right)  \overline{e_{j}\left(
w\right)  }.
\]
\end{theorem}

\textit{Proof.} The annoying part of the proof is the convergence issues. Once
this is done, verifying the formula for $K$ is fairly easy. So on a first
reading you should skip to the last paragraph of the proof.

For any $f\in\mathcal{H}L^{2}\left(  U,\alpha\right)  ,$ Parseval's Theorem
says that
\[
\sum_{j}\left|  \left\langle f,e_{j}\right\rangle \right|  ^{2}=\left\|
f\right\|  ^{2}.
\]
Then for any $f,g\in\mathcal{H}L^{2}\left(  U,\alpha\right)  ,$ consider the
Schwarz inequality in the space $l^{2}$ of square-summable sequences, applied
to the sequences $\left|  \left\langle f,e_{j}\right\rangle \right|  $ and
$\left|  \left\langle g,e_{j}\right\rangle \right|  .$ This gives
\[
\sum_{j}\left|  \left\langle f,e_{j}\right\rangle \left\langle e_{j}%
,g\right\rangle \right|  \leq\left\|  f\right\|  \left\|  g\right\|  .
\]
Taking $f=\phi_{z}$ and $g=\phi_{w}$ we get
\[
\sum_{j}\left|  e_{j}\left(  z\right)  \overline{e_{j}\left(  w\right)
}\right|  \leq\left\|  \phi_{z}\right\|  \left\|  \phi_{w}\right\|  <\infty.
\]
So the sum is absolutely convergent for each $z$ and $w.$

Now think of the partial sums of $\sum_{j}e_{j}\left(  z\right)
\overline{e_{j}\left(  w\right)  }$ as functions of $w$ with $z$ fixed. Then
the series is orthogonal and
\begin{align*}
\sum_{j}\left\|  e_{j}\left(  z\right)  \overline{e_{j}\left(  w\right)
}\right\|  _{L^{2}\left(  w\right)  }^{2}  & =\sum_{j}\left\|  \left\langle
\phi_{z},e_{j}\right\rangle e_{j}\right\|  ^{2}\\
& =\sum_{j}\left|  \left\langle \phi_{z},e_{j}\right\rangle \right|
^{2}=\left\|  \phi_{z}\right\|  ^{2}<\infty.
\end{align*}
So the series is actually $L^{2}$ convergent as a function of $w$ for fixed
$z.$ This shows (by the obvious analog of Theorem \ref{first.thm} for
anti-holomorphic functions) that the sum is anti-holomorphic as a function of
$w$ for each fixed $z.$ Arguing in a similar way with the roles of $z$ and $w$
reversed shows that the sum is holomorphic as a function of $z$ for each fixed $w.$

So now that the unpleasant convergence issues are settled, let's prove the
formula for $K.$ In essence Theorem \ref{rep.on} says that any $F\in
\mathcal{H}L^{2}\left(  U,\alpha\right)  $ is the sum of its projections onto
the orthonormal basis elements $e_{j}.$ If you prefer, you can verify the
theorem first just for $F\left(  z\right)  =e_{k}\left(  z\right)  $ and then
extend by linearity to arbitrary $F.$

For any $F\in\mathcal{H}L^{2}\left(  U,\alpha\right)  $ we have
\begin{align*}
F\left(  z\right)   & =\left\langle \phi_{z},F\right\rangle =\sum
_{j}\left\langle \phi_{z},e_{j}\right\rangle \left\langle e_{j},F\right\rangle
\\
& =\sum_{j}e_{j}\left(  z\right)  \int_{U}\overline{e_{j}\left(  w\right)
}F\left(  w\right)  \,\alpha\left(  w\right)  \,dw\\
& =\int_{U}\left[  \sum_{j}e_{j}\left(  z\right)  \overline{e_{j}\left(
w\right)  }\right]  F\left(  w\right)  \,\alpha\left(  w\right)  \,dw.
\end{align*}
In the first line we have used the basic property of the $\phi_{z}$'s and
Parseval's Theorem. In the second line we have used the basic property of
$\phi_{z}$ to evaluate $\left\langle \phi_{z},e_{j}\right\rangle ,$ and we
have written out $\left\langle e_{j},F\right\rangle $ as an integral. In the
third line we have interchanged the sum and integral, as justified by the
$L^{2}$ convergence of the sum. Finally, then, by Point 6 of the last theorem
we conclude that the quantity in square brackets must be $K\left(  z,w\right)
.$ \qed

\textit{Remark}. Most of this time, this formula for the reproducing kernel is
not especially useful, since (1) you can't usually find explicitly an
orthonormal basis, and (2) even if you could, you probably couldn't compute
the sum. But it will give explicit formulas for the reproducing kernel in
certain important cases.

\subsection{Exercises}

\begin{exercise}
Show that for all $z\in U,$ there exist constants $c_{z,k}$ such that
\[
\left|  \frac{\partial F}{\partial z_{k}}\left(  z\right)  \right|  _{{}}%
^{2}\leq c_{z,k}\left\|  F\right\|  ^{2}.
\]
(You may do this just in the case $d=1$ if it makes things easier.)
\end{exercise}

\begin{exercise}
\label{leb.zero}Show that $\mathcal{H}L^{2}\left(  \mathbb{C},1\right)
=\left\{  0\right\}  .$

Hint: Suppose $F\in\mathcal{H}L^{2}\left(  \mathbb{C},1\right)  .$ Use Theorem
\ref{first.thm} to show that $F$ must be bounded.
\end{exercise}

\begin{exercise}
\label{badex.ex}* Consider the measure $\mu$ on the plane with the property
that for all bounded measurable functions $f,$%
\[
\int_{\mathbb{C}}f\,d\mu=\int_{\mathbb{R}}f\left(  x,0\right)  \,\,dx.
\]
(So this measure is concentrated on the real axis.) Show that $\mathcal{H}%
\left(  \mathbb{C}\right)  \cap L^{2}\left(  \mathbb{C},\mu\right)  $ is dense
in $L^{2}\left(  \mathbb{C},\mu\right)  .$ In this case $\mathcal{H}\left(
\mathbb{C}\right)  \cap L^{2}\left(  \mathbb{C},\mu\right)  $ is not a Hilbert
space and pointwise evaluation is not continuous. (This is why we consider
only measures that have a positive density with respect to Lebesgue measure on
$\mathbb{C}^{d}.$)
\end{exercise}

\section{Examples of holomorphic function spaces\label{examples.sect}}

\subsection{The weighted Bergman spaces}

\begin{definition}
The \textbf{weighted Bergman spaces} are the spaces
\[
\mathcal{H}L^{2}\left(  \mathbb{D},\left(  1-\left|  z\right|  ^{2}\right)
^{a}\right)  ,\quad a>-1,
\]
where $\mathbb{D}$ is the unit disk,
\[
\mathbb{D}=\left\{  z\in\mathbb{C}\left|  \,\left|  z\right|  <1\right.
\right\}  .
\]
\end{definition}

Here the restriction $a>-1$ is needed to get a non-zero space. The weighted
Bergman spaces are important in operator theory and in representation theory.
We will compute the reproducing kernel for the weighted Bergman spaces just in
the case $a=0,$ in which case the space is the \textbf{standard Bergman
space}. We will denote this space $\mathcal{H}L^{2}\left(  \mathbb{D}\right)
,$ with the weight 1 being understood.

\textit{Step 1}. \textit{Show that }$\left\{  z^{n}\right\}  _{n=0}^{\infty}$
\textit{is an orthogonal basis for }$\mathcal{H}L^{2}\left(  \mathbb{D}%
\right)  .$

We first check orthogonality, computing the integral in polar coordinates.
\begin{align*}
\left\langle z^{n},z^{m}\right\rangle  & =\int_{0}^{2\pi}\int_{0}^{1}%
r^{n}e^{-in\theta}r^{m}e^{im\theta}r\,dr\,d\theta\\
& =\int_{0}^{1}r^{n+m+1}\int_{0}^{2\pi}e^{i\left(  m-n\right)  \theta
}\,d\theta\,dr\\
& =0\quad\left(  n\neq m\right)  .
\end{align*}
We now need to show that the $z^{n}$'s span a dense subspace of $\mathcal{H}%
L^{2}\left(  \mathbb{D}\right)  .$ It suffices to show that if $F\in
\mathcal{H}L^{2}\left(  \mathbb{D}\right)  $ and $\left\langle z^{n}%
,F\right\rangle =0$ for all $n,$ then $F=0.$ So suppose $F\in\mathcal{H}%
L^{2}\left(  \mathbb{D}\right)  . $ We expand $F$ in a power series
\begin{equation}
F\left(  z\right)  =\sum_{n=0}^{\infty}c_{n}z^{n}.\label{f.series}%
\end{equation}
This series converges uniformly on compact subsets of $\mathbb{D}.$ Now
compute
\begin{align*}
\left\langle z^{m},F\right\rangle  & =\int_{0}^{1}\int_{0}^{2\pi}%
r^{m}e^{-im\theta}F\left(  re^{i\theta}\right)  \,r\,dr\,d\theta\\
& =\lim_{a\rightarrow1}\int_{0}^{a}\int_{0}^{2\pi}r^{m}e^{-im\theta}F\left(
re^{i\theta}\right)  \,r\,dr\,d\theta
\end{align*}
where the last equality is by Dominated Convergence. But now the series
(\ref{f.series}) converges uniformly on the set $r\leq a,$ and so we may
interchange integral and sum to get
\begin{align*}
\left\langle z^{m},F\right\rangle  & =\lim_{a\rightarrow1}\sum_{n=0}^{\infty
}\int_{0}^{a}\int_{0}^{2\pi}r^{m}e^{-im\theta}c_{n}r^{n}e^{in\theta
}\,r\,dr\,d\theta\\
& =\lim_{a\rightarrow1}\sum_{n=0}^{\infty}c_{n}\int_{0}^{a}r^{n+m+1}\int
_{0}^{2\pi}e^{i\left(  n-m\right)  \theta}\,d\theta\,dr.
\end{align*}
But the $\theta$ integral gives zero except when $n=m.$ So only one term in
the sum survives, and we can then let $a$ tend to 1 to get
\begin{align*}
\left\langle z^{m},F\right\rangle  & =2\pi c_{m}\int_{0}^{1}r^{2m+1}\,dr\\
& =2\pi c_{m}\frac{1}{2m+2}=\frac{\pi c_{m}}{m+1}.
\end{align*}

So if $\left\langle z^{m},F\right\rangle =0$ for all $m,$ then $c_{m}=0$ for
all $m,$ in which case $F$ is identically zero. So $\left\{  z^{m}\right\}  $
is a basis.

\textit{Step 2. Normalize.}

Compute that
\begin{align*}
\left\|  z^{n}\right\|  ^{2}  & =\int_{0}^{1}\int_{0}^{2\pi}r^{2n}%
\,r\,dr\,d\theta\\
& =2\pi\frac{1}{2n+2}=\frac{\pi}{n+1}.
\end{align*}
So
\[
\left\{  z^{n}\sqrt{\frac{n+1}{\pi}}\right\}  _{n=0}^{\infty}
\]
is an ortho\textit{normal} basis for $\mathcal{H}L^{2}\left(  \mathbb{D}%
\right)  . $

\textit{Step 3. Compute reproducing kernel.}

According to our theorem, we may now compute the reproducing kernel as
\begin{align}
K\left(  z,w\right)   & =\sum_{n=0}^{\infty}z^{n}\sqrt{\frac{n+1}{\pi}}\bar
{w}^{n}\sqrt{\frac{n+1}{\pi}}\nonumber\\
& =\frac{1}{\pi}\sum_{n=0}^{\infty}\left(  n+1\right)  \left(  z\bar
{w}\right)  ^{n}.\label{k.berg}%
\end{align}
So now let us consider the function
\begin{align*}
f\left(  \xi\right)   & =\sum_{n=0}^{\infty}\left(  n+1\right)  \xi^{n}\\
& =\sum_{n=0}^{\infty}\frac{d}{d\xi}\xi^{n+1}\\
& =\frac{d}{d\xi}\sum_{n=0}^{\infty}\xi^{n+1}=\frac{d}{d\xi}\left(  \xi
+\xi^{2}+\xi^{3}+\cdots\right)  .
\end{align*}
Adding a 1 inside the derivative (harmless since the derivative of 1 is zero)
we get
\[
f\left(  \xi\right)  =\frac{d}{d\xi}\frac{1}{1-\xi}=\frac{1}{\left(
1-\xi\right)  ^{2}}.
\]
Now by (\ref{k.berg}), $K\left(  z,w\right)  =f\left(  z\bar{w}\right)  /\pi.$
So we have the following.

\begin{conclusion}
The reproducing kernel for the standard Bergman space is
\[
K\left(  z,w\right)  =\frac{1}{\pi}\frac{1}{\left(  1-z\bar{w}\right)  ^{2}}.
\]
Thus in particular,
\[
\left|  F\left(  z\right)  \right|  ^{2}\leq\frac{1}{\pi\left(  1-\left|
z\right|  ^{2}\right)  ^{2}}\left\|  F\right\|  ^{2},
\]
for all $F\in\mathcal{H}L^{2}\left(  \mathbb{D}\right)  $ and all
$z\in\mathbb{D}.$
\end{conclusion}

\subsection{The Segal-Bargmann spaces}

\begin{definition}
The Segal-Bargmann spaces are the holomorphic function spaces
\[
\mathcal{H}L^{2}\left(  \mathbb{C}^{d},\mu_{t}\right)  ,
\]
where
\[
\mu_{t}\left(  z\right)  =\left(  \pi t\right)  ^{-d}e^{-\left|  z\right|
^{2}/t}.
\]
Here $\left|  z\right|  ^{2}=\left|  z_{1}\right|  ^{2}+\cdots+\left|
z_{d}\right|  ^{2}$ and $t$ is a positive number.
\end{definition}

We will now compute the reproducing kernel for the Segal-Bargmann space. We
consider at first just the case $d=1.$

\textit{Step 1. Show that }$\left\{  z^{n}\right\}  _{n=0}^{\infty}$
\textit{is a basis for the Segal-Bargmann space, with} $d=1.$

The proof of this is nearly the same as the proof of Step 1 in the computation
of the reproducing kernel for the standard Bergman space, and is omitted.

\textit{Step 2. Normalize.}

We compute $\left\|  z^{n}\right\|  ^{2}$ by induction on $n.$ For $n=0$ we
observe that (with $d=1$)
\begin{align*}
\int_{\mathbb{C}}\left(  1\right)  ^{2}\mu_{t}\left(  z\right)  \,dz  &
=\frac{1}{\pi t}\int_{0}^{2\pi}\int_{0}^{\infty}e^{-r^{2}/t}\,r\,dr\,d\theta\\
& =\frac{2\pi}{\pi t}\left(  -\frac{t}{2}\right)  \lim_{A\rightarrow\infty
}\left.  e^{-r^{2}/t}\right|  _{0}^{A}\\
& =-\lim_{A\rightarrow\infty}\left[  e^{-A^{2}/2}-1\right]  =1.
\end{align*}
Next we compute that for $n>0,$%
\begin{align*}
\left\|  z^{n}\right\|  ^{2}  & =\int_{\mathbb{C}}\left|  z^{n}\right|
^{2}\mu_{t}\left(  z\right)  \,dz=\frac{1}{\pi t}\int_{0}^{2\pi}\int
_{0}^{\infty}e^{-r^{2}/t}r^{2n+1}\,dr\,d\theta\\
& =\frac{2}{t}\int_{0}^{\infty}r^{2n}\left(  e^{-r^{2}/t}r\right)  \,dr.
\end{align*}
Integrating by parts gives
\begin{align*}
\left\|  z^{n}\right\|  ^{2}  & =-\frac{2}{t}\int_{0}^{\infty}\left(
2nr^{2n-1}\right)  \left(  -\frac{t}{2}e^{-r^{2}/t}\right)  \,dr\\
& =\frac{2}{t}\left(  nt\right)  \int_{0}^{\infty}e^{-r^{2}/t}r^{2\left(
n-1\right)  +1}\,dr\\
& =nt\left\|  z^{n-1}\right\|  ^{2}.
\end{align*}
Thus we will have
\[
\left\|  z^{n}\right\|  ^{2}=n!t^{n},
\]
and so
\[
\left\{  \frac{z^{n}}{\sqrt{n!t^{n}}}\right\}  _{n=0}^{\infty}
\]
is an orthonormal basis for $\mathcal{H}L^{2}\left(  \mathbb{C}^{d},\mu
_{t}\right)  .$

\textit{Step 3. Compute the reproducing kernel.}

Our formula for the reproducing kernel is now
\begin{align*}
K\left(  z,w\right)   & =\sum_{n=0}^{\infty}\frac{z^{n}}{\sqrt{n!t^{n}}}%
\frac{\overline{w^{n}}}{\sqrt{n!t^{n}}}\\
& =\sum_{n=0}^{\infty}\frac{1}{n!}\left(  \frac{z\bar{w}}{t}\right)
^{n}=e^{z\bar{w}/t}.
\end{align*}

So we have computed the reproducing kernel explicitly for the case $d=1.$ For
general $d$ we have the following result.

\begin{theorem}
For all $d\geq1,$ the reproducing kernel for the space $\mathcal{H}%
L^{2}\left(  \mathbb{C}^{d},\mu_{t}\right)  $ is given by
\[
K\left(  z,w\right)  =e^{z\cdot\bar{w}/t},
\]
where $z\cdot\bar{w}=z_{1}\bar{w}_{1}+\cdots+z_{d}\bar{w}_{d}.$ In particular,
we have the pointwise bounds
\[
\left|  F\left(  z\right)  \right|  ^{2}\leq e^{\left|  z\right|  ^{2}%
/t}\left\|  F\right\|  ^{2}
\]
for all $F\in\mathcal{H}L^{2}\left(  \mathbb{C}^{d},\mu_{t}\right)  $ and all
$z\in\mathbb{C}^{d}.$
\end{theorem}

Note that the bounds are reasonable, since $\left|  F\left(  z\right)
\right|  ^{2}$ is required to be square-integrable against the density
$e^{-\left|  z\right|  ^{2}/t}.$ We will derive the reproducing kernel of the
Segal-Bargmann in two other ways, one in Section \ref{unitary.sect} and one in
Section \ref{sbt.sect}.

\textit{Proof}. We have already proved this for the case $d=1.$ For general
$d$ we note first of all that $\overline{K\left(  z,w\right)  },$ with $K$ as
given in the theorem, is certainly holomorphic and square-integrable against
$\mu_{t}$ as a function of $w$ for each fixed $z.$ (The function
$\overline{K\left(  z,w\right)  }$ grows only exponentially with $w$ for each
fixed $z,$ and so it is square-integrable against $e^{-\left|  z\right|
^{2}/t}.$) Note that the $d$-dimensional density $\mu_{t}\left(  z\right)  $
just factors as a product of the 1-dimensional densities in each variable.
Thus
\[
\int_{\mathbb{C}^{d}}e^{z\cdot\bar{w}/t}F\left(  w\right)  \mu_{t}\left(
w\right)  \,dw=\int_{\mathbb{C}}\cdots\int_{\mathbb{C}}e^{z_{1}\bar{w}_{1}%
/t}\cdots e^{z_{d}\bar{w}_{d}/t}F\left(  w_{1},\cdots,w_{d}\right)
\frac{dw_{1}}{\pi t}\cdots\frac{dw_{d}}{\pi t}.
\]
Now, $F\left(  w_{1},\cdots,w_{d}\right)  $ is holomorphic in each variable
with the others fixed. So \textit{provided} that $F$ is square-integrable with
respect to $\left(  \pi t\right)  ^{-1}e^{-\left|  w_{i}\right|  ^{2}%
/t}\,dw_{i}$ with respect to each $w_{i}$ with the other variables fixed, then
we may apply the one-dimensional result $d$ times, which gives simply
\begin{equation}
\int_{\mathbb{C}^{d}}e^{z\cdot\bar{w}/t}F\left(  w\right)  \mu_{t}\left(
w\right)  \,dw=F\left(  z_{1},\cdots,z_{d}\right)  .\label{int.ez}%
\end{equation}
This, by Point 6 of Theorem \ref{rep.thm}, would show that $K\left(
z,w\right)  =e^{z\cdot\bar{w}/t},$ as claimed.

So let us assume at first that $F$ is a polynomial. Then $F$ is a polynomial
in each variable with the others fixed, and so there is no trouble with
square-integrability. But a Taylor series argument, similar to the density of
polynomials in the standard Bergman space, shows that polynomials are dense in
$\mathcal{H}L^{2}\left(  \mathbb{C}^{d},\mu_{t}\right)  .$ So since
$e^{z\cdot\bar{w}/t}$ is $\mu_{t}$-square-integrable as a function of $w,$ if
(\ref{int.ez}) holds on a dense set, it must hold for all $F\in\mathcal{H}%
L^{2}\left(  \mathbb{C}^{d},\mu_{t}\right)  .$ \qed

\subsection{The Hardy space}

This space is not quite an example of the sort considered in Section
\ref{general.sect}, but is an important space which I therefore wish to
introduce. It is also interesting to contrast this example with Exercise
\ref{badex.ex} of Section \ref{general.sect}.

\begin{definition}
The \textbf{Hardy space} is the space of holomorphic functions $F$ on the unit
disk $\mathbb{D}$ such that
\[
\sup_{r<1}\int_{0}^{2\pi}\left|  F\left(  re^{i\theta}\right)  \right|
^{2}\,d\theta<\infty.
\]
\end{definition}

So this is almost like an $L^{2}$ space of holomorphic functions on
$\mathbb{D}$ with respect to a measure $\mu,$ except that the measure is not a
measure on $\mathbb{D},$ but is a measure on the boundary of $\mathbb{D}.$ It
is not too hard, using Taylor series, to show that for any $F\in
\mathcal{H}\left(  \mathbb{D}\right)  ,$ $\int_{0}^{2\pi}\left|  F\left(
re^{i\theta}\right)  \right|  ^{2}\,d\theta$ is an increasing function of $r.$
Thus the supremum in the definition is equal to the limit as $r$ approaches
one. We then define a norm and an inner product on the Hardy space by
defining
\[
\left\|  F\right\|  ^{2}=\lim_{r\rightarrow1}\int_{0}^{2\pi}\left|  F\left(
re^{i\theta}\right)  \right|  ^{2}\,d\theta
\]
and
\[
\left\langle F,G\right\rangle =\lim_{r\rightarrow1}\int_{0}^{2\pi}%
\overline{F\left(  re^{i\theta}\right)  }G\left(  re^{i\theta}\right)
\,d\theta.
\]
Using Taylor series again, it is not too hard to show that the limit defining
the inner product exists whenever $F$ and $G$ are in the Hardy space.

Even though it is not a holomorphic function space of the sort considered in
Section \ref{general.sect}, nevertheless the Hardy space has all the same
properties of those spaces: pointwise evaluation is continuous, there is a
reproducing kernel with all the properties of Theorem \ref{rep.thm}
(substituting the Hardy space itself for $\mathcal{H}L^{2}\left(
U,\alpha\right)  $ everywhere), the Hardy space is a Hilbert space. I will not
give the proofs here, but they require no more than the Cauchy integral
formula and Taylor series. See Exercise \ref{hardy.ex}.

\subsection{Exercises}

\begin{exercise}
\label{verify.0}Verify directly that the formula for the reproducing kernel of
the standard Bergman space is correct when $z=0.$ (Recall Point 6 of Theorem
\ref{rep.thm}.) Do the same for the Segal-Bargmann space with $d=1$.
\end{exercise}

\begin{exercise}
\label{hardy.ex}Show that for $F\in\mathcal{H}\left(  \mathbb{D}\right)  ,$
$\int_{0}^{2\pi}\left|  F\left(  re^{i\theta}\right)  \right|  ^{2}\,d\theta$
is increasing with $r.$ Show that for all $z\in\mathbb{D},$ there is a
constant $c_{z}$ such that
\[
\left|  F\left(  z\right)  \right|  ^{2}\leq c_{z}\lim_{r\rightarrow1}\int
_{0}^{2\pi}\left|  F\left(  re^{i\theta}\right)  \right|  ^{2}\,d\theta
\]
for all $F$ in the Hardy space.
\end{exercise}

\begin{exercise}
Compute the reproducing kernel for the Hardy space. You may assume that the
standard formula for the reproducing kernel holds, even though the Hardy space
is not a ``standard'' holomorphic function space.
\end{exercise}

\begin{exercise}
Compute the reproducing kernel for the weighted Bergman spaces.
\end{exercise}

\section{A special property of the Segal-Bargmann and weighted Bergman
spaces\label{unitary.sect}}

One may well ask why we consider the examples we did. That is, why use the
particular densities that appear in the weighted Bergman and Segal-Bargmann
spaces? Why not some other densities? We will examine here one special
property of the Segal-Bargmann spaces that holds (essentially) only for those
spaces. Something similar holds for the weighted Bergman spaces, which we will
touch on briefly.

\subsection{Unitarized translations on the Segal-Bargmann space}

\begin{definition}
Consider the Segal-Bargmann space $\mathcal{H}L^{2}\left(  \mathbb{C}^{d}%
,\mu_{t}\right)  ,$ for some $t>0.$ Now for each $a\in\mathbb{C}^{d},$ define
a linear transformation $T_{a}:\mathcal{H}L^{2}\left(  \mathbb{C}^{d},\mu
_{t}\right)  \rightarrow\mathcal{H}L^{2}\left(  \mathbb{C}^{d},\mu_{t}\right)
$ by
\[
T_{a}F\left(  z\right)  =e^{-\left|  a\right|  ^{2}/2t}e^{\bar{a}\cdot
z/t}F\left(  z-a\right)  .
\]
\end{definition}

At the moment it is not obvious that $T_{a}$ actually maps the Segal-Bargmann
space into itself. But in fact $T_{a}$ is unitary for each $a. $

\begin{theorem}
\label{ta.thm}

\begin{enumerate}
\item  For all $a\in\mathbb{C}^{d},$ $T_{a}$ is unitary on $\mathcal{H}%
L^{2}\left(  \mathbb{C}^{d},\mu_{t}\right)  .$

\item  For all $a,b\in\mathbb{C}^{d},$%
\[
T_{a}T_{b}=e^{i\operatorname{Im}\left(  a\cdot\bar{b}\right)  /t}\,T_{a+b}.
\]
\end{enumerate}
\end{theorem}

Let us discuss this theorem before proving it. To prove that $T_{a}$ is
isometric, we need only check that
\[
\left|  e^{-\left|  a\right|  ^{2}/2t}e^{\bar{a}\cdot z/t}\right|  ^{2}%
=\frac{\mu_{t}\left(  z-a\right)  }{\mu_{t}\left(  z\right)  },
\]
which is an easy calculation. (See the proof.) So $T_{a}$ is a ``unitarized
translation''; that is, it first translates $F$ by $a,$ and then multiplies by
something which makes the transformation unitary. Note that translation itself
is not unitary, since our measure $\mu_{t}$ is not translation-invariant. And
we may not use the translation-invariant measure (Lebesgue measure) because
then the only square-integrable holomorphic function would be 0. (Recall
Exercise \ref{leb.zero}.)

If we were working with ordinary (non-holomorphic) $L^{2}$ spaces, then we
could define unitarized translations for \textit{any} strictly positive
density $\alpha$, simply by taking as our ``multiplier'' (i.e., the thing that
we multiply $F\left(  z-a\right)  $ by) to be given by $\left[  \alpha\left(
z-a\right)  /\alpha\left(  z\right)  \right]  ^{1/2}.$ But if we want to map
the holomorphic subspace into itself, then \textit{the multiplier must be
holomorphic.} The special property of the Segal-Bargmann space is that there
exists a holomorphic function $\phi_{a}$ such that $\left|  \phi_{a}\left(
z\right)  \right|  ^{2}=\mu_{t}\left(  z-a\right)  /\mu_{t}\left(  z\right)
.$ (If you take just any old positive function $\gamma\left(  z\right)  ,$
then there will usually not exist any holomorphic function $\phi$ with
$\left|  \phi\left(  z\right)  \right|  ^{2}=\gamma\left(  z\right)  .$ See
Exercise \ref{log.harmonic}.) This special property holds only for spaces
which are holomorphically equivalent to one of the Segal-Bargmann spaces. See
Section \ref{unitary.equiv}.

Point 2 of the theorem says that the $T_{a}$'s multiply the way ordinary
translations do, \textit{modulo a constant}. You might think that we could
alter the definition of the $T_{a}$'s by a constant to make them multiply
exactly as ordinary translations, but this is impossible. After all, Point 2
implies that in general $T_{a}$ fails to commute with $T_{b},$ which means
that a constant times $T_{a}$ will fail to commute with a constant times
$T_{b},$ which means that we cannot have
\[
\left(  \alpha T_{a}\right)  \left(  \beta T_{b}\right)  =\gamma
T_{a+b}=\gamma T_{b+a}=\left(  \beta T_{b}\right)  \left(  \alpha
T_{a}\right)
\]
in general. Point 2 says that the $T_{a}$'s constitute a \textit{projective
unitary representation} of the additive group $\mathbb{C}^{d}.$

[A classic paper of Bargmann \cite{B4} determines for which groups it is
always possible to choose constants so that a projective unitary
representation becomes an ordinary unitary representation. We see that
$\mathbb{C}^{d}$ is not such a group. The rotation group $SO\left(  3\right)
$ is not such a group either, but $SU\left(  2\right)  $ is. This theory helps
to explain the physical significance of ``spin'' in quantum physics.]

\textit{Proof of Theorem \ref{ta.thm}.} (1) Recall as in the formula for the
reproducing kernel that $\bar{a}\cdot z=\bar{a}_{1}z_{1}+\cdots+\bar{a}%
_{d}z_{d}.$ This is a complex-valued quantity, whose real part is
$\operatorname{Re}\bar{a}\cdot z=\sum_{k=1}^{d}\operatorname{Re}%
a_{k}\operatorname{Re}z_{k}+\operatorname{Im}a_{k}\operatorname{Im}z_{k}.$
Thus
\begin{align*}
\left|  z-a\right|  ^{2}  & =\sum_{k=1}^{d}\overline{\left(  z_{k}%
-a_{k}\right)  }\left(  z_{k}-a_{k}\right) \\
& =\sum_{k=1}^{d}\left(  \bar{z}_{k}z_{k}+\bar{a}_{k}a_{k}-\bar{a}_{k}%
z_{k}-\bar{z}_{k}a_{k}\right) \\
& =\left|  z\right|  ^{2}+\left|  a\right|  ^{2}-2\operatorname{Re}\bar
{a}\cdot z.
\end{align*}

Now by brute force calculation we find that
\begin{align*}
\left|  e^{-\left|  a\right|  ^{2}/2t}e^{\bar{a}\cdot z/t}F\left(  z-a\right)
\right|  ^{2}e^{-\left|  z\right|  ^{2}/t}  & =e^{-\left|  a\right|  ^{2}%
/t}e^{2\operatorname{Re}\left(  \bar{a}\cdot z\right)  /t}e^{-\left|
z\right|  ^{2}/t}\left|  F\left(  z-a\right)  \right|  ^{2}\\
& =e^{-\left|  z-a\right|  ^{2}/t}\left|  F\left(  z-a\right)  \right|  ^{2}.
\end{align*}
Multiplying by $\left(  \pi t\right)  ^{-d}$ and integrating shows that
$\left\|  T_{a}F\right\|  ^{2}=\left\|  F\right\|  ^{2}.$ Thus $T_{a}$ is an
isometric map of $\mathcal{H}L^{2}\left(  \mathbb{C}^{d},\mu_{t}\right)  $ to
itself. The invertibility of $T_{a}$ will follow from Point 2, which implies
(with $b=-a$) that $\left(  T_{a}\right)  ^{-1}$ is $T_{-a}.$

(2) We compute that
\begin{align*}
T_{b}F\left(  z\right)   & =e^{-\left|  b\right|  ^{2}/2t}e^{\bar{b}\cdot
z/t}F\left(  z-b\right) \\
T_{a}T_{b}F\left(  z\right)   & =e^{-\left|  a\right|  ^{2}/2t}e^{\bar{a}\cdot
z/t}e^{-\left|  b\right|  ^{2}/2t}e^{\bar{b}\cdot\left(  z-a\right)
/t}F\left(  z-a-b\right) \\
& =e^{-\left|  a\right|  ^{2}/2t}e^{-\left|  b\right|  ^{2}/2t}e^{-\bar
{b}\cdot a/t}e^{\left(  \bar{a}+\bar{b}\right)  \cdot z/t}F\left(  z-\left(
a+b\right)  \right)  .
\end{align*}
But $\left|  a+b\right|  ^{2}=\left|  a\right|  ^{2}+\left|  b\right|
^{2}+2\operatorname{Re}\bar{b}\cdot a.$ Thus the first three factors in the
expression for $T_{a}T_{b}$ will combine to give $\exp\left(  -\left|
a+b\right|  ^{2}/2t\right)  ,$ with a leftover factor of $\exp\left(
-i\operatorname{Im}\left(  a\cdot\bar{b}\right)  /t\right)  .$ So
\[
T_{a}T_{b}F\left(  z\right)  =e^{-i\operatorname{Im}\left(  a\cdot\bar
{b}\right)  /t}T_{a+b}F\left(  z\right)  ,
\]
which is what we want to prove. \qed

One can use these unitarized translations to give another derivation of the
reproducing kernel of the Segal-Bargmann space, as follows. I do this for the
case $d=1$; the general case can be reduced to this case as in Section
\ref{examples.sect}. Using polar coordinates and Taylor series one may easily
prove (Exercise \ref{verify.0} of Section \ref{examples.sect}) that for any
$F$ in the Segal-Bargmann space
\[
F\left(  0\right)  =\int_{\mathbb{C}}1\cdot F\left(  w\right)  \,\mu
_{t}\left(  w\right)  \,dw,
\]
or equivalently,
\[
F\left(  0\right)  =\left\langle \mathbf{1},F\right\rangle
\]
where $\mathbf{1}$ denotes the constant function identically equal to one, and
the inner product is in $\mathcal{H}L^{2}\left(  \mathbb{C}^{d},\mu
_{t}\right)  .$

But then
\begin{align*}
\left(  T_{-a}F\right)  \left(  0\right)   & =\left\langle \mathbf{1}%
,T_{-a}F\right\rangle \\
& =\left\langle T_{a}\mathbf{1},T_{a}T_{-a}F\right\rangle \\
& =\left\langle T_{a}\mathbf{1},F\right\rangle ,
\end{align*}
where we have used the unitarity of $T_{a}$ and Point 2 of Theorem
\ref{ta.thm} (with $b=-a$) to show that $T_{a}=\left(  T_{-a}\right)  ^{-1}.$
Thus we have
\[
e^{-\left|  a\right|  ^{2}/2t}e^{-\bar{a}\cdot z/t}F\left(  z+a\right)
_{z=0}=e^{-\left|  a\right|  ^{2}/2t}F\left(  a\right)  =\left\langle
T_{a}\mathbf{1},F\right\rangle .
\]
So
\[
F\left(  a\right)  =\left\langle e^{+\left|  a\right|  ^{2}/2t}T_{a}%
\mathbf{1},F\right\rangle .
\]
By Point 6 of Theorem \ref{rep.thm} we have
\begin{align*}
K\left(  a,u\right)   & =e^{+\left|  a\right|  ^{2}/2t}T_{a}\mathbf{1}\left(
u\right)  =e^{\left|  a\right|  ^{2}/2t}e^{-\left|  a\right|  ^{2}/2t}%
e^{\bar{a}\cdot u}\\
& =e^{\bar{a}\cdot u}.
\end{align*}

\subsection{Unitarized transformations of the weighted Bergman
spaces\label{su11.sect}}

There is an analogous theory for the weighted Bergman spaces. Recall the
definition of fractional linear transformations. These are transformations of
the form
\begin{equation}
z\rightarrow\frac{az+b}{cz+d},\label{fract.trans}%
\end{equation}
with $ad-bc=1.$ (You could allow constants with $ad-bc\neq1,$ but you don't
get any new transformations this way.) We wish to consider those fractional
linear transformations that map the unit disk $\mathbb{D}$ onto itself. These
are precisely the fractional linear transformations where $a,b,c,d$ form a
matrix of the form
\begin{equation}
\left\{  \left.  \left(
\begin{array}
[c]{cc}%
a & b\\
\bar{b} & \bar{a}%
\end{array}
\right)  \right|  \left|  a\right|  ^{2}-\left|  b\right|  ^{2}=1\right\}
.\label{su11}%
\end{equation}
The set of matrices of this form make up a group denoted $SU\left(
1,1\right)  .$ If $g$ is a matrix of the form (\ref{su11}) we will let $g\cdot
z$ denote the result of the corresponding fractional linear transformation,
namely, $\left(  az+b\right)  /\left(  \bar{b}z+\bar{a}\right)  . $

\begin{theorem}
Fix a number $a>-1$ and consider the weighted Bergman spaces
\[
\mathcal{H}L^{2}\left(  \mathbb{D},\left(  1-\left|  z\right|  ^{2}\right)
^{a}\right)  .
\]
For each $g\in SU\left(  1,1\right)  $ there exists a holomorphic function
$\phi_{g},$ unique up to a constant, such that the map
\[
U_{g}F\left(  z\right)  =\phi_{g}\left(  z\right)  F\left(  g^{-1}\cdot
z\right)
\]
is unitary on $\mathcal{H}L^{2}\left(  \mathbb{D},\left(  1-\left|  z\right|
^{2}\right)  ^{a}\right)  .$ For all $g,h\in SU\left(  1,1\right)  $ there
exists a real number $\theta$ such that
\[
U_{g}U_{h}=e^{i\theta}U_{gh}.
\]
\end{theorem}

Here $g^{-1}$ denotes a matrix inverse and $gh$ denotes a matrix product. I
will not prove this theorem. A similar theory holds for weighted Bergman
spaces on the unit ball in $\mathbb{C}^{d},$ with the group $SU\left(
n,1\right)  $ replacing $SU\left(  1,1\right)  ,$ and more generally for
``bounded symmetric domains.''

[The operators $U_{g}$ constitute a projective unitary representation of
$SU\left(  1,1\right)  .$ For certain discrete values of $a,$ this projective
unitary representation can be made into an ordinary unitary representation.
That is, for certain values of $a$ one can choose constants so that
$U_{g}U_{h}=U_{gh}$ for all $g$ and $h.$ The resulting unitary representations
of $SU\left(  1,1\right)  $ are called the \textit{holomorphic discrete
series}, first described by (who else?) Bargmann \cite{B5}.]

\subsection{Holomorphic equivalence\label{unitary.equiv}}

Let us return briefly to the general setting $\mathcal{H}L^{2}\left(
U,\alpha\right)  $. Let $\phi$ be a nowhere-zero holomorphic function on $U. $
Then
\[
\int_{U}\left|  F\left(  z\right)  \right|  ^{2}\alpha\left(  z\right)
\,dz=\int_{U}\left|  \phi\left(  z\right)  F\left(  z\right)  \right|
^{2}\frac{1}{\left|  \phi\left(  z\right)  \right|  ^{2}}\alpha\left(
z\right)  \,dz.
\]
So the map $F\rightarrow\phi F$ is a unitary map of $\mathcal{H}L^{2}\left(
U,\alpha\right)  $ onto $\mathcal{H}L^{2}\left(  U,\alpha/\left|  \phi\right|
^{2}\right)  ,$ whose inverse is the map $G\rightarrow\frac{1}{\phi}G.$

\begin{definition}
The holomorphic function spaces $\mathcal{H}L^{2}\left(  U,\alpha\right)  $
and $\mathcal{H}L^{2}\left(  U,\beta\right)  $ are said to be
\textbf{holomorphically equivalent} if there exists a nowhere zero holomorphic
function $\phi$ on $U$ such that
\[
\beta\left(  z\right)  =\frac{\alpha\left(  z\right)  }{\left|  \phi\left(
z\right)  \right|  ^{2}}
\]
for all $z\in U.$ The \textbf{holomorphic equivalence} between $\mathcal{H}%
L^{2}\left(  U,\alpha\right)  $ and $\mathcal{H}L^{2}\left(  U,\beta\right)  $
is the unitary map $F\rightarrow\phi F.$
\end{definition}

\begin{theorem}
\label{equiv.thm1}Suppose $\alpha$ is a strictly positive smooth function on
$\mathbb{C}$ such that:

\begin{enumerate}
\item $\mathcal{H}L^{2}\left(  \mathbb{C},\alpha\right)  $ contains at least
one non-zero function, and

\item  For all $a\in\mathbb{C}$ there exists a holomorphic function $\phi_{a}$
such that the map
\[
T_{a}F\left(  z\right)  =\phi_{a}\left(  z\right)  F\left(  z-a\right)
\]
is unitary on $\mathcal{H}L^{2}\left(  \mathbb{C},\alpha\right)  .$
\end{enumerate}

Then $\mathcal{H}L^{2}\left(  \mathbb{C},\alpha\right)  $ is holomorphically
equivalent to one of the Segal-Bargmann spaces.
\end{theorem}

The proof of this theorem is left as an exercise. You should use Exercise
\ref{log.harmonic}. Then let $\beta\left(  z\right)  =\log\alpha\left(
z\right)  .$ If the hypotheses of the theorem hold, show that $\Delta
\beta\left(  z\right)  =c$ (a constant). Here $\Delta$ is the standard
Laplacian operator, $\Delta=\partial^{2}/\partial x^{2}+\partial^{2}/\partial
y^{2}. $

\begin{theorem}
\label{equiv.thm2}Suppose $\alpha$ is a strictly positive smooth function on
$\mathbb{D}$ such that

\begin{enumerate}
\item $\mathcal{H}L^{2}\left(  \mathbb{D},\alpha\right)  $ contains at least
one non-zero function, and

\item  For all $g\in SU\left(  1,1\right)  $ there exists a holomorphic
function $\phi_{g}$ such that the map
\[
U_{g}F\left(  z\right)  =\phi_{g}\left(  z\right)  F\left(  g^{-1}\cdot
z\right)
\]
is unitary on $\mathcal{H}L^{2}\left(  \mathbb{D},\alpha\right)  .$
\end{enumerate}

Then $\mathcal{H}L^{2}\left(  \mathbb{D},\alpha\right)  $ is holomorphically
equivalent to one of the weighted Bergman spaces.
\end{theorem}

The proof is a starred exercise. The proof is similar to that of Theorem
\ref{equiv.thm1}, except that you need to think ``hyperbolically.'' This means
that you should express things in terms of the hyperbolic volume measure
$\left(  1-\left|  z\right|  ^{2}\right)  ^{-2}\,dz,$ which is invariant under
the action of $SU\left(  1,1\right)  ,$ and in terms of the hyperbolic
Laplacian, $\Delta_{H}=\left(  1-\left|  z\right|  ^{2}\right)  ^{2}\Delta,$
which commutes with the action of $SU\left(  1,1\right)  .$ You may assume
these properties of the hyperbolic volume measure and the hyperbolic Laplacian.

\subsection{Exercises}

\begin{exercise}
\label{log.harmonic}Let $U$ be a an open, simply connected set in
$\mathbb{C}^{1},$ and let $\alpha$ be a strictly positive smooth function on
$U.$ Show that there exists a holomorphic function $\phi$ with $\left|
\phi\right|  ^{2}=\alpha$ if and only if $\log\alpha$ is harmonic.
\end{exercise}

\begin{exercise}
Prove Theorem \ref{equiv.thm1}, using the hints given after the statement of
the theorem.
\end{exercise}

\begin{exercise}
*Prove Theorem \ref{equiv.thm2}, using the hints given after the statement of
the theorem.
\end{exercise}

\section{Canonical commutation relations\label{ccr.sect}}

\subsection{The standard form of the canonical commutation relations}

Let us make a brief digression from things holomorphic to consider the matter
of the ``canonical commutation relations,'' which Bargmann used to derive the
Segal-Bargmann transform. The transform itself will make its entrance in the
next section. So let us consider the Hilbert space $L^{2}\left(
\mathbb{R},dx\right)  ,$ and two (unbounded) linear operators on it, denoted
$X$ and $P$ and given by
\begin{align*}
Xf\left(  x\right)   & =xf\left(  x\right) \\
Pf\left(  x\right)   & =-i\hbar\frac{df}{dx}.
\end{align*}
Here $\hbar$ (pronounced ``aitch-bar'') is Planck's constant, which is a
positive constant. We will say more about $\hbar$ in Section
\ref{quantum.sect}. Note that $X$ and $P$ are not defined on all of
$L^{2}\left(  \mathbb{R},dx\right)  ,$ since $L^{2}$ functions are not
necessarily differentiable, and since $xf\left(  x\right)  $ may not be in
$L^{2},$ even if $f\left(  x\right)  $ is. Thus properly speaking $X$ and $P$
are defined on certain domains, which are dense subspaces of $L^{2}.$ However,
I am not going to worry (much) about such domain issues here, since I want to
convey the basic ideas without becoming bogged down in functional-analytic
technicalities. Ignoring domain issues, it is easily seen that $X$ and $P$ are
self-adjoint operators on $L^{2}\left(  \mathbb{R},dx\right)  .$ (The classic
book of Reed and Simon \cite{RS} is a good place to start on such matters.)

Let us now compute the commutator of $X$ and $P.$ (Recall that the commutator
of two operators $A$ and $B,$ denoted $\left[  A,B\right]  ,$ is defined by
$\left[  A,B\right]  =AB-BA.$ The commutator measures the extent to which $A$
and $B$ fail to commute.) So
\begin{align*}
\left[  X,P\right]  f  & =x\left(  -i\hbar\frac{df}{dx}\right)  +i\hbar
\frac{d}{dx}\left(  xf\left(  x\right)  \right) \\
& =-i\hbar x\frac{df}{dx}+i\hbar\left(  \frac{dx}{dx}f\left(  x\right)
+x\frac{df}{dx}\right) \\
& =i\hbar f\left(  x\right)  .
\end{align*}
That is,
\begin{equation}
\left[  X,P\right]  =i\hbar I,\label{ccr1}%
\end{equation}
where $I$ is the identity operator. The relation (\ref{ccr1}) is called the
\textbf{canonical commutation relation}, abbreviated CCR. The quantum
mechanical idea behind this relation will be explained in Section
\ref{quantum.sect}. The operator $X$ is called the \textbf{position operator}
and the operator $P$ is called the \textbf{momentum operator}, for reasons to
be explained in Section \ref{quantum.sect}. (Of course, we know that CCR
stands for Creedence Clearwater Revival, but we will allow it also to stand
for canonical commutation relation.)

There are position and momentum operators in $L^{2}\left(  \mathbb{R}%
^{d},dx\right)  ,$ $X_{k}$ and $P_{k},$ $k=1,\cdots,d,$ given by
\begin{align}
X_{k}f\left(  x\right)   & =x_{k}f\left(  x\right) \nonumber\\
P_{k}f\left(  x\right)   & =-i\hbar\frac{\partial f}{\partial x_{k}%
},\label{xkpk}%
\end{align}
where $x=\left(  x_{1},\cdots,x_{d}\right)  .$ The relations among these are
\begin{align}
\left[  X_{k},X_{l}\right]   & =0\nonumber\\
\left[  P_{k},P_{l}\right]   & =0\label{ccr2}\\
\left[  X_{k},P_{l}\right]   & =i\hbar\delta_{k,l}I,\nonumber
\end{align}
where $\delta_{k,l}$ is the Kronecker delta function, given by $\delta
_{k,l}=0$ if $k\neq l$ and $\delta_{k,l}=1$ if $k=l.$ These are the
$d$-dimensional version of the canonical commutation relations. Of course
these imply that $\left[  P_{l},X_{k}\right]  =-i\hbar\delta_{k,l}.$

It is convenient to re-write the canonical commutation relations in terms of
the so-called \textbf{creation and annihilation operators} (also called
\textbf{raising and lowering operators}) defined by
\begin{align*}
a_{k}  & =\frac{X_{k}+iP_{k}}{\sqrt{2}}\\
a_{k}^{\ast}  & =\frac{X_{k}-iP_{k}}{\sqrt{2}}.
\end{align*}
Note that since $X_{k}$ and $P_{k}$ are self-adjoint, then $a_{k}^{\ast}$ is
the adjoint of $a_{k},$ which is what the star is supposed to mean. We compute
that
\begin{align*}
\left[  a_{k},a_{l}\right]   & =\frac{1}{2}\left[  X_{k}+iP_{k},X_{l}%
+iP_{l}\right] \\
& =\frac{1}{2}\left(  \left[  X_{k},X_{l}\right]  +i\left[  P_{k}%
,X_{l}\right]  +i\left[  X_{k},P_{l}\right]  -\left[  P_{k},P_{l}\right]
\right) \\
& =\frac{1}{2}\left(  0+\hbar\delta_{k,l}I-\hbar\delta_{k,l}I+0\right)  =0,
\end{align*}
and similarly that $\left[  a_{k}^{\ast},a_{l}^{\ast}\right]  =0.$ Meanwhile,
\begin{align*}
\left[  a_{k},a_{l}^{\ast}\right]   & =\frac{1}{2}\left[  X_{k}+iP_{k}%
,X_{l}-iP_{l}\right] \\
& =\frac{1}{2}\left(  \left[  X_{k},X_{l}\right]  +i\left[  P_{k}%
,X_{l}\right]  -i\left[  X_{k},P_{l}\right]  +\left[  P_{k},P_{l}\right]
\right) \\
& =\frac{1}{2}\left(  0+\hbar\delta_{k,l}I+\hbar\delta_{k,l}I+0\right)
=\hbar\delta_{k,l}I.
\end{align*}

So the canonical commutation relations take the form
\begin{align}
\left[  a_{k},a_{l}\right]   & =0\nonumber\\
\left[  a_{k}^{\ast},a_{l}^{\ast}\right]   & =0\label{ccr3}\\
\left[  a_{k},a_{l}^{\ast}\right]   & =\hbar\delta_{k,l}I.\nonumber
\end{align}
Sometimes only the third of these is written, with other commutators being
understood to be zero. I have defined the creation and annihilation operators
in a slightly unconventional way by not absorbing the $\hbar$ into the
definition of the operators. That is, conventionally the $\sqrt{2}$ in the
denominator of $a$ and $a^{\ast}$ is replaced by $\sqrt{2\hbar}.$ This would
eliminate the factor of $\hbar$ in the CCRs. I would prefer to dispense as
well with the factor of $\sqrt{2}$ in the denominator, for reasons of my own,
but I don't want to break too much with tradition.

We now turn to the famous Stone-von Neumann Theorem, which explains the
significance of these canonical commutation relations. The idea is to consider
\textit{any} Hilbert space $H$ and \textit{any} collection $\left\{
a_{1},\cdots,a_{d}\right\}  $ of operators, together with their adjoints
$\left\{  a_{1}^{\ast},\cdots,a_{d}^{\ast}\right\}  $, which satisfy the CCRs
in the form (\ref{ccr3}). Unfortunately, the statement of this theorem usually
given in physics is false, because of those annoying domain issues which we
are ignoring. Nevertheless, I want to state the theorem in this imprecise way
first, since this way is easier to grasp. But keep in mind that the actual
theorem is different. See \cite[Chap. VIII.5, Ex. 2]{RS} for a counter-example.

\begin{claim}
[Stone-von Neumann Non-Theorem]\label{svn1}Let $H$ be a Hilbert space, let
$a_{1},\cdots,a_{d}$ be possibly unbounded operators on $H,$ and let
$a_{1}^{\ast},\cdots,a_{d}^{\ast}$ be the adjoints of the $a_{k}$'s. Suppose that

\begin{enumerate}
\item (CCRs) For all $k,l,$ $\left[  a_{k},a_{l}\right]  =\left[  a_{k}^{\ast
},a_{l}^{\ast}\right]  =0$ and $\left[  a_{k},a_{l}^{\ast}\right]
=\hbar\delta_{k,l}I,$ and

\item (Irreducibility) If $V$ is a closed subspace of $H$ which is invariant
under all the $a_{k}$'s and $a_{k}^{\ast}$'s, then either $V=\left\{
0\right\}  $ or $V=H.$
\end{enumerate}

Then there exists a unitary map (unique up to a constant) $U:H\rightarrow
L^{2}\left(  \mathbb{R}^{d},dx\right)  $ such that
\begin{align*}
Ua_{k}U^{-1}  & =\frac{X_{k}+iP_{k}}{\sqrt{2}}\\
Ua_{k}^{\ast}U^{-1}  & =\frac{X_{k}-iP_{k}}{\sqrt{2}},
\end{align*}
where $X_{k}$ and $P_{k}$ are given by (\ref{xkpk}).
\end{claim}

This non-theorem (we will have a real theorem momentarily) says that up to
unitary equivalence there is only one irreducible representation of the CCRs.
This result is important in quantum mechanics because it helps to justify the
use of the Hilbert space $L^{2}\left(  \mathbb{R}^{d},dx\right)  $ and the
operators (\ref{xkpk}). After all, if you had any other Hilbert space with
operators satisfying the CCRs (and irreducibility) it would be unitarily
equivalent to $L^{2}\left(  \mathbb{R}^{d},dx\right)  $ in a way that took
these operators to the standard creation and annihilation operators.

\subsection{The exponentiated form of the canonical commutation relations}

To get the correct form of the Stone-von Neumann theorem one needs to consider
\textit{exponentiated} operators, which are bounded, meaning that you don't
have to worry about domain issues. So let $X_{k}$ and $P_{k}$ be the operators
in (\ref{xkpk}), and consider $e^{isX_{k}/\hbar}$ and $e^{itP_{l}/\hbar}.$ I
won't go into the details, but these are everywhere-defined unitary operators
for each $s$ and $t.$ These can be computed explicitly as follows
\begin{align}
e^{irX_{k}/\hbar}f\left(  x\right)   & =e^{irx_{k}/\hbar}f\left(  x\right)
\label{exp1}\\
e^{isP_{l}/\hbar}f\left(  x\right)   & =f\left(  x_{1},\cdots,x_{l-1}%
,x_{l}+s,x_{l+1},\cdots,x_{d}\right)  .\label{exp2}%
\end{align}
Equation (\ref{exp1}) is clear, at least formally. To understand (\ref{exp2})
formally, expand the right side in a Taylor series in powers of $s$ (without
worrying about convergence). The terms in this expansion will precisely match
the formal power series of the left side obtained by writing $e^{isP_{l}%
/\hbar}=\sum_{n=0}^{\infty}\left(  isP_{l}/\hbar\right)  ^{n}/n!. $ Let's see
what relations hold for these exponentiated operators. I will take $d=1$ for
simplicity. Then
\[
e^{irX/\hbar}e^{isP/\hbar}f\left(  x\right)  =e^{irx/\hbar}f\left(
x+s\right)
\]
and
\begin{align*}
e^{isP/\hbar}e^{irX/\hbar}  & =e^{isP/\hbar}\left[  e^{irx/\hbar}f\left(
x\right)  \right] \\
& =e^{ir\left(  x+s\right)  /\hbar}f\left(  x+s\right) \\
& =e^{irs}e^{irX/\hbar}e^{isP/\hbar}f\left(  x\right)  .
\end{align*}
There are no domain issues here; this holds unambiguously for \textit{all} $f
$ in $L^{2}.$

We see that
\[
e^{irX/\hbar}e^{isP/\hbar}=e^{-irs/\hbar}e^{isP/\hbar}e^{irX/\hbar}.
\]
In $d$ dimensions we have
\begin{align}
e^{irX_{k}/\hbar}e^{isX_{l}/\hbar}  & =e^{isX_{l}/\hbar}e^{irX_{k}/\hbar
}\nonumber\\
e^{irP_{k}/\hbar}e^{isP_{l}/\hbar}  & =e^{isP_{l}/\hbar}e^{irP_{k}/\hbar
}\label{ccr4}\\
e^{irX_{k}/\hbar}e^{isP_{l}/\hbar}  & =e^{-irs\delta_{k,l}/\hbar}%
e^{isP_{l}/\hbar}e^{irX_{k}/\hbar}\nonumber
\end{align}
The equations (\ref{ccr4}) are the exponentiated form of the CCRs. One can
\textit{formally} derive the exponentiated form of the CCRs from the original
form (\ref{ccr2}) (Exercise \ref{ccr.exp}). However, this derivation is only
formal and the conclusion is not actually correct. That is, (\ref{ccr2}) does
not actually imply (\ref{ccr4}), without making additional domain assumptions.
But this should be regarded as a technicality.

We may now state the correct form of the Stone-von Neumann Theorem. (A
slightly different form of this is given in \cite[Thm. VIII.14]{RS}.)

\begin{theorem}
[Stone-von Neumann Theorem]\label{svn2}Suppose that $A_{1},\cdots,A_{d}$ and
$B_{1},\cdots,B_{d}$ are (possibly unbounded) self-adjoint operators on a
Hilbert space $H$ satisfying:

\begin{enumerate}
\item (CCRs) For all $k$ and $l$ and $r$ and $s,$ $e^{irA_{k}/\hbar}$ commutes
with $e^{isA_{l}/\hbar},$ $e^{irB_{k}/\hbar}$ commutes with $e^{isB_{l}/\hbar
},$ and
\[
e^{irA_{k}/\hbar}e^{isB_{l}/\hbar}=e^{-irs\delta_{k,l}/\hbar}e^{isB_{l}/\hbar
}e^{irA_{k}/\hbar}.
\]

\item (Irreducibility) If $V$ is a closed subspace of $H$ invariant under
$e^{irA_{k}/\hbar}$ and $e^{isB_{l}/\hbar}$ for all $k$ and $l$ and $s$ and
$t,$ then $V=\left\{  0\right\}  $ or $V=H.$
\end{enumerate}

Then there is a unitary map (unique up to a constant) $U:H\rightarrow
L^{2}\left(  \mathbb{R}^{d},dx\right)  $ such that
\begin{align*}
Ue^{irA_{k}/\hbar}U^{-1}  & =e^{irX_{k}/\hbar}\\
Ue^{isB_{l}/\hbar}U^{-1}  & =e^{isP_{l}/\hbar},
\end{align*}
where $e^{irX_{k}/\hbar}$ and $e^{isP_{l}/\hbar}$ are given by (\ref{exp1})
and (\ref{exp2}).
\end{theorem}

\subsection{Exercises}

\begin{exercise}
Let $\tilde{X}_{k}=P_{k}$ and $\tilde{P}_{k}=-X_{k}.$ Show that $\tilde{X}%
_{k}$ and $\tilde{P}_{k}$ satisfy the CCRs.
\end{exercise}

\begin{exercise}
In the notation of the previous exercise, describe the unitary transformation
$U:L^{2}\left(  \mathbb{R}^{d}\right)  \rightarrow L^{2}\left(  \mathbb{R}%
^{d}\right)  $ such that $U\tilde{X}_{k}U^{-1}=X_{k}$ and $U\tilde{P}%
_{k}U^{-1}=P_{k}.$
\end{exercise}

\begin{exercise}
Suppose $A$ and $B$ are $n\times n$ matrices, for some (finite) number $n.$
Suppose that
\[
\left[  A,B\right]  =cI
\]
for some constant $c.$ Show that $c$ must equal zero. So the CCRs cannot hold
for matrices.
\end{exercise}

\begin{exercise}
\label{ccr.exp}* a) Suppose that $A$ and $B$ are $n\times n$ matrices such
that
\[
\left[  A,\left[  A,B\right]  \right]  =\left[  B,\left[  A,B\right]  \right]
=0.
\]
(That is, $A$ and $B$ commute with their commutator.) Show that
\[
e^{A}e^{B}=e^{A+B+\frac{1}{2}\left[  A,B\right]  }.
\]
Hint: Show that $e^{sA}e^{sB}e^{-s^{2}\left[  A,B\right]  /2}$ satisfies the
same ordinary differential equation as $e^{s\left(  A+B\right)  }.$

b) If $A$ and $B$ are as in (a) show that
\[
e^{irA/\hbar}e^{isB/\hbar}=e^{-irs/\hbar}e^{isB/\hbar}e^{irA/\hbar}.
\]

c) Explain why this gives a \textbf{formal} argument that the CCRs in form
(\ref{ccr2}) should imply the exponentiated CCRs (\ref{ccr4}).

Of course, (\ref{ccr2}) does not actually imply (\ref{ccr4}) without
additional domain conditions.
\end{exercise}

\begin{exercise}
\label{svn.proof}* This exercise guides you through a \textbf{formal} argument
for the conventional (non-theorem) form of Stone-von Neumann. (This argument
cannot be made rigorous without additional domain assumptions.) We will
consider only $d=1,$ though the general case is nearly the same. In all cases,
``show'' should be understood as ``show ignoring domain issues.'' So let $H$
be any Hilbert space and let $a$ and $a^{\ast}$ satisfy the conditions of
Claim \ref{svn1}.

a) Show that $E:=a^{\ast}a$ is self-adjoint and positive.

b) Let $v\neq0$ be any eigenvector for $E$ with eigenvalue $\lambda.$ Show
that
\[
Eav=\left(  \lambda-1\right)  av.
\]

c) Show that there exists $k\geq0$ such that $a^{k}v\neq0$ but $a^{k+1}v=0. $
Hint: $E$ is positive.

d) Let $u=a^{k}v,$ so that $u\neq0$ but $au=0$ and so also $Eu=0.$ Show
inductively that
\[
E\left(  a^{\ast}\right)  ^{n}u=n\left(  a^{\ast}\right)  ^{n}u,\quad
n=0,1,\cdots.
\]

e) Show inductively that
\[
a\left(  a^{\ast}\right)  ^{n}u=n\left(  a^{\ast}\right)  ^{n-1}u,\quad
n=0,1,\cdots.
\]

f) Show that
\begin{align*}
\left\langle \left(  a^{\ast}\right)  ^{n}u,\left(  a^{\ast}\right)
^{n}u\right\rangle  & =n!\left\langle u,u\right\rangle ,\quad n=0,1,\cdots\\
\left\langle \left(  a^{\ast}\right)  ^{n}u,\left(  a^{\ast}\right)
^{m}u\right\rangle  & =0,\quad n\neq m.
\end{align*}

g) Show that the closed span of $\left\{  \left(  a^{\ast}\right)
^{n}u\right\}  $ is all of $H.$

h) If $\left(  K,b,b^{\ast}\right)  $ is any other irreducible representation
of the CCRs, show that $K$ is unitarily equivalent to $H.$ Hint: let
$\tilde{u}$ be the analog of $u$ in $K$ (chosen to have the same norm as $u$).
Define $U:K\rightarrow H$ so that $U\tilde{u}=u$ and $U\left(  b^{\ast
}\right)  ^{n}\tilde{u}=\left(  a^{\ast}\right)  ^{n}u.$
\end{exercise}

\section{The Segal-Bargmann transform\label{sbt.sect}}

\subsection{Bargmann's extension of Fock's observation}

Let us consider for now just the $d=1$ case of the CCRs. Fock (1928) made the
following observation. Consider the space $\mathcal{H}\left(  \mathbb{C}%
\right)  $ of holomorphic functions on $\mathbb{C}.$ Consider the operators
$z$ and $\hbar\,d/dz$ on $\mathcal{H}\left(  \mathbb{C}\right)  ,$ where $z$
denotes multiplication by $z.$ Fock observed that
\begin{align*}
\left[  \hbar\frac{d}{dz},z\right]  f  & =\hbar\frac{d}{dz}\left(  zf\left(
z\right)  \right)  -\hbar z\frac{df}{dz}\\
& =\hbar f\left(  z\right)  +\hbar z\frac{df}{dz}\left(  z\right)  -\hbar
z\frac{df}{dz}\\
& =\hbar f\left(  z\right)  .
\end{align*}
That is,
\[
\left[  \hbar\frac{d}{dz},z\right]  =\hbar I.
\]
Thus $\hbar\,d/dz$ and multiplication by $z$ have the same commutation
relations as the annihilation and creation operators.

However, this does not constitute a representation of the canonical
commutation relations. After all, the CCRs require that we have a Hilbert
space $H$ and operators $a$ and $a^{\ast}$ \textit{that are adjoints of one
another} satisfying $\left[  a,a^{\ast}\right]  =\hbar I.$ Bargmann in
\cite{B1} sought an inner product on $\mathcal{H}\left(  \mathbb{C}\right)  $
which would make $z$ and $\hbar\,d/dz$ adjoints of one another. It is not too
hard to work out what such an inner product would have to be (see \cite{B1});
it turns out to be the inner product on the Segal-Bargmann space
$\mathcal{H}L^{2}\left(  \mathbb{C},\mu_{\hbar}\right)  .$ Note here that we
are identifying the parameter $t$ in the Segal-Bargmann space with Planck's
constant $\hbar.$ So we have the following result, stated as usual without
specifying domains. (See \cite{B1} for a precise statement.)

\begin{theorem}
\label{adjoint.thm}In the Segal-Bargmann space $\mathcal{H}L^{2}\left(
\mathbb{C}^{d},\mu_{\hbar}\right)  ,$%
\[
\left(  z_{k}\right)  ^{\ast}=\hbar\frac{\partial}{\partial z_{k}},
\]
where $z_{k}$ denotes multiplication by $z_{k}$ and * denotes the adjoint with
respect to the inner product on $\mathcal{H}L^{2}\left(  \mathbb{C}^{d}%
,\mu_{\hbar}\right)  .$
\end{theorem}

\textit{Remark.} In the theory of spherical harmonics one needs an inner
product on the space $\mathcal{P}\left(  \mathbb{R}^{d}\right)  $ of
polynomials in $\mathbb{R}^{d}$ with the property that $\left\langle \left(
x_{1}^{2}+\cdots x_{d}^{2}\right)  p,q\right\rangle =\left\langle p,\Delta
q\right\rangle $ for all polynomials $p$ and $q.$ Such an inner product is
obtained by analytically continuing $p$ and $q$ to $\mathbb{C}^{d}$ and then
using the inner product on $L^{2}\left(  \mathbb{C}^{d},\mu_{1}\right)  .$

\textit{Proof (modulo domain issues).} An alternative proof is given in
Exercise \ref{adjoint.ex}. We will for notational simplicity consider only the
$d=1$ case. Recall the definition of the operators $\partial/\partial z$ and
$\partial/\partial\bar{z},$ acting on not-necessarily-holomorphic functions on
$\mathbb{C}:$%
\begin{align*}
\frac{\partial}{\partial z}  & =\frac{1}{2}\left(  \frac{\partial}{\partial
x}-i\frac{\partial}{\partial y}\right) \\
\frac{\partial}{\partial\bar{z}}  & =\frac{1}{2}\left(  \frac{\partial
}{\partial x}+i\frac{\partial}{\partial y}\right)  .
\end{align*}
A $\mathcal{C}^{1}$ function on $\mathbb{C}$ is holomorphic if and only if
$\partial f/\partial\bar{z}=0.$ (This is equivalent to the Cauchy-Riemann
equations.) If $f$ is holomorphic, then the usual complex derivative $df/dz$
coincides with $\partial f/\partial z.$

So now assume that $F$ and $G$ are in $\mathcal{H}L^{2}\left(  \mathbb{C}%
,\mu_{\hbar}\right)  ,$ and consider
\begin{align*}
\left\langle \frac{\partial F}{\partial z},G\right\rangle  & =\frac{1}{2}%
\int_{\mathbb{C}}\overline{\left(  \frac{\partial F}{\partial x}%
-i\frac{\partial F}{\partial y}\right)  }G\left(  z\right)  \mu_{\hbar}\left(
z\right)  \,dz\\
& =\frac{1}{2}\int_{\mathbb{C}}\left(  \frac{\partial\bar{F}}{\partial
x}+i\frac{\partial\bar{F}}{\partial y}\right)  G\left(  z\right)  \mu_{\hbar
}\left(  z\right)  \,dz.
\end{align*}
We now want to integrate by parts. I will assume that $F$ and $G$ grow slowly
enough at infinity that the rapid decay of $\mu_{\hbar}$ will kill off the
boundary terms in the integration by parts. In that case we get
\begin{align*}
\left\langle \frac{\partial F}{\partial z},G\right\rangle  & =-\frac{1}{2}%
\int_{\mathbb{C}}\bar{F}\left(  z\right)  \left(  \frac{\partial}{\partial
x}+i\frac{\partial}{\partial y}\right)  \left(  G\left(  z\right)  \mu_{\hbar
}\left(  z\right)  \right)  \,dz\\
& =-\int_{\mathbb{C}}\bar{F}\left(  z\right)  \frac{\partial}{\partial\bar{z}%
}\left(  G\left(  z\right)  \mu_{\hbar}\left(  z\right)  \right)  \,dz\\
& =-\int_{\mathbb{C}}\bar{F}\left(  z\right)  \frac{\partial G}{\partial
\bar{z}}\mu_{\hbar}\left(  z\right)  \,dz-\int_{\mathbb{C}}\bar{F}\left(
z\right)  G\left(  z\right)  \frac{\partial\mu_{\hbar}\left(  z\right)
}{\partial\bar{z}}\,dz.
\end{align*}

Now, since $G$ is holomorphic, $\partial G/\partial\bar{z}=0.$ Meanwhile we
compute that
\begin{align*}
\frac{\partial\mu_{\hbar}}{\partial\bar{z}}  & =\frac{\partial}{\partial
\bar{z}}\left(  \pi\hbar\right)  ^{-1}e^{-z\bar{z}/\hbar}\\
& =\left(  -\frac{z}{\hbar}\right)  \left(  \pi\hbar\right)  ^{-1}e^{-z\bar
{z}/\hbar}.
\end{align*}
So we get just
\begin{align*}
\left\langle \frac{\partial F}{\partial z},G\right\rangle  & =\int
_{\mathbb{C}}\bar{F}\left(  z\right)  G\left(  z\right)  \frac{z}{\hbar}%
\mu_{\hbar}\left(  z\right)  \,dz\\
& =\frac{1}{\hbar}\left\langle F,zG\right\rangle .
\end{align*}
That is, $\left(  \partial/\partial z\right)  ^{\ast}=\left(  1/\hbar\right)
z,$ or equivalently, $\left(  z\right)  ^{\ast}=\hbar\,\partial/\partial z.$
\qed

So the canonical commutation relations hold with $a=\hbar\,\partial/\partial
z$ and $a^{\ast}=z.$ If we assume irreducibility and that the exponentiated
form of the CCRs hold, then the Stone-von Neumann theorem will tell us that
there is a unitary map (unique up to a constant) from $\mathcal{H}L^{2}\left(
\mathbb{C},\mu_{\hbar}\right)  $ to $L^{2}\left(  \mathbb{R},dx\right)  $
which turns these operators into the standard creation and annihilation
operators. Such a map does indeed exist and is the \textit{Segal-Bargmann
transform}, which we consider in the next subsection. In fact, it is not hard
to show that the exponentiated CCRs do hold--see Exercise \ref{ccr.expex}. The
exponentiated position and momentum operators are expressed in terms of the
unitarized translation operators $T_{a}.$

We can easily extend this analysis to the $d$-dimensional case, by considering
the Segal-Bargmann space $\mathcal{H}L^{2}\left(  \mathbb{C}^{d},\mu_{\hbar
}\right)  $ and considering the operators $z_{k}$ (multiplication by $z_{k}$)
and $\hbar\partial/\partial z_{k}.$

\subsection{The transform}

The calculations of the previous subsection suggest (almost prove) that there
should be a unitary map $A_{\hbar}:L^{2}\left(  \mathbb{R}^{d},dx\right)
\rightarrow\mathcal{H}L^{2}\left(  \mathbb{C}^{d},\mu_{\hbar}\right)  $ that
intertwines the usual creation and annihilation operators with the operators
$z_{k}$ and $\hbar\,d/dz_{k}.$ There is indeed such a map, as described in the
following theorem.

\begin{theorem}
\label{at.thm}Consider the map $A_{\hbar}:L^{2}\left(  \mathbb{R}%
^{d},dx\right)  \rightarrow\mathcal{H}\left(  \mathbb{C}^{d},\mu_{\hbar
}\right)  $ given by
\begin{equation}
A_{\hbar}f\left(  z\right)  =\left(  \pi\hbar\right)  ^{-d/4}\int
_{\mathbb{R}^{d}}e^{\left(  -z^{2}+2\sqrt{2}x\cdot z-x^{2}\right)  /2\hbar
}f\left(  x\right)  \,dx.\label{at.form}%
\end{equation}

\begin{enumerate}
\item  For all $f\in L^{2}\left(  \mathbb{R}^{d},dx\right)  ,$ the integral is
convergent and is a holomorphic function of $z\in\mathbb{R}^{d}.$

\item  The map $A_{\hbar}$ is a unitary map of $L^{2}\left(  \mathbb{R}%
^{d},dx\right)  $ onto $\mathcal{H}L^{2}\left(  \mathbb{C}^{d},\mu_{\hbar
}\right)  .$

\item  For $k=1,\cdots,d$%
\begin{align}
A_{\hbar}\left(  \frac{X_{k}+iP_{k}}{\sqrt{2}}\right)  A_{\hbar}^{-1}  &
=\hbar\frac{\partial}{\partial z_{k}}\label{inter1}\\
A_{\hbar}\left(  \frac{X_{k}-iP_{k}}{\sqrt{2}}\right)  A_{\hbar}^{-1}  &
=z_{k}.\label{inter2}%
\end{align}
\end{enumerate}
\end{theorem}

There are many different ways to prove this theorem. However, it is reasonable
to prove it in a way that makes use of the canonical commutation relations. At
the same time, we do not want to rely on the ``non-theorem'' form of the
Stone-von Neumann theorem, and we do not want to have to check irreducibility,
which we would need to do to make use of either form of Stone-von Neumann. So
we will follow the following strategy. First, we verify that the intertwining
formulas (\ref{inter1}) and (\ref{inter2}) hold at least on ``nice''
functions. Second, we use (\ref{inter1}) and (\ref{inter2}) to show that
$A_{\hbar}$ maps a known orthonormal basis for $L^{2}\left(  \mathbb{R}%
^{d}\right)  $ onto a known orthonormal basis for $\mathcal{H}L^{2}\left(
\mathbb{C}^{d},\mu_{\hbar}\right)  $. (See also the argument in \cite{B1},
which is a very readable paper. Bargmann normalizes $\hbar$ to be 1.)

\textit{Proof of Theorem \ref{at.thm}.} As usual we will do the proof in the
case $d=1,$ though the general case is entirely analogous. The integral
converges because the function $e^{-(z^{2}+2\sqrt{2}xz-x^{2})/2\hbar}$ is
square-integrable as a function of $x$ for each fixed $z.$ Holomorphicity can
be proved using Morera's Theorem.

We do not yet know that $A_{\hbar}$ is invertible. So we will prove
(\ref{inter1}) and (\ref{inter2}) in the form
\begin{align}
A_{\hbar}\frac{X+iP}{\sqrt{2}}  & =\hbar\frac{\partial}{\partial z}A_{\hbar
}\nonumber\\
A_{\hbar}\frac{X-iP}{\sqrt{2}}  & =zA_{\hbar}.\label{inter3}%
\end{align}
I will assume that we apply $A_{\hbar}$ only to ``nice'' functions $f$, namely
ones that are smooth and decay rapidly at infinity, so that we may freely
integrate by parts and differentiate under the integral sign.

Then we compute that
\begin{align*}
\frac{d}{dz}\left(  A_{\hbar}f\right)  \left(  z\right)   & =\left(  \pi
\hbar\right)  ^{-1/4}\int_{\mathbb{R}}\frac{d}{dz}e^{\left(  -z^{2}+2\sqrt
{2}xz-x^{2}\right)  /2\hbar}f\left(  x\right)  \,dx\\
& =\left(  \pi\hbar\right)  ^{-1/4}\int_{\mathbb{R}}\left(  -\frac{z}{\hbar
}+\frac{\sqrt{2}x}{\hbar}\right)  e^{\left(  -z^{2}+2\sqrt{2}xz-x^{2}\right)
/2\hbar}f\left(  x\right)  \,dx\\
& =-\frac{z}{\hbar}A_{\hbar}f\left(  z\right)  +\frac{\sqrt{2}}{\hbar}%
A_{\hbar}\left[  xf\left(  x\right)  \right]  \left(  z\right)  .
\end{align*}
So
\begin{equation}
\frac{d}{dz}A_{\hbar}=-\frac{z}{\hbar}A_{\hbar}+\frac{\sqrt{2}}{\hbar}%
A_{\hbar}X,\label{calc1}%
\end{equation}
where as usual $X$ means multiplication by $x.$ Next compute using integration
by parts:
\begin{align*}
A_{\hbar}\left[  \frac{df}{dx}\right]  \left(  z\right)   & =\left(  \pi
\hbar\right)  ^{-1/4}\int_{\mathbb{R}}e^{\left(  -z^{2}+2\sqrt{2}%
xz-x^{2}\right)  /2\hbar}\frac{df}{dx}\,dx\\
& =-\left(  \pi\hbar\right)  ^{-1/4}\int_{\mathbb{R}}\left(  \frac{\sqrt{2}%
z}{\hbar}-\frac{x}{\hbar}\right)  e^{\left(  -z^{2}+2\sqrt{2}xz-x^{2}\right)
/2\hbar}f\left(  x\right)  \,dx,
\end{align*}
which tells us that
\begin{equation}
A_{\hbar}\frac{d}{dx}=-\frac{\sqrt{2}}{\hbar}zA_{\hbar}+\frac{1}{\hbar
}A_{\hbar}X.\label{calc2}%
\end{equation}
We may solve (\ref{calc2}) for $zA_{\hbar}$ to get one part of (\ref{inter3}),
and then substitute this expression for $zA_{\hbar}$ into (\ref{calc1}) and
solve for $\left(  d/dz\right)  A_{\hbar}$ to get the other part of
(\ref{inter3}). The algebra is left to the reader.

Having established (\ref{inter3}), we now prove the unitarity of $A_{\hbar}$
in a way that is similar to Exercise \ref{svn.proof} of the previous section.
Consider the ``ground state'' function
\[
f_{0}\left(  x\right)  =\left(  \pi\hbar\right)  ^{-1/4}e^{-x^{2}/2\hbar}.
\]
This is the unique (up to a constant) function with the property that
$af_{0}=0,$ where $a=2^{-1/2}\left(  x+\hbar\,d/dx\right)  $ is the
annihilation operator. Applying the Segal-Bargmann operator $A_{\hbar}$ to
$f_{0}$ we get with a little algebra
\begin{align}
\left(  A_{\hbar}f_{0}\right)  \left(  z\right)   & =\left(  \pi\hbar\right)
^{-1/2}\int_{\mathbb{R}}e^{-\left(  z-\sqrt{2}x\right)  ^{2}/2\hbar
}dx\label{int.gauss}\\
& =1.\nonumber
\end{align}
That is, $A_{\hbar}f_{0}$ is the constant function $\mathbf{1}.$ To evaluate
(\ref{int.gauss}) first observe that for $z\in\mathbb{R},$ a change of
variable shows that the integral is independent of $z.$ Since $A_{\hbar}f_{0}$
is holomorphic, if it is constant on $\mathbb{R}$ then it is constant on
$\mathbb{C}.$ The evaluation of the constant is a standard Gaussian integral.

Once it is established that $A_{\hbar}f_{0}=\mathbf{1},$ the intertwining
properties (\ref{inter3}) show that
\[
A_{\hbar}\left(  \left(  a^{\ast}\right)  ^{n}f_{0}\right)  =z^{n}%
\mathbf{1}=z^{n}.
\]
But the functions $\left(  a^{\ast}\right)  ^{n}f_{0}$ are the Hermite
functions, which are known to form an orthogonal basis for $L^{2}\left(
\mathbb{R}\right)  ,$ with $\left\|  \left(  a^{\ast}\right)  ^{n}%
f_{0}\right\|  _{L^{2}\left(  \mathbb{R}\right)  }^{2}=\hbar^{n}n!.$ (That
they are orthogonal with the indicated norms is proved using the canonical
commutation relations as in Exercise \ref{svn.proof} of the previous section.)
Meanwhile, we computed in Section 3 that the functions $z^{n}$ form an
orthogonal basis for $\mathcal{H}L^{2}\left(  \mathbb{C},\mu_{\hbar}\right)  $
with $\left\|  z^{n}\right\|  ^{2}=\hbar^{n}n!.$ Since $A_{\hbar}$ takes an
orthogonal basis to an orthogonal basis with the same norms, $A_{\hbar}$ is
unitary. \qed

I now want to describe a slightly different form of the Segal-Bargmann
transform, obtained by making the ``ground state transformation.'' This
transformation is necessary if one is going to take the infinite-dimensional
limit (as in Segal), and is often convenient even in finite dimensions.
Consider the unitary map
\[
G_{\hbar}:L^{2}\left(  \mathbb{R}^{d},dx\right)  \rightarrow L^{2}\left(
\mathbb{R}^{d},\left(  \pi\hbar\right)  ^{-d/2}e^{-x^{2}/\hbar}\,dx\right)
\]
given by
\begin{equation}
G_{\hbar}f\left(  x\right)  =\frac{f\left(  x\right)  }{f_{0}\left(  x\right)
}=\frac{f\left(  x\right)  }{\left(  \pi\hbar\right)  ^{-d/4}e^{-x^{2}/2\hbar
}},\label{gs}%
\end{equation}
where $f_{0}\left(  x\right)  :=\left(  \pi\hbar\right)  ^{-d/4}%
e^{-x^{2}/2\hbar}$ is the ground state in $\mathbb{R}^{d}.$ Note that the
measure in the image space is the measure $f_{0}\left(  x\right)  ^{2}dx,$ and
that the unitarity of $G_{\hbar}$ is a very elementary calculation. This is
called the ground state transformation since we are dividing each function by
the ground state $f_{0}.$ Note that $G_{\hbar}f_{0}$ is the constant function
$\mathbf{1}.$ So the effect of $G_{\hbar}$ is to turn the ground state into
the constant function $\mathbf{1}$ and to convert from Lebesgue measure to the
measure $f_{0}\left(  x\right)  ^{2}dx.$

I leave it as an exercise to calculate that
\begin{align*}
Ga_{k}G^{-1}  & =\frac{\hbar}{\sqrt{2}}\frac{\partial}{\partial x_{k}}\\
Ga_{k}^{\ast}G^{-1}  & =\sqrt{2}x_{k}-\frac{\hbar}{\sqrt{2}}\frac{\partial
}{\partial x_{k}}.
\end{align*}
It is now convenient to make an additional change of variable by letting
$y=\sqrt{2}x,$ and then renaming our variable back to $x.$ The resulting
creation and annihilation operators then take the form
\begin{align}
\tilde{a}_{k}  & =\hbar\frac{\partial}{\partial x_{k}}\nonumber\\
\tilde{a}_{k}^{\ast}  & =x_{k}-\hbar\frac{\partial}{\partial x_{k}%
}.\label{gs.ops}%
\end{align}
Our Hilbert space becomes (after this change) $L^{2}\left(  \mathbb{R}%
^{d},\rho_{\hbar}\left(  x\right)  \,dx\right)  ,$ where
\begin{equation}
\rho_{\hbar}\left(  x\right)  =\left(  2\pi\hbar\right)  ^{-d/2}%
e^{-x^{2}/2\hbar}.\label{rho.hbar}%
\end{equation}
Our goal, then, is to find a unitary map from $L^{2}\left(  \mathbb{R}%
^{d},\rho_{\hbar}\left(  x\right)  \,dx\right)  $ to $\mathcal{H}L^{2}\left(
\mathbb{C}^{d},\mu_{\hbar}\right)  $ which will convert the operators in
(\ref{gs.ops}) to the operators $\hbar\,\partial/\partial z_{k}$ and $z_{k}.$
This operator can be obtained by undoing the change of variable $x\rightarrow
\sqrt{2}x$ and the ground state transformation, and then applying the
Segal-Bargmann transform $A_{\hbar}.$ I\ will spare the reader the calculation
and simply record the result. Alternatively, one can prove Theorem
\ref{bt.thm} directly in a way similar to the proof of Theorem \ref{at.thm}.
See Exercise \ref{bt.ex}. In (\ref{bt1}) I make use of the fact that the
function $\rho_{\hbar}$ in (\ref{rho.hbar}) has an analytic continuation
(still called $\rho_{\hbar}$) from $\mathbb{R}^{d}$ to $\mathbb{C}^{d}.$

\begin{theorem}
\label{bt.thm}For all $\hbar>0,$ consider the map $B_{\hbar}:L^{2}\left(
\mathbb{R}^{d},\rho_{\hbar}\left(  x\right)  \,dx\right)  \rightarrow
\mathcal{H}\left(  \mathbb{C}^{d}\right)  $ given by
\begin{align}
B_{\hbar}f\left(  z\right)   & =\int_{\mathbb{R}^{d}}\rho_{\hbar}\left(
z-x\right)  f\left(  x\right)  \,dx\label{bt1}\\
& =\left(  2\pi\hbar\right)  ^{-d/2}\int_{\mathbb{R}^{d}}e^{-\left(
z-x\right)  ^{2}/2\hbar}f\left(  x\right)  \,dx.\label{bt2}%
\end{align}

\begin{enumerate}
\item  For all $f\in L^{2}\left(  \mathbb{R}^{d},\rho_{\hbar}\left(  x\right)
\,dx\right)  $ this integral is absolutely convergent and gives a holomorphic
function of $z\in\mathbb{C}^{d}.$

\item  The map $B_{\hbar}$ is a unitary map of $L^{2}\left(  \mathbb{R}%
^{d},\rho_{\hbar}\left(  x\right)  \,dx\right)  $ onto $\mathcal{H}%
L^{2}\left(  \mathbb{C}^{d},\mu_{\hbar}\right)  .$

\item  For all $k=1,\cdots,d$%
\begin{align*}
B_{\hbar}\tilde{a}_{k}B_{\hbar}^{-1}  & =\hbar\frac{\partial}{\partial z_{k}%
}\\
B_{\hbar}\tilde{a}_{k}^{\ast}B_{\hbar}^{-1}  & =z_{k},
\end{align*}
where $\tilde{a}_{k}$ and $\tilde{a}_{k}^{\ast}$ are given by (\ref{gs.ops}).
\end{enumerate}
\end{theorem}

\textit{Remarks.} 1) This form of the Segal-Bargmann transform is very close
to the finite-dimensional version of what Segal described in \cite[Thm. 5]{S3}
(or \cite[Thm. 1.14]{BSZ}). (See Exercise \ref{bt.form2}.) The only
differences are that Segal uses anti-holomorphic instead of holomorphic
functions, and that he does not make the change of variable $y=\sqrt{2}x,$ so
that there are some factors of $\sqrt{2}$ left in his formulas.

2) I have made the change of variable $y=\sqrt{2}x$ so that the transform will
have the form of a convolution, namely, an integral of $f\left(  x\right)  $
against a function of $z-x,$ as in (\ref{bt1}). Although this change is only a
convenience in the $\mathbb{R}^{d}$ case, it is necessary to a generalization
of the Segal-Bargmann transform to compact Lie groups, as described in Section 9.

The unitarity of the Segal-Bargmann transform (in either the $A_{\hbar}$ or
$B_{\hbar}$ form) can be used to give yet another derivation of the
reproducing kernel for the Segal-Bargmann space. See Exercise
\ref{compute.kernel}.

\subsection{The ``invariant'' form of the Segal-Bargmann transform}

I wish now to describe briefly another form of the Segal-Bargmann transform
that is technically advantageous. It expresses the transform as a convolution
as in the $B_{\hbar}$ form but has as its domain Hilbert space $L^{2}\left(
\mathbb{R}^{d},dx\right)  .$ Define a density $\nu_{\hbar}$ on $\mathbb{C}%
^{d}$ by
\[
\nu_{\hbar}\left(  z\right)  =\left(  \pi\hbar\right)  ^{-d/2}e^{-\left(
\operatorname{Im}z\right)  ^{2}/\hbar}.
\]
We then have the associated holomorphic function space $\mathcal{H}%
L^{2}\left(  \mathbb{C}^{d},\nu_{\hbar}\right)  .$ Recall the density
$\rho_{\hbar}$ from the previous subsection given by
\[
\rho_{\hbar}\left(  x\right)  =\left(  2\pi\hbar\right)  ^{-d/2}%
e^{-x^{2}/2\hbar}.
\]
Recall that this function admits an entire analytic continuation to
$\mathbb{C}^{d},$ also called $\rho_{\hbar}.$

\begin{theorem}
\label{ch.thm1}Consider the map $C_{\hbar}:L^{2}\left(  \mathbb{R}%
^{d},dx\right)  \rightarrow\mathcal{H}L^{2}\left(  \mathbb{C}^{d},\nu_{\hbar
}\right)  $ given by
\begin{align*}
C_{\hbar}f\left(  z\right)   & =\int_{\mathbb{R}^{d}}\rho_{\hbar}\left(
z-x\right)  f\left(  x\right)  \,dx\\
& =\left(  2\pi\hbar\right)  ^{-d/2}\int_{\mathbb{R}^{d}}e^{-\left(
z-x\right)  ^{2}/2\hbar}f\left(  x\right)  \,dx.
\end{align*}

\begin{enumerate}
\item  For all $f\in L^{2}\left(  \mathbb{R}^{d},dx\right)  $ the integral
defining $C_{\hbar}f\left(  z\right)  $ is absolutely convergent and
holomorphic in $z\in\mathbb{C}^{d}.$

\item  The map $C_{\hbar}$ is a unitary map of $L^{2}\left(  \mathbb{R}%
^{d},dx\right)  $ onto $\mathcal{H}L^{2}\left(  \mathbb{C}^{d},\nu_{\hbar
}\right)  .$

\item  For $k=1,\cdots,d,$%
\begin{align*}
C_{\hbar}\left(  X_{k}-iP_{k}\right)  C_{\hbar}^{-1}  & =z_{k}\\
C_{\hbar}\left(  X_{k}+iP_{k}\right)  C_{\hbar}^{-1}  & =z_{k}+2\hbar
\frac{\partial}{\partial z_{k}}%
\end{align*}
where $z_{k}$ denotes multiplication by $z_{k}.$
\end{enumerate}
\end{theorem}

\textit{Remarks}. 1) This form of the Segal-Bargmann transform is not truly
different from the conventional forms, just a convenient alternative
normalization. See Exercise \ref{sb.equiv}. In fact, we can relate, say, the
$A_{\hbar}$ and $C_{\hbar}$ forms of the transform as follows:
\begin{equation}
\left[  C_{\hbar}f\right]  \left(  z\right)  =\left(  4\pi\hbar\right)
^{-d/4}e^{-z^{2}/4\hbar}\left[  A_{\hbar}f\right]  \left(  \frac{z}{\sqrt{2}%
}\right)  .\label{at.vsct}%
\end{equation}
This is by direct calculation using the formulas defining $A_{\hbar}$ and
$C_{\hbar}.$ I will not prove the above theorem, since it is similar to the
proofs for the $A_{\hbar}$ and $B_{\hbar}$ forms of the transform. One can
also deduce the unitarity of $C_{\hbar}$ from that of $A_{\hbar}$ using the
relation (\ref{at.vsct}).

2) The formula for $C_{\hbar}$ is precisely the same as that for $B_{\hbar}.$
However, when considering $B_{\hbar}$ we were using different measures on both
the $\mathbb{R}^{d}$ side and the $\mathbb{C}^{d}$ side.

3) I am considering a modified set of creation and annihilation operators
$\hat{a}_{k}^{\ast}=X_{k}-iP_{k}$ and $\hat{a}_{k}=X_{k}+iP_{k},$ without the
usual factors of $\sqrt{2}$ in the denominator. In $\mathcal{H}L^{2}\left(
\mathbb{C}^{d},\nu_{\hbar}\right)  $ these correspond to the operators $z_{k}$
and $z_{k}+2\hbar\,\partial/\partial z_{k}.$ These satisfy the relation
\[
\left[  X_{k}+iP_{k},X_{l}-iP_{l}\right]  =2\hbar\delta_{k,l}
\]
and correspondingly
\[
\left[  z_{k}+2\hbar\frac{\partial}{\partial z_{k}},z_{l}\right]
=2\hbar\delta_{k,l},
\]
as is easily calculated. Also, in $\mathcal{H}L^{2}\left(  \mathbb{C}^{d}%
,\nu_{\hbar}\right)  $ the operators $z_{k}$ and $z_{k}+2\hbar\,\partial
/\partial z_{k}$ are adjoints of one another, as I invite you to verify.

4) There is an unfortunate minus sign in this business, namely that the
operator $z_{k}$ corresponds on the $\mathbb{R}^{d}$ side to $X_{k}-iP_{k}$
instead of $X_{k}+iP_{k}.$ However, this is an improvement over the standard
Segal-Bargmann transform, where $z_{k}$ corresponds to $\left(  X_{k}%
-iP_{k}\right)  /\sqrt{2}.$ The minus sign could be fixed either by redefining
the conventions of quantum mechanics or (as in Segal) working with
anti-holomorphic functions instead of holomorphic ones.

\subsection{Historical remarks}

In the summer of 1960 there was a conference in Boulder, Colorado, attended by
both Valentine Bargmann and Irving Segal. At this conference Segal gave a talk
which described the infinite-dimensional ($d=\infty$) version of the
Segal-Bargmann space, but not of the transform. After Segal's talk Bargmann
told Segal that he (Bargmann) had been working on a similar theory but in the
finite-dimensional case. Bargmann then published a paper \cite{B1} in 1961 in
Communications on Pure and Applied Mathematics describing the
finite-dimensional Segal-Bargmann space, the associated transform, and various
other interesting results. Bargmann's 1961 paper has a footnote (p. 191)
acknowledging that ``the Hilbert space defined here has already been used by
I.E. Segal for a representation of the quantum mechanical canonical
operators,'' as described in Segal's 1960 talk in Boulder.

Meanwhile, Segal's work on the infinite-dimensional theory was published in
the proceedings of the Boulder conference \cite{S1}, which appeared in 1963,
and in a paper in the Illinois Journal of Mathematics \cite{S2} that appeared
in 1962. The paper \cite{S2} has a footnote (p. 520) acknowledging that at the
Boulder meeting ``Professor Bargmann informed us that in the case of systems
of a finite number of degrees of freedom he had independently studied aspects
of the representation.'' Segal's conference proceeding article does not
discuss a transform. The Illinois Journal paper proves (Corollary 4.1) that
the holomorphic function representation is unitarily equivalent to a real
function representation, but does not describe this equivalence (which would
be the Segal-Bargmann transform) explicitly.

After finishing his 1961 paper Bargmann had a serious illness. After
recuperating he wrote a short paper that appeared in the Proceedings of the
National Academy of Sciences (U.S.A) in 1962 \cite{B2}. In this paper Bargmann
described the infinite-dimensional version of the Segal-Bargmann space, having
forgotten that Segal had already treated this case in his 1960 talks. Once
Bargmann realized his mistake he published an acknowledgment \cite{B3}, noting
that his oversight had resulted from ``exceptional circumstances.''

Finally, in 1978 Segal published a paper \cite{S3} that gives a technically
better description of the Segal-Bargmann space in infinitely many degrees of
freedom and describes explicitly the corresponding transform (similar to what
I call $B_{\hbar}$). In this paper Segal cites Bargmann's 1961 paper \cite{B1}
and the acknowledgment to Bargmann's 1962 paper \cite{B2}, but not the 1962
paper itself. This has fostered some confusion (reflected in the citations of
other authors) as to whether the acknowledgment refers to Bargmann's 1961
paper or to his 1962 paper. See Section 10 for a description of the
infinite-dimensional transform similar to that of \cite{S3}.

I should also mention John Klauder, who published a paper \cite{K} in Annals
of Physics in 1960 that described certain states (now called coherent states)
and a ``resolution of the identity'' which is equivalent to the isometricity
of the Segal-Bargmann transform. Although Klauder does not explicitly
introduce either the Segal-Bargmann transform or the corresponding holomorphic
function space, both of these objects are implicit in his resolution of the
identity. There has been much work lately on coherent states, which are
closely related to the Segal-Bargmann transform, but with a slightly different
point of view. See the discussion of coherent states in Section 10.

\subsection{Exercises}

\begin{exercise}
\label{adjoint.ex}This exercise gives an alternative proof of Theorem
\ref{adjoint.thm}. We know that the functions $\psi_{n}\left(  z\right)
:=z^{n}/\sqrt{\hbar^{n}n!},$ $n=0,1,2,\cdots$ form an orthonormal basis for
$\mathcal{H}L^{2}\left(  \mathbb{C},\mu_{\hbar}\right)  .$ Given $F_{1}%
,F_{2}\in\mathcal{H}L^{2}\left(  \mathbb{C},\mu_{\hbar}\right)  ,$ we can
write $F_{1}\left(  z\right)  =\sum a_{n}\psi_{n}\left(  z\right)  $ and
$F_{2}\left(  z\right)  =\sum b_{n}\psi_{n}\left(  z\right)  .$ Compute how
$z$ and $\hbar\,d/dz$ act on the $\psi_{n}$'s and use this to show that
$\left\langle zF_{1},F_{2}\right\rangle =\left\langle F_{1},\hbar
dF_{2}/dz\right\rangle .$
\end{exercise}

\begin{exercise}
\label{ccr.expex}Recall the unitarized translation operators $T_{a}$ on
$\mathcal{H}L^{2}\left(  \mathbb{C},\mu_{\hbar}\right)  $ ($a\in\mathbb{C}$)
described in Theorem \ref{ta.thm} (with the parameter $t$ in that theorem now
called $\hbar$).

a) A collection $\left\{  U_{s}\right\}  _{s\in\mathbb{R}}$ of unitary
operators is called a \textbf{one-parameter unitary group} if $U_{0}=I$ and
$U_{r+s}=U_{r}U_{s}$ for all $r,s\in\mathbb{R}.$ Let
\begin{align*}
V_{r}  & =T_{-ir/\sqrt{2}}\\
W_{s}  & =T_{-s/\sqrt{2}}.
\end{align*}
Show that $V_{r}$ and $W_{s}$ are one-parameter unitary groups.

b) Show that
\[
V_{r}W_{s}=e^{-irs/\hbar}W_{s}V_{r}
\]
for all $r,s\in\mathbb{R}.$

c) A standard functional-analytic result (Stone's Theorem) states that every
(strongly continuous) one-parameter unitary group $U_{s}$ may be expressed
uniquely in the form $U_{s}=e^{isA/\hbar},$ where $A$ is a self-adjoint
operator given (on a suitable domain) by
\[
A=-i\hbar\left.  \frac{d}{ds}\right|  _{s=0}U_{s}.
\]
The operator $A$ is called the generator of $U_{s}.$

Show that the generator of $V_{r}$ is
\[
A=\frac{1}{\sqrt{2}}\left(  \hbar\frac{d}{dz}+z\right)  =\frac{1}{\sqrt{2}%
}\left(  a+a^{\ast}\right)
\]
and that the generator of $W_{s}$ is
\[
B=\frac{1}{i\sqrt{2}}\left(  \hbar\frac{d}{dz}-z\right)  =\frac{1}{i\sqrt{2}%
}\left(  a-a^{\ast}\right)  .
\]
Conclude that $A$ and $B$ satisfy the exponentiated canonical commutation relations.
\end{exercise}

\begin{exercise}
\label{gs.ex}Suppose $\psi\in L^{2}\left(  \mathbb{R},dx\right)  $ satisfies
$a\psi=0,$ where $a=2^{-1/2}(x+\hbar\,d/dx)$ is the annihilation operator.
Show that $\psi\left(  x\right)  =c\,\exp\left(  -x^{2}/2\hbar\right)  .$
\end{exercise}

\begin{exercise}
\label{bt.ex}Verify Point 3 of Theorem \ref{bt.thm} in the case $d=1$, by
imitating the proof of Theorem \ref{at.thm}.
\end{exercise}

\begin{exercise}
\label{bt.form2}Show that $B_{\hbar}$ can be computed as
\[
B_{\hbar}f\left(  z\right)  =e^{-z^{2}/2\hbar}\int_{\mathbb{R}^{d}}e^{z\cdot
x/\hbar}f\left(  x\right)  \,\rho_{\hbar}\left(  x\right)  dx.
\]
\end{exercise}

\begin{exercise}
\label{sb.equiv}Show that the space $\mathcal{H}L^{2}\left(  \mathbb{C}^{d},\nu_{\hbar
}\right)  $ is holomorphically equivalent to $\mathcal{H}L^{2}\left(
\mathbb{C}^{d},\mu_{2\hbar}\right)  .$
\end{exercise}

\begin{exercise}
\label{compute.kernel}*Regard $A_{\hbar}$ (with $d=1$) as an isometric map of
$L^{2}\left(  \mathbb{R},dx\right)  $ \textbf{into} the full $L^{2}$ space
$L^{2}\left(  \mathbb{C},\mu_{\hbar}\right)  ,$ whose image is the holomorphic
subspace $\mathcal{H}L^{2}\left(  \mathbb{C},\mu_{\hbar}\right)  .$

a) Show that $A_{\hbar}^{\ast}A_{\hbar}=I$ on $L^{2}\left(  \mathbb{R}%
,dx\right)  $ and that $A_{\hbar}^{{}}A_{\hbar}^{\ast}=P,$ where $P$ is the
orthogonal projection from $L^{2}\left(  \mathbb{C},\mu_{\hbar}\right)  $ onto
$\mathcal{H}L^{2}\left(  \mathbb{C},\mu_{\hbar}\right)  .$

b) Compute $A_{\hbar}^{\ast}.$ Hint: How do you compute the adjoint of an
integral operator?

c) Compute $A_{\hbar}A_{\hbar}^{\ast}$ and show that your formula agrees with
our formula for the reproducing kernel of the Segal-Bargmann space.
\end{exercise}

\section{Quantum mechanics and quantization\label{quantum.sect}}

\subsection{A brief survey of classical mechanics}

See, for example, the book of Thirring \cite{Th} for more information. We
begin as usual with the one-dimensional case, which means that we consider a
particle moving, say, along a wire. The motion of such a particle is governed
by Newton's equation $F=ma.$ This means that if $x\left(  t\right)  $ is the
position of the particle, so that $\ddot{x}\left(  t\right)  $ (the second
derivative with respect to time) is the acceleration, then $m\ddot{x}=F,$
where $F$ is the force. (Here $m$ is the particle's mass.) Frequently $F $
depends only on position, and can be expressed in the form $F\left(  x\right)
=-V^{\prime}\left(  x\right)  ,$ where $V$ is the potential energy and
$V^{\prime}$ denotes the derivative with respect to $x.$ So our equation of
motion becomes
\begin{equation}
m\ddot{x}=-V^{\prime}\left(  x\right)  .\label{eom1}%
\end{equation}

Now, this is a second-order equation. It is convenient to re-write this as a
system of two first-order equations. Let
\[
p=m\dot{x}
\]
be the particle's momentum. We want to express our equations in terms of $x$
and $p.$ So $\dot{x}=p/m$ and $\dot{p}=m\ddot{x}=-V^{\prime}\left(  x\right)
.$ So in first-order form our equations of motion become
\begin{align}
\dot{x}  & =\frac{p}{m}\nonumber\\
\dot{p}  & =-V^{\prime}\left(  x\right)  .\label{eom2}%
\end{align}
Notice that (\ref{eom1}) is a second-order equation on the line, whereas
(\ref{eom2}) is a first-order equation on the plane. We have the following
notation
\begin{align*}
\text{Line (}x\text{-space) }  & =\text{\textbf{configuration space}}\\
\text{Plane (}\left(  x,p\right)  \text{-space) }  & =\text{\textbf{phase
space}.}%
\end{align*}
In mechanics you must always be very clear about the distinction between the
configuration space and the phase space.

We now consider the \textbf{Hamiltonian form} of mechanics. Let $H:\mathbb{R}%
^{2}\rightarrow\mathbb{R}$ be a smooth function (of the variables $x$ and
$p$). Then define equations of motion on $\mathbb{R}^{2}$ by
\begin{align}
\dot{x}  & =\frac{\partial H}{\partial p}\nonumber\\
\dot{p}  & =-\frac{\partial H}{\partial x}.\label{eom3}%
\end{align}
These are called \textbf{Hamilton's equations}. More explicitly these
equations mean that we want trajectories $\left(  x\left(  t\right)  ,p\left(
t\right)  \right)  $ satisfying
\begin{align*}
\frac{d}{dt}x\left(  t\right)   & =\frac{\partial H}{\partial p}\left(
x\left(  t\right)  ,p\left(  t\right)  \right) \\
\frac{d}{dt}p\left(  t\right)   & =-\frac{\partial H}{\partial x}\left(
x\left(  t\right)  ,p\left(  t\right)  \right)  .
\end{align*}
The function $H$ is called the (classical) Hamiltonian function, and
physically is the energy of the system.

Although $H$ can be any smooth function, it is often of the form
\[
H\left(  x,p\right)  =\frac{p^{2}}{2m}+V\left(  x\right)  ,
\]
where the first term is the kinetic energy and the second is the potential
energy. (Note that since $p=m\dot{x},$ $p^{2}/2m=\frac{1}{2}m(\dot{x})^{2},$
the usual expression for kinetic energy.) In that case, (\ref{eom3}) becomes
\begin{align*}
\dot{x}  & =\frac{\partial H}{\partial p}=\frac{p}{m}\\
\dot{p}  & =-\frac{\partial H}{\partial x}=-V^{\prime}\left(  x\right)  ,
\end{align*}
which agrees with (\ref{eom2}).

We need one more piece of fanciness to allow us to write classical mechanics
in a way that allows a reasonable comparison with quantum mechanics.

\begin{definition}
If $f_{1},f_{2}$ are smooth, real-valued function on $\mathbb{R}^{2},$ define
the \textbf{Poisson bracket} of $f_{1}$ and $f_{2},$ denoted $\left\{
f_{1},f_{2}\right\}  ,$ by
\[
\left\{  f_{1},f_{2}\right\}  =\frac{\partial f_{1}}{\partial x}\frac{\partial
f_{2}}{\partial p}-\frac{\partial f_{1}}{\partial p}\frac{\partial f_{2}%
}{\partial x},
\]
so that $\left\{  f_{1},f_{2}\right\}  $ is another smooth function on
$\mathbb{R}^{2}.$
\end{definition}

\begin{theorem}
\label{eom4}If $f$ is any smooth function on $\mathbb{R}^{2}$ and $\left(
x\left(  t\right)  ,p\left(  t\right)  \right)  $ is a solution to Hamilton's
equations (\ref{eom3}) then
\[
\frac{d}{dt}f\left(  x\left(  t\right)  ,p\left(  t\right)  \right)  =\left\{
f,H\right\}  \left(  x\left(  t\right)  ,p\left(  t\right)  \right)  ,
\]
or more concisely,
\[
\frac{df}{dt}=\left\{  f,H\right\}  .
\]
\end{theorem}

\textit{Proof}. We use the chain rule, Hamilton's equations, and the
definition of the Poisson bracket:
\begin{align*}
\frac{df}{dt}  & =\frac{\partial f}{\partial x}\frac{dx}{dt}+\frac{\partial
f}{\partial p}\frac{dp}{dt}\\
& =\frac{\partial f}{\partial x}\frac{\partial H}{\partial p}-\frac{\partial
f}{\partial p}\frac{\partial H}{\partial x}\\
& =\left\{  f,H\right\}  .
\end{align*}
\qed

Let us consider two examples. First let us take $f=H$ itself. Then
\[
\frac{dH}{dt}=\left\{  H,H\right\}  =\frac{\partial H}{\partial x}%
\frac{\partial H}{\partial p}-\frac{\partial H}{\partial p}\frac{\partial
H}{\partial x}=0.
\]
So energy is conserved! Next take $f$ to be the coordinate functions $x$ and
$p.$ Then we get
\begin{align*}
\frac{dx}{dt}  & =\left\{  x,H\right\}  =\frac{\partial x}{\partial x}%
\frac{\partial H}{\partial p}-\frac{\partial x}{\partial p}\frac{\partial
H}{\partial x}\\
& =\frac{\partial H}{\partial p}%
\end{align*}
and
\begin{align*}
\frac{dp}{dt}  & =\left\{  p,H\right\}  =\frac{\partial p}{\partial x}%
\frac{\partial H}{\partial p}-\frac{\partial p}{\partial p}\frac{\partial
H}{\partial x}\\
& =-\frac{\partial H}{\partial x}.
\end{align*}
Thus Theorem \ref{eom4} contains Hamilton's equation as a special case.

All of this can be done just as easily in $d$ dimensions. In that case the
configuration space is $\mathbb{R}^{d}$ and the phase space is $\mathbb{R}%
^{2d},$ with coordinates $x_{1},\cdots,x_{d},p_{1},\cdots,p_{d}.$ The
Hamiltonian is then a smooth function $H\left(  x,p\right)  $ on
$\mathbb{R}^{2d},$ which is typically (but not always) of the form
\begin{equation}
H\left(  x,p\right)  =\frac{1}{2m}\sum_{k=1}^{d}p_{k}^{2}+V\left(  x\right)
.\label{h.form}%
\end{equation}
Hamilton's equations take the form
\begin{align}
\frac{dx_{k}}{dt}  & =\frac{\partial H}{\partial p_{k}}\nonumber\\
\frac{dp_{k}}{dt}  & =-\frac{\partial H}{\partial x_{k}}.\label{eom5}%
\end{align}
The Poisson bracket is defined as
\begin{equation}
\left\{  f_{1},f_{2}\right\}  =\sum_{k=1}^{d}\frac{\partial f_{1}}{\partial
x_{k}}\frac{\partial f_{2}}{\partial p_{k}}-\frac{\partial f_{1}}{\partial
p_{k}}\frac{\partial f_{2}}{\partial x_{k}},\label{pbracket}%
\end{equation}
and Theorem \ref{eom4} holds:
\begin{equation}
\frac{df}{dt}=\left\{  f,H\right\}  .\label{eom6}%
\end{equation}

\subsection{A very brief survey of quantum mechanics}

It takes some time and effort to understand what quantum mechanics is all
about. Although I will try to explain enough to be comprehensible, I cannot
possibly convey the full physical meaning of the theory in just the few pages
I have here. I will describe the structure of quantum mechanics by analogy to
the preceding description of classical mechanics. But quantum mechanics is not
supposed to be the same as classical mechanics; this is merely an analogy, not
an equivalence.

Let us recall the structures that we had in classical mechanics.

\begin{itemize}
\item  Phase space $\mathbb{R}^{2d}.$

\item  Points $\left(  x,p\right)  $ in phase space.

\item  Real-valued functions $f$ on phase space.

\item  The value $f\left(  x,p\right)  \in\mathbb{R}$ of a function $f$ at a
point $\left(  x,p\right)  .$

\item  The Poisson bracket $\left\{  f_{1},f_{2}\right\}  .$

\item  The dynamics (equations of motion): $\frac{df}{dt}=\left\{
f,H\right\}  .$
\end{itemize}

The fourth point may seem too obvious to be worth mentioning, but the quantum
analog is not so obvious. Recall that the equation $df/dt=\left\{
f,H\right\}  $ is implied by Hamilton's equations and contains them as a
special case. I will now write down the corresponding structures in quantum mechanics.

\begin{itemize}
\item  A complex Hilbert space $\mathcal{H}.$

\item  Unit vectors $\psi$ in $\mathcal{H},$ called ``states.''

\item  Self-adjoint linear operators $A$ on $\mathcal{H}.$

\item  The expected value of an operator $A$ in the state $\psi,$ defined to
be $\left\langle \psi,A\psi\right\rangle .$

\item  The analog of the Poisson bracket for operators $A_{1}$ and $A_{2}$:
\[
\frac{1}{i\hbar}\left[  A_{1},A_{2}\right]
\]

\item  The dynamics
\[
\frac{dA}{dt}=\frac{1}{i\hbar}\left[  A,\hat{H}\right]  .
\]
\end{itemize}

Several points here require explanation. The most important point is that
\textit{functions on the classical phase space correspond to operators on the
quantum Hilbert space}. We will discuss this correspondence in detail later.
Note also that there is a parameter $\hbar$ in the quantum theory, which is
called Planck's constant and which has no classical analog. We will regard
$\hbar$ as merely a parameter, although physically it is an experimentally
determined constant, whose numerical value depends on the system of units. (It
is small compared to the scale of everyday life.) The process of converting
from the classical picture to the quantum picture is called ``quantization.''
There is in general no hard-and-fast rule for how to quantize things, although
there are well-established procedures in certain important cases.

The dynamics in the quantum picture requires a whole discussion itself. The
quantity $\hat{H}$ is an operator on the quantum Hilbert space, called the
(quantum) Hamiltonian operator. By analogy with the classical picture we
assume that there is some distinguished self-adjoint operator $\hat{H}$
(determined by theoretical or experimental considerations) which governs the
dynamics. Then the dynamics of the \textit{states }$\psi\in\mathcal{H}$ is
assumed to satisfy the \textbf{Schr\"{o}dinger equation}
\begin{equation}
i\hbar\frac{d\psi}{dt}=\hat{H}\psi.\label{schrodinger.eq}%
\end{equation}
We can then use this to determine how the expectation value of some operator
$A$ varies in time in a state $\psi\left(  t\right)  $ satisfying the
Schr\"{o}dinger equation. We compute
\begin{align*}
\frac{d}{dt}\left\langle \psi\left(  t\right)  ,A\psi\left(  t\right)
\right\rangle  & =\left\langle \frac{d\psi}{dt},A\psi\right\rangle
+\left\langle \psi,A\frac{d\psi}{dt}\right\rangle \\
& =\left\langle \frac{1}{i\hbar}\hat{H}\psi,A\psi\right\rangle +\left\langle
\psi,A\frac{1}{i\hbar}\hat{H}\psi\right\rangle \\
& =-\frac{1}{i\hbar}\left\langle \psi,\hat{H}A\psi\right\rangle +\frac
{1}{i\hbar}\left\langle \psi,A\hat{H}\psi\right\rangle \\
& =\frac{1}{i\hbar}\left\langle \psi,\left[  A,\hat{H}\right]  \psi
\right\rangle .
\end{align*}
Between the second and third lines I have used the fact that $\hat{H}$ is
self-adjoint. So
\[
\frac{d}{dt}\left\langle \psi\left(  t\right)  ,A\psi\left(  t\right)
\right\rangle =\left\langle \psi\left(  t\right)  ,\frac{1}{i\hbar}\left[
A,\hat{H}\right]  \psi\left(  t\right)  \right\rangle ,
\]
or (suppressing the dependence on the state as in the classical theory)
\begin{equation}
\frac{dA}{dt}=\frac{1}{i\hbar}\left[  A,\hat{H}\right]  .\label{heisenberg.eq}%
\end{equation}
This is the \textbf{Heisenberg form of the Schr\"{o}dinger equation}. This
form emphasizes the analogy with classical mechanics.

\textit{Example: }$\mathcal{H}=L^{2}\left(  \mathbb{R}^{d},dx\right)  .$

In the ``standard'' quantization scheme the quantum Hilbert space is
$L^{2}\left(  \mathbb{R}^{d},dx\right)  .$ Note that this is $L^{2}$ of
configuration space, not $L^{2}$ of phase space! The classical function
$x_{k}$ corresponds the operator $X_{k}$ (multiplication by $x_{k}$) and the
classical function $p_{k}$ corresponds to the operator $P_{k}=-i\hbar
\,\partial/\partial x_{k}.$ A classical function of the form $H\left(
x,p\right)  =p^{2}/2m+V\left(  x\right)  $ corresponds to the operator
\begin{align*}
\hat{H}  & =\frac{P^{2}}{2m}+V\left(  X\right) \\
& =-\frac{\hbar^{2}}{2m}\Delta+V\left(  X\right)  .
\end{align*}
Here $V\left(  X\right)  $ means multiplication by $V\left(  x\right)  $ and
$P^{2}=P_{1}^{2}+\cdots+P_{k}^{2}.$

Let us compare the classical Poisson bracket to the commutator of the
corresponding operators.
\begin{align}
\left\{  x_{k},p_{l}\right\}   & =\sum_{j}\frac{\partial x_{k}}{\partial
x_{j}}\frac{\partial p_{l}}{\partial p_{j}}-\frac{\partial x_{k}}{\partial
p_{j}}\frac{\partial p_{l}}{\partial x_{j}}\nonumber\\
& =\delta_{k,l}\label{xp1}%
\end{align}
and
\begin{align}
\frac{1}{i\hbar}\left[  X_{k},P_{l}\right]   & =\frac{1}{i\hbar}i\hbar
\delta_{k,l}I\nonumber\\
& =\delta_{k,l}I.\label{xp2}%
\end{align}
A comparison of (\ref{xp1}) and (\ref{xp2}) explains in part the canonical
commutation relations.

Let's try another example.
\begin{align}
\left\{  x_{k},H\right\}   & =\sum_{j}\frac{\partial x_{k}}{\partial x_{j}%
}\frac{\partial H}{\partial p_{j}}-\frac{\partial x_{k}}{\partial p_{j}}%
\frac{\partial H}{\partial x_{j}}\nonumber\\
& =2p_{k}.\label{xh1}%
\end{align}
And
\begin{align*}
\frac{1}{i\hbar}\left[  X_{k},\hat{H}\right]   & =\frac{1}{i\hbar}\left[
X_{k},\frac{1}{2m}P^{2}+V\right] \\
& =\frac{1}{i\hbar}\frac{1}{2m}\left[  X_{k},P_{k}^{2}\right]  ,
\end{align*}
since multiplication by $x_{k}$ commutes with multiplication by $V\left(
x\right)  $ and with $\partial/\partial x_{l}$ ($k\neq l$). Using Point (d) of
Exercise \ref{qbracket.ex} $\left[  X_{k},P_{k}^{2}\right]  =\left[
X_{k},P_{k}\right]  P_{k}+P_{k}\left[  X_{k},P_{k}\right]  =2i\hbar P_{k}.$
So
\begin{equation}
\frac{1}{i\hbar}\left[  X_{k},\hat{H}\right]  =2P_{k}.\label{xh2}%
\end{equation}

One more encouraging example before things start to get problematical.
\begin{align}
\left\{  p_{k},H\right\}   & =\sum_{j}\frac{\partial p_{k}}{\partial x_{j}%
}\frac{\partial H}{\partial p_{j}}-\frac{\partial p_{k}}{\partial p_{j}}%
\frac{\partial H}{\partial x_{j}}\nonumber\\
& =-\frac{\partial V\left(  x\right)  }{\partial x_{k}}.\label{ph1}%
\end{align}
and
\begin{align}
\frac{1}{i\hbar}\left[  P_{k},\hat{H}\right]   & =\frac{1}{i\hbar}\left[
P_{k},V\left(  X\right)  \right] \nonumber\\
& =-\frac{\partial}{\partial x_{k}}V\left(  x\right)  +V\left(  x\right)
\frac{\partial}{\partial x_{k}}\label{ph2}\\
& =-\frac{\partial V\left(  x\right)  }{\partial x_{k}}.\nonumber
\end{align}

In the three previous examples the classical Poisson bracket seems to
correspond exactly with the quantum commutator. (Compare (\ref{xp1}) and
(\ref{xp2}), (\ref{xh1}) and (\ref{xh2}), and (\ref{ph1}) and (\ref{ph2}).)
But things are not always so simple. Skipping the algebra I will record that
(in $d=1$)
\begin{align*}
\left\{  x^{3},p^{2}\right\}   & =6x^{2}p\\
\frac{1}{i\hbar}\left[  X^{3},P^{2}\right]   & =3\left(  X^{2}P+PX^{2}\right)
.
\end{align*}
Is $3\left(  X^{2}P+PX^{2}\right)  $ the operator which corresponds to the
classical function $6x^{2}p$? It's not so clear is it? Although classically
$x_{{}}^{2}p=px^{2}=xpx,$ on the quantum side $X^{2}P,$ $PX^{2},$ and $XPX$
are all different, so it is not evident what the quantum operator should be.
We need some sort of systematic theory here, which will come in the next subsection.

\subsection{Quantization schemes}

Dirac, one of the founders of quantum mechanics, proposed the following
axiomatic approach to quantization. Dirac wanted a Hilbert space $\mathcal{H}
$ and a map
\[
Q:\text{functions on the phase space }\mathbb{R}^{2d}\rightarrow\text{
operators on }\mathcal{H}
\]
with the following properties.

\begin{enumerate}
\item $Q$ is linear and $Q\left(  1\right)  =I.$

\item $Q\left(  \left\{  f,g\right\}  \right)  =\frac{1}{i\hbar}\left[
Q\left(  f\right)  ,Q\left(  g\right)  \right]  .$

\item $\mathcal{H}$ is irreducible under the action of $Q\left(  x_{k}\right)
$ and $Q\left(  p_{k}\right)  $

\item[3$^{\prime}$] $\mathcal{H}=L^{2}\left(  \mathbb{R}^{d},dx\right)
,\,Q\left(  x_{k}\right)  =$ multiplication by $x_{k},$ and $Q\left(
p_{k}\right)  =$ $-i\hbar\,\partial/\partial x_{k}.$

\item[4.] If $f$ is real-valued, then $Q\left(  f\right)  $ is self-adjoint.
\end{enumerate}

Note that the canonical commutation relations are implied by (1) and (2), in
light of the fact that $\left\{  x_{k},p_{l}\right\}  =\delta_{k,l}.$ Note
also that (3) and (3$^{\prime}$) are more or less equivalent in the presence
of (1) and (2), by the Stone-von Neumann Theorem. Also, the map $Q$ clearly
depends on $\hbar$ (in light of (2)), but I am suppressing this dependence in
my notation. This is a reasonable axiomatic system, except for the following result.

\begin{theorem}
[Groenewold-van Hove Theorem]There is no map $Q$ satisfying (1), (2), (3), and (4).
\end{theorem}

I should emphasize that I am being very seriously imprecise in the statement
of this theorem. The real theorem has precise domain conditions, and they are
necessary to get a real theorem. See Theorem 4.59 and the following discussion
in \cite{F}, or \cite{Go}.

The ``prequantization'' map of geometric quantization \cite{W} satisfies (1),
(2), and (4), but not (3). However, this is not supposed to be quantization,
but only prequantization. That is, condition (3) is important and cannot
simply be abandoned.

So we have a bit of problem, and the sad fact of the matter is that there is
no neat mathematical principle that tells us how to proceed. Ultimately the
test of whether we have the right ``quantization scheme'' comes from
experiment. But it is generally accepted that we should keep (1) and (3), and
require (2) to hold at least for the coordinate functions $x_{k}$ and $p_{k}.$
Thus we require that the canonical commutation relations hold! This explains
the importance of the canonical commutation relations from the point of view
of quantum mechanics.

The Stone-von Neumann Theorem says that if you accept (1) and (3) and the CCRs
(a special case of (2)), then up to unitary equivalence we may as well take
(3$^{\prime}$) as well. This then determines what our Hilbert space should be
and what $Q\left(  x_{k}\right)  $ and $Q\left(  p_{k}\right)  $ should be.
But then we have nothing that tells us what, say, $Q\left(  x_{k}p_{k}\right)
$ should be. There are several different ways of defining $Q$ assuming (1) and
(3$^{\prime}$). I will describe these in the case $d=1,$ though all can be
extended to arbitrary $d.$

(a) \textit{Put all the }$X$\textit{'s to the left and all the }$P$\textit{'s
to the right.}

This is called the (standard) pseudodifferential operator quantization. So
this map should send the function $x^{m}p^{n}$ to $X^{n}P^{m}.$ For general
functions $f\left(  x,p\right)  $ this quantization can be described in terms
of the Fourier transform as follows.
\begin{equation}
Q\left(  f\right)  \psi\left(  x\right)  =\frac{1}{2\pi}\int_{\mathbb{R}%
}f\left(  x,\hbar\xi\right)  \hat{\psi}\left(  \xi\right)  \,d\xi
,\label{psido}%
\end{equation}
where $\psi\in L^{2}\left(  \mathbb{R},dx\right)  $ and $\hat{\psi}\left(
\xi\right)  =\int e^{-i\xi x}\psi\left(  x\right)  \,dx$ is the Fourier
transform of $\psi.$ There is a large theory relating properties of $f$ to
properties of $Q\left(  f\right)  $ in this quantization scheme. However, the
pseudo-differential quantization does \textit{not} satisfy property (4) and so
is not a good candidate for the physical quantization map. (Consider just the
example $f\left(  x,p\right)  =xp.$) It is nevertheless an important map in
the theory of differential equations.

(b) \textit{Put all the }$X$\textit{'s to the right and all the }$P$\textit{'s
to the left.}

This is similar of course to (a).

(c) \textit{Symmetric or Weyl ordering.}

This scheme and the two remaining ones satisfy property (4).

The Weyl quantization is probably the best candidate physically for the right
quantization scheme for general functions. It has the property that
\begin{equation}
Q\left(  x^{n}p^{m}\right)  =\frac{1}{\left(  n+m\right)  !}\sum_{\sigma\in
S_{n+m}}\sigma\cdot\left(  X^{n}P^{m}\right)  .\label{weyl1}%
\end{equation}
Here $S_{n+m}$ is the permutation group on $n+m$ objects, and $\sigma
\cdot\left(  X^{n}P^{m}\right)  $ is schematic notation for what you get by
permuting the $n+m$ factors in $X^{n}P^{m}.$ So for example if $n=2,$ $m=1,$
then we have permutations of 3 objects, listable as $\left(  1,2,3\right)  ,$
$\left(  1,3,2\right)  ,$ $\left(  2,1,3\right)  ,$ $\left(  2,3,1\right)  ,$
$\left(  3,1,2\right)  ,$ $\left(  3,2,1\right)  .$ So applying these
permutations to $XXP$ gives
\begin{align*}
Q\left(  x^{2}p\right)   & =\frac{1}{6}\left[  XXP+XPX+XXP+XPX+PXX+PXX\right]
\\
& =\frac{1}{3}\left[  X^{2}P+XPX+PX^{2}\right]  .
\end{align*}
Of course, many of the terms you get will be the same; actually you need just
one instance of each \textit{distinct} ordering of $n$ $X$'s and $m$ $P$'s,
but I couldn't think of any compact way of writing this.

The Weyl quantization is characterized by the fact that
\[
Q\left(  e^{iax+ibp}\right)  =e^{iaX+ibP}.
\]
Thus for general $f$ we have
\begin{equation}
Q\left(  f\right)  =\left(  \frac{1}{2\pi}\right)  ^{2}\int_{\mathbb{R}^{2}%
}e^{iaX+ibP}\,\hat{f}\left(  a,b\right)  \,da\,db.\label{weyl2}%
\end{equation}
It is not obvious but true that (\ref{weyl2}) implies (\ref{weyl1}). See
\cite{F} Chapter 2.1, especially Eq. (2.20) and the following paragraph.

(d) \textit{The Wick ordering or normal ordering.}

This quantization scheme is very important in quantum field theory. Recall the
annihilation and creation operators
\begin{align*}
a  & =\frac{X+iP}{\sqrt{2}}\\
a^{\ast}  & =\frac{X-iP}{\sqrt{2}}.
\end{align*}
Wick ordering puts the annihilation operators to the right (acting first) and
the creation operators to the left. That is,
\begin{align}
Q\left(  \left(  x-ip\right)  ^{n}\left(  x+ip\right)  ^{m}\right)   &
=\left(  X-iP\right)  ^{n}\left(  X+iP\right)  ^{m}\nonumber\\
& =2^{\left(  m+n\right)  /2}\left(  a^{\ast}\right)  ^{m}a^{n}.\label{wick}%
\end{align}

As an example let's consider the function $\frac{1}{2}\left(  x^{2}%
+p^{2}\right)  =\frac{1}{2}\left(  x-ip\right)  \left(  x+ip\right)  .$ This
quantizes to
\begin{align*}
\frac{1}{2}\left(  X-iP\right)  \left(  X+iP\right)   & =\frac{1}{2}\left(
X^{2}+P^{2}+i\left(  XP-PX\right)  \right) \\
& =\frac{1}{2}\left(  X^{2}+P^{2}+i\left(  i\hbar\right)  \right) \\
& =\frac{1}{2}\left(  X^{2}+P^{2}\right)  -\frac{\hbar}{2}.
\end{align*}
I will leave it as an exercise to calculate that in this scheme
\[
Q\left(  x^{2}\right)  =X^{2}-\frac{\hbar}{2}.
\]
So in contrast to (a), (b), and (c), in this quantization scheme $Q\left(
x^{n}\right)  \neq Q\left(  x\right)  ^{n}.$

(e) \textit{The anti-Wick or anti-normal ordering.}

This is the reverse of Wick ordering, namely,
\begin{align}
Q\left(  \left(  x-ip\right)  ^{n}\left(  x+ip\right)  ^{m}\right)   &
=\left(  X+iP\right)  ^{m}\left(  X-iP\right)  ^{n}\nonumber\\
& =2^{\left(  m+n\right)  /2}a^{n}\left(  a^{\ast}\right)  ^{m}%
.\label{antiwick}%
\end{align}
So in this ordering the creation operators go to the right (acting first).
Imitating the above computation shows that in this ordering
\begin{align*}
Q\left(  \frac{1}{2}\left(  x^{2}+p^{2}\right)  \right)   & =\frac{1}%
{2}\left(  X^{2}+P^{2}\right)  +\frac{\hbar}{2}\\
Q\left(  x^{2}\right)   & =X^{2}+\frac{\hbar}{2}.
\end{align*}
The anti-Wick ordering can be described very naturally in terms of Toeplitz
operators, as we shall see in Section \ref{toeplitz.sect}. The anti-Wick
ordering also has nice properties that none of the other orderings have.

\subsection{The significance of the Segal-Bargmann
representation\label{sb.sig}}

It may be worthwhile at this point to make some remarks about the significance of
the Segal-Bargmann space and the associated transform. Although the
Segal-Bargmann space is naturally connected to the anti-Wick ordering (as we
shall see in Section \ref{toeplitz.sect}), it is also useful with other
quantization schemes, for example the Weyl ordering. I think of the
Segal-Bargmann space as simply a \textit{different but unitarily equivalent
representation of the canonical commutation relations}. So it is not so much a
different quantization as a unitarily equivalent realization of the same
quantization. The value of the Segal-Bargmann transform lies in the fact that
this unitary transformation makes certain problems easier to work with.
Certainly for semiclassical analysis (in which one tries to relate quantum
theory to classical theory) it is very natural to use the Segal-Bargmann
representation, because it is a Hilbert space of functions on the
\textit{phase space }$\mathbb{R}^{2d}=\mathbb{C}^{d}$ rather than on the
configuration space $\mathbb{R}^{d}.$ Since classical mechanics is naturally
formulated in phase space, this is a big advantage. In WKB theory (which is an
important part of semiclassical analysis), the Segal-Bargmann representation
also has technical advantages over the traditional $L^{2}\left(
\mathbb{R}^{d}\right)  $ representation. (See Voros \cite{V}, Paul-Uribe
\cite{PU}, Graffi-Paul \cite{GP}, Thomas-Wassell \cite{TW}, and
Borthwick-Paul-Uribe \cite{BPU}, \cite{Bo}.) The Segal-Bargmann space is also
the natural home for the Husimi function, which is gaining popularity in
physics and which we will describe in the next section, and for the Wick and
anti-Wick quantization schemes.

\subsection{Exercises}

\begin{exercise}
\label{pbracket.ex}Show that the Poisson bracket on $\mathbb{R}^{2d}$ has the
following properties:

a) (Skew-symmetry) $\left\{  f_{1},f_{2}\right\}  =-\left\{  f_{2}%
,f_{1}\right\}  .$

b) (Bilinearity) $\left\{  f_{1},f_{2}\right\}  $ is linear with respect to
$f_{1}$ with $f_{2}$ fixed, and vice versa.

c) (Jacobi identity) $\left\{  f_{1},\left\{  f_{2},f_{3}\right\}  \right\}
+\left\{  f_{2},\left\{  f_{3},f_{1}\right\}  \right\}  +\left\{
f_{3},\left\{  f_{1},f_{2}\right\}  \right\}  =0.$

d) $\left\{  f_{1},f_{2}f_{3}\right\}  =\left\{  f_{1},f_{2}\right\}
f_{3}+f_{2}\left\{  f_{1},f_{3}\right\}  .$

Point (d) says that $\left\{  f_{1},\cdot\right\}  $ is a derivation; that it,
it is a Leibniz-type product rule for the Poisson bracket.
\end{exercise}

\begin{exercise}
\label{qbracket.ex}Show that the commutator of operators $\left[  A,B\right]
:=AB-BA$ has the following properties:

a) (Skew-symmetry) $\left[  A,B\right]  =-\left[  B,A\right]  .$

b) (Bilinearity) $\left[  A,B\right]  $ is linear with respect to $A$ with $B$
fixed and vice versa.

c) (Jacobi identity) $\left[  A,\left[  B,C\right]  \right]  +\left[
B,\left[  C,A\right]  \right]  +\left[  C,\left[  A,B\right]  \right]  =0.$

d) $\left[  A,BC\right]  =\left[  A,B\right]  C+B\left[  A,C\right]  .$
\end{exercise}

\begin{exercise}
Verify that the pseudodifferential quantization, as given by (\ref{psido}),
satisfies $Q\left(  x^{n}p^{m}\right)  =X^{n}P^{m}.$
\end{exercise}

\section{Toeplitz operators, anti-Wick ordering, and phase space probability
densities\label{toeplitz.sect}}

The anti-Wick ordering can be expressed in an analytically nice way in terms
of Toeplitz operators on the Segal-Bargmann space. This will lead us to the
notion of phase space probability densities, one of which (the Husimi
function) also is most naturally expressed in terms of the Segal-Bargmann space.

\subsection{General theory of Toeplitz operators}

Let us return to the general setting of holomorphic function spaces,
$\mathcal{H}L^{2}\left(  U,\alpha\right)  .$ Recall that $\mathcal{H}%
L^{2}\left(  U,\alpha\right)  $ is a closed subspace of $L^{2}\left(
U,\alpha\right)  ,$ and therefore there is an orthogonal projection operator
$P:L^{2}\left(  U,\alpha\right)  \rightarrow\mathcal{H}L^{2}\left(
U,\alpha\right)  .$ We showed that this projection is given in terms of the
reproducing kernel as
\[
PF\left(  z\right)  =\int_{U}K\left(  z,w\right)  F\left(  w\right)
\,\alpha\left(  w\right)  \,dw
\]
for all $F\in L^{2}\left(  U,\alpha\right)  .$

Now suppose that $\phi$ is any bounded measurable function on $U.$ Define a
linear operator
\[
T_{\phi}:\mathcal{H}L^{2}\left(  U,\alpha\right)  \rightarrow\mathcal{H}%
L^{2}\left(  U,\alpha\right)
\]
by
\[
T_{\phi}F=P\left(  \phi F\right)  ,\quad F\in\mathcal{H}L^{2}\left(
U,\alpha\right)  .
\]
This is called the \textbf{Toeplitz operator with symbol }$\phi.$ So a
Toeplitz operator is one of the form ``multiply then project,'' that is,
multiply by $\phi$ and then project back into the holomorphic subspace.

\begin{theorem}
Toeplitz operators have the following properties.

\begin{enumerate}
\item $\left\|  T_{\phi}\right\|  \leq\left\|  \phi\right\|  _{L^{\infty}}.$

\item $T_{\phi}$ is linear as a function of $\phi$.

\item $T_{1}=I.$

\item $T_{\bar{\phi}}=\left(  T_{\phi}\right)  ^{\ast}.$ In particular, if
$\phi$ is real then $T_{\phi}$ is self-adjoint.

\item  For all $F_{1,}F_{2}\in\mathcal{H}L^{2}\left(  U,\alpha\right)  ,$%
\[
\left\langle F_{1},T_{\phi}F_{2}\right\rangle =\left\langle F_{1},\phi
F_{2}\right\rangle .
\]
\end{enumerate}
\end{theorem}

\textit{Proof.} We can think of $T_{\phi}$ as an operator on all of
$L^{2}\left(  U,\alpha\right)  $ by setting it to zero on the orthogonal
complement of the holomorphic subspace. In that case we can write
\[
T_{\phi}=PM_{\phi}P
\]
where $M_{\phi}$ denotes multiplication by $\phi.$ Thus
\begin{align*}
\left\|  T_{\phi}\right\|   & \leq\left\|  P\right\|  \left\|  M_{\phi
}\right\|  \left\|  P\right\| \\
& =\left\|  \phi\right\|  _{L^{\infty}}%
\end{align*}
and
\begin{align*}
\left(  T_{\phi}\right)  ^{\ast}  & =P^{\ast}\left(  M_{\phi}\right)  ^{\ast
}P^{\ast}\\
& =PM_{\bar{\phi}}P=T_{\bar{\phi}}.
\end{align*}
This establishes Point 1 and Point 4. Points 2 and 3 are clear.

For Point 5 we observe that
\begin{align*}
\left\langle F_{1},T_{\phi}F_{2}\right\rangle  & =\left\langle F_{1},P\phi
F_{2}\right\rangle \\
& =\left\langle PF_{1},\phi F_{2}\right\rangle \\
& =\left\langle F_{1},\phi F_{2}\right\rangle
\end{align*}
since $P$ is self-adjoint and $F_{1}$ is assumed to be holomorphic. \qed

Now if $\phi$ is an unbounded function, then we can still define the Toeplitz
operator $T_{\phi}$ in the same way, except that $T_{\phi}$ may be unbounded.
I will assume that $\phi$ is nice enough that $T_{\phi}$ is at least densely
defined. We expect that Point 4 of the theorem still holds modulo domain
issues (which we are not worrying about in these notes). The following result
is asserted and proved without worrying about domain issues.

\begin{theorem}
\label{order.thm}Suppose $\phi_{1},\cdots,\phi_{n}$ and $\psi_{1},\cdots
,\psi_{m}$ are holomorphic. Then
\[
T_{\bar{\psi}_{1}\cdots\bar{\psi}_{m}\phi_{1}\cdots\phi_{n}}=T_{\bar{\psi}%
_{1}}\cdots T_{\bar{\psi}_{m}}T_{\phi_{1}}\cdots T_{\phi_{n}}.
\]
\end{theorem}

\textit{Proof.} If $\phi_{1},\cdots,\phi_{n}$ are holomorphic then
\[
T_{\phi_{1}}\cdots T_{\phi_{n}}=PM_{\phi_{1}}PPM_{\phi_{2}}P\cdots
PM_{\phi_{n}}P.
\]
But all the projections except the first are unnecessary, since the $\phi$'s
are holomorphic. So
\begin{align*}
T_{\phi_{1}}\cdots T_{\phi_{n}}  & =PM_{\phi_{1}}\cdots M_{\phi_{n}}P\\
& =T_{\phi_{1}\cdots\phi_{n}}.
\end{align*}
Taking the adjoint of this we see that
\[
T_{\bar{\psi}_{1}\cdots\bar{\psi}_{m}}=T_{\bar{\psi}_{1}}\cdots T_{\bar{\psi
}_{m}}.
\]
Finally,
\begin{align*}
T_{\bar{\psi}_{1}\cdots\bar{\psi}_{m}\phi_{1}\cdots\phi_{n}}  & =PM_{\bar
{\psi}_{1}\cdots\bar{\psi}_{m}}M_{\phi_{1}\cdots\phi_{n}}P\\
& =PM_{\bar{\psi}_{1}\cdots\bar{\psi}_{m}}PM_{\phi_{1}\cdots\phi_{n}}P\\
& =T_{\bar{\psi}_{1}\cdots\bar{\psi}_{m}}T_{\phi_{1}\cdots\phi_{n}}\\
& =T_{\bar{\psi}_{1}}\cdots T_{\bar{\psi}_{m}}T_{\phi_{1}}\cdots T_{\phi_{n}}.
\end{align*}
\qed

\subsection{Toeplitz operators on the Segal-Bargmann space}

If we consider Toeplitz operators in the Segal-Bargmann space, $\mathcal{H}%
L^{2}\left(  \mathbb{C},\mu_{\hbar}\right)  ,$ then
\[
T_{z}=z
\]
(the projection being unnecessary since $z$ is holomorphic) and
\[
T_{\bar{z}}=\left(  T_{z}\right)  ^{\ast}=\hbar\frac{d}{dz}.
\]
These are the creation and annihilation operators. So Theorem \ref{order.thm}
says that
\[
T_{\bar{z}^{n}z^{m}}=\left(  \hbar\frac{d}{dz}\right)  ^{n}z^{m}.
\]
Note that the $z^{m}$'s, which are the creation operators, are to the right
and the $\left(  \hbar\,d/dz\right)  $'s, which are the annihilation
operators, are to the left. So this is clearly closely related to the
anti-Wick ordering. But there are some pesky minus signs and $\sqrt{2}$'s to
be dealt with. The creation operator, recall, is the operator
\[
\frac{X-iP}{\sqrt{2}}=a^{\ast}\Rightarrow T_{z},
\]
whereas the annihilation operator is
\[
\frac{X+iP}{\sqrt{2}}=a\Rightarrow T_{\bar{z}},
\]
where $\Rightarrow$ means ``corresponds to under the Segal-Bargmann
transform.'' So $x+ip$ does not correspond to $T_{z}.$ To fix this we need a
minus sign and a $\sqrt{2}.$

\begin{theorem}
\label{aw.thm1}Given a function $\phi$ on $\mathbb{C}^{d},$ define another
function $\phi^{\prime}$ on $\mathbb{C}^{d}$ by
\[
\phi^{\prime}\left(  z\right)  =\phi(\sqrt{2}\bar{z}).
\]
For each $\phi,$ consider the Toeplitz operator $T_{\phi^{\prime}}$ as an
operator on the Segal-Bargmann space $\mathcal{H}L^{2}\left(  \mathbb{C}%
^{d},\mu_{\hbar}\right)  .$ Then the map $\phi\rightarrow T_{\phi^{\prime}}$
is unitarily equivalent to the anti-Wick ordering. More precisely, for any
$\phi,$ the operator $A_{\hbar}^{-1}T_{\phi^{\prime}}A_{\hbar}$ on
$L^{2}\left(  \mathbb{R}^{d}\right)  $ is the same as the anti-Wick
quantization of $\phi.$
\end{theorem}

If we use the ``invariant'' form of the Segal-Bargmann space, $\mathcal{H}%
L^{2}\left(  \mathbb{C}^{d},\nu_{\hbar}\right)  ,$ then we get a similar
theorem without the factor of $\sqrt{2}.$ But even then it is necessary to
include the complex conjugate. If one is willing to use anti-holomorphic
functions (as in Segal) instead of holomorphic functions, then one can get rid
of the complex conjugate as well.

\subsection{Wigner function and Husimi function}

In the conventional quantization, in which the Hilbert space is $L^{2}\left(
\mathbb{R}^{d},dx\right)  ,$ if $\psi$ is a unit vector, then we interpret
$\left|  \psi\left(  x\right)  \right|  ^{2}$ as the ``position probability
density.'' This means that in quantum theory the particle does not have a
definite position, but only a probability distribution for the position, whose
density is given by $\left|  \psi\left(  x\right)  \right|  ^{2}.$ By using a
Fourier transform, one can also associate to each unit vector $\psi$ a
``momentum probability density.'' A natural next step is to ask, can you
define something like a \textit{joint} distribution of position and momentum,
which we would call a ``phase space probability density''? A little exposure
to quantum mechanics will convince you that there is no notion of a phase
space probability density that has all the properties you would like.
(Certainly taking just the product of the position and momentum distributions
is much too simplistic.) Still, we can still try for something that has enough
reasonable properties to be useful.

On reasonable way to try to define a phase space probability density is to
pick a Hilbert space $\mathcal{H}$ and a quantization scheme $Q.$ Then for
each unit vector $\psi\in\mathcal{H}$ we can look for a probability density
$p_{\psi}$ on the phase space $\mathbb{R}^{2d}$ satisfying
\begin{equation}
\int_{\mathbb{R}^{2d}}f\left(  x,p\right)  \,p_{\psi}\left(  x,p\right)
\,dx\,dp=\left\langle \psi,Q\left(  f\right)  \psi\right\rangle _{\mathcal{H}%
}\label{prob.def}%
\end{equation}
for all nice functions $f.$ This condition is reasonable because in
probabilistic language the left side is the expectation value of the function
$f$ with respect to the probability density $p_{\psi}.$ Meanwhile the right
side is what we have called the expectation value of the corresponding
operator $Q\left(  f\right)  $ in the state $\psi.$

Note that (\ref{prob.def}) serves to \textit{define} $p_{\psi}$ as a
distribution (or generalized function), provided only that $\psi$ is in the
domain of $Q\left(  f\right)  $ whenever $f$ is a $\mathcal{C}^{\infty}$
function of compact support. (For most quantization schemes, $Q\left(
f\right)  $ is bounded whenever $f$ is $\mathcal{C}^{\infty}$ and of compact
support.) Furthermore, taking $f\equiv1$ we see that the total integral of
$p_{\psi}$ is one, since $Q\left(  1\right)  =I$ and $\psi$ is a unit vector.
Unfortunately, $p_{\psi}$ will not be positive in general.

The first example of interest is the Weyl quantization, in which case the
associated $p_{\psi}$ is called the \textbf{Wigner function}. This is the most
natural thing to try, since the Weyl quantization is the most physically
natural quantization scheme. Unfortunately, the Wigner function is not always
positive. That is, for most unit vectors $\psi\in L^{2}\left(  \mathbb{R}%
^{d},dx\right)  ,p_{\psi}$ is negative at certain points. In fact, the
\textit{only} functions $\psi$ for which $p_{\psi}$ is everywhere non-negative
are ones of the form $\psi\left(  x\right)  =const.\,\exp(-\left(  x-a\right)
^{2}/b)\exp\left(  ix\cdot c\right)  ,$ with $a,c\in\mathbb{R}^{d}$ and
$b\in\left(  0,\infty\right)  .$ The Wigner function is therefore in general
called a \textit{pseudo-probability density}.

The other main example of interest is the anti-Wick quantization, in which
case the associated $p_{\psi}$ is called the \textbf{Husimi function}.
Although this is not as natural sounding as the Wigner function because the
anti-Wick quantization is not as natural, the Husimi function has the very
nice property that it is \textit{always positive} (that is, non-negative). We
see this explicitly in the following theorem.

\begin{theorem}
For $\psi\in L^{2}\left(  \mathbb{R}^{d},dx\right)  $ with $\left\|
\psi\right\|  =1,$ the Husimi function of $\psi,$ denoted $H_{\psi},$ is given
by
\[
H_{\psi}\left(  x,p\right)  =\left|  A_{\hbar}\psi\right|  ^{2}\left(
\frac{x-ip}{\sqrt{2}}\right)  \frac{e^{-\left|  z\right|  ^{2}/2\hbar}%
}{\left(  2\pi\hbar\right)  ^{d}}.
\]
where $A_{\hbar}$ is the Segal-Bargmann transform.

In terms of the invariant form of the Segal-Bargmann transform we have
\[
H_{\psi}\left(  x,p\right)  =\left|  C_{\hbar}\psi\right|  ^{2}\left(
x-ip\right)  \frac{e^{-\left(  \operatorname{Im}z\right)  ^{2}/\hbar}}{\left(
\pi\hbar\right)  ^{-d/2}}.
\]
\end{theorem}

This result follows almost immediately from Theorem \ref{aw.thm1} and the
analogous result for $C_{\hbar}.$ The reader may verify using (\ref{at.vsct})
from Section 6 that the two expressions for $H_{\psi}$ are indeed equal.

The following theorem describes the relationship between the anti-Wick and
Weyl quantizations, and correspondingly between the Husimi function and the
Wigner function.

\begin{theorem}
\label{aw.thm2}

\begin{enumerate}
\item  For all bounded measurable functions $f,$%
\[
Q_{anti-Wick}\left(  f\right)  =Q_{Weyl}\left(  e^{\hbar\Delta/4}f\right)
\]
where $\Delta$ is the standard Laplacian on $\mathbb{R}^{2d}$ and
$e^{\hbar\Delta/4}f$ is given explicitly by
\[
e^{\hbar\Delta/4}f\left(  z\right)  =\left(  \pi\hbar\right)  ^{-d}%
\int_{\mathbb{C}^{d}}e^{-\left|  z-u\right|  ^{2}/\hbar}f\left(  u\right)
\,du.
\]

\item  For all $\psi\in L^{2}\left(  \mathbb{R}^{d},dx\right)  $ with norm
one,
\[
H_{\psi}\left(  z\right)  =\left(  \pi\hbar\right)  ^{-d}\int_{\mathbb{C}^{d}%
}e^{-\left|  z-u\right|  ^{2}/\hbar}W_{\psi}\left(  u\right)  \,du,
\]
where $H_{\psi}$ is the Husimi function and $W_{\psi}$ is the Wigner function.
\end{enumerate}
\end{theorem}

Point 2 of the theorem is often described by saying that the Husimi function
is obtained by ``smearing out'' the Wigner function, by convolving it with a
Gaussian (whose ``width'' is proportional to $\sqrt{\hbar}.$) It is
interesting that this smearing is just enough to make the Husimi function
always positive, even when the Wigner function is not. A proof of Point 1 is
found in \cite{F}; Point 2 then follows.

Let us compare further the Husimi and Wigner functions. Once nice property of
the Wigner function is that it properly reproduces the ``marginal
distributions'' of $x$ and $p.$ That is, if you take $W_{\psi}\left(
x,p\right)  $ and integrate out the $p$-dependence, you obtain the standard
position probability density, and similarly with $x$ and $p$ reversed. That
is, we have the following result.

\begin{theorem}
For $\psi\in L^{2}\left(  \mathbb{R}^{d},dx\right)  $ with $\left\|
\psi\right\|  =1,$ the Wigner function $W_{\psi}$ satisfies
\[
\int_{\mathbb{R}^{d}}W_{\psi}\left(  x,p\right)  \,dp=\left|  \psi\left(
x\right)  \right|  ^{2}
\]
and
\[
\int_{\mathbb{R}^{d}}W_{\psi}\left(  x,p\right)  \,dx=\left|  \tilde{\psi
}\left(  p\right)  \right|  ^{2},
\]
where $\tilde{\psi}$ is the $\hbar$-scaled Fourier transform:
\[
\tilde{\psi}\left(  p\right)  =\left(  2\pi\hbar\right)  ^{-d/2}%
\int_{\mathbb{R}^{d}}e^{ip\cdot x/\hbar}\psi\left(  x\right)  \,dx.
\]
\end{theorem}

Note that the marginal distributions of $W_{\psi}$ are positive even if
$W_{\psi}$ is not positive. The reason for this result is that the Weyl
quantization has the property that $Q\left(  x^{n}\right)  =X^{n}$ and
$Q\left(  p^{n}\right)  =P^{n}$ (and more generally $Q\left(  f\left(
x\right)  \right)  =f\left(  X\right)  $ and $Q\left(  f\left(  p\right)
\right)  =f\left(  P\right)  $). The anti-Wick quantization does not have this
property, so it does not have the desired marginal distributions. But there is
a compensating benefit, namely a result that gives the position and momentum
\textit{wave functions} in terms of the ``phase space wave function,'' namely,
the Segal-Bargmann transform, where the Husimi function is essentially just
the absolute value squared of the Segal-Bargmann transform. It is easiest to
state this in terms of the invariant form $C_{\hbar}$ of the Segal-Bargmann transform.

\begin{theorem}
\label{inv.thm}If $\psi\in L^{2}\left(  \mathbb{R}^{d},dx\right)  $ then
\[
\psi\left(  x\right)  =\left(  2\pi\hbar\right)  ^{-d/2}\int_{\mathbb{R}^{d}%
}C_{\hbar}\psi\left(  x+ip\right)  e^{-p^{2}/2\hbar}\,dp
\]
and
\[
\tilde{\psi}\left(  p\right)  =\left(  2\pi\hbar\right)  ^{-d/2}%
e^{-p^{2}/2\hbar}\int_{\mathbb{R}^{d}}C_{\hbar}\psi\left(  x+ip\right)  \,dp
\]
where $\tilde{\psi}$ is as in the previous theorem.
\end{theorem}

In both theorems I am glossing over convergence issues. If you assume that
$\psi$ is nice enough then all formulas can be taken literally. But for
general $\psi\in L^{2},$ $\psi$ can diverge at certain points and there must
correspondingly be some divergences in the integrals. In the second theorem,
for example, one can deal with this by integrating over a ball of radius $R$
and then taking an $L^{2}$ limit as $R\rightarrow\infty.$

\subsection{Exercises}

\begin{exercise}
Verify Point 1 of Theorem \ref{aw.thm2} in the case $f\left(  x,p\right)
=\frac{1}{2}\left(  x^{2}+p^{2}\right)  .$ Hint: if $f$ is a polynomial, then
$e^{\hbar\Delta/4}$ can be computed by expanding it in a power series.
\end{exercise}

\begin{exercise}
*Verify Point 1 of Theorem \ref{aw.thm2} if $f\left(  x,p\right)  =x^{n}.$
\end{exercise}

\begin{exercise}
If $\phi$ is a positive function, show that the Toeplitz operator $T_{\phi} $
is a positive operator. Use this to explain why the Husimi function is always positive.
\end{exercise}

\begin{exercise}
\label{husimi.ex}Show that for all unit vectors $\psi,$ the Husimi function
$H_{\psi}$ satisfies
\[
H_{\psi}\left(  z\right)  \leq\left(  2\pi\hbar\right)  ^{-d}
\]
for all $z\in\mathbb{C}^{d}.$ This is a form of the uncertainty principle,
namely a limit on how concentrated a state can be in phase space. (After all,
$H_{\psi}$ integrates to one. So if it can't be too big at any one point, it
must be fairly spread out.)
\end{exercise}

\begin{exercise}
The functions $\psi_{z}$ which give equality in the previous problem for a
given value of $z$ are called the coherent states. Compute the Husimi function
of the coherent states.
\end{exercise}

\section{The Segal-Bargmann transform for compact Lie groups}

\subsection{Beyond the Canonical Commutation Relations}

I have introduced the ordinary Segal-Bargmann transform from the point of view
of the canonical commutation relations, a point of view that fits well with
the way Segal and Bargmann described the transform. I now want to describe a
generalization of the Segal-Bargmann transform in which the configuration
space $\mathbb{R}^{d}$ is replaced by a compact Lie group. In this generalized
setting \textit{there are no canonical commutation relations}. So having
described the ordinary Segal-Bargmann transform entirely in terms of the CCRs,
I am now going to describe a generalization of the Segal-Bargmann transform
that does not involve CCRs at all! Although this may seem strange, there is a
good reason for abandoning the CCRs, namely that in a more general setting
there seems to be no good candidate for what the CCRs ought to be. Recall that
in the $\mathbb{R}^{d}$ case, the CCRs are the quantum-mechanical analog of
the Poisson bracket relations $\left\{  x_{k},p_{l}\right\}  =\delta_{k,l}.$
We are now going to replace the configuration space $\mathbb{R}^{d}$ by a
compact Lie group $K.$ (This will be explained below.) Correspondingly we
replace the phase space $\mathbb{R}^{2d}$ by the cotangent bundle of $K,$
$T^{\ast}\!\!\left(  K\right)  .$ But on $T^{\ast}\!\!\left(  K\right)  $
there is no distinguished class of functions that could play the role of
$x_{k}$ and $p_{k}$ and thus tell us what the canonical commutation relations
ought to be.

So if one considers general classical-mechanical systems, one will not have a
preferred space of functions on the phase space that have simple relations
under the Poisson bracket. As a result, when quantizing such systems, one will
not have a simple set of commutation relations that could determine what the
quantum operators should be. So instead of using commutation relations as our
method of quantization we try some more geometrical construction of the
quantum Hilbert space, which should have the property that if this
construction is applied in the $\mathbb{R}^{d}$ case it produces one of the
familiar quantizations of $\mathbb{R}^{d}.$ In the $\mathbb{R}^{d}$ case, even
though the classical phase space $\mathbb{R}^{2d}$ is $2d$-dimensional, the
two (equivalent) possibilities we discussed for the quantum Hilbert space
consist of spaces of functions of only $d$ variables. That is, in the
``position'' or ``Schr\"{o}dinger'' representation $L^{2}\left(
\mathbb{R}^{d}\right)  ,$ our functions depend on the $d$ variables
$x_{1},\cdots,x_{d}$ but not on $p_{1},\cdots,p_{d},$ and in the
Segal-Bargmann space our functions depend on $z_{1},\cdots,z_{d}$ but not on
$\bar{z}_{1},\cdots,\bar{z}_{d}.$ We will consider similar possibilities in
the group case. (More generally, the theory of geometric quantization \cite{W}
proceeds by choosing a ``polarization'' on a symplectic manifold, which is
roughly a choice of $d$ variables out of $2d$ on which the functions in the
quantum Hilbert space should depend.)

In quantizing general classical systems we must simply accept that there are
no canonical commutation relations. Even so, when constructing, say, a
Segal-Bargmann transform, we should ask whether we have the ``right'' set-up.
In the next subsection I will describe a version of the Segal-Bargmann
transform for a compact Lie group, and after doing so I\ will discuss some
things that seem ``right'' about this construction.

Before doing this, let me mention that there are some classical systems
besides $\mathbb{R}^{2d}$ which \textit{do} have a distinguished set of
functions that allow one to define something like the CCRs. (The cotangent
bundle of a compact Lie group is \textit{not} such a system.) Usually such
functions arise in connection with some symmetry of the system. In the case of
$\mathbb{R}^{2d}$ the functions $x_{k}$ and $p_{k}$ have to do with the
translational symmetry of $\mathbb{R}^{2d}.$ What this means is that if you
consider Hamilton's equations with the Hamiltonian function $H\left(
x,p\right)  =p_{k},$ then the solutions are precisely the trajectories of the
form
\begin{align*}
x\left(  t\right)   & =x_{0}+te_{k}\\
p\left(  t\right)   & =p_{0},
\end{align*}
where $e_{k}$ is the vector $\left(  0,\cdots,0,1,0,\cdots0\right)  ,$ with
the 1 in the $k$th spot. This may be expressed as saying, ``$p_{k}$ is the
generator of translations in the $x_{k}$ direction.'' Similarly, $x_{k}$
generates translations in the negative $p_{k}$ direction.

More generally, we may consider a symplectic manifold $M$, that is, a manifold
equipped with some reasonable notion of Poisson bracket (satisfying the same
properties as in Exercise \ref{pbracket.ex}). If a Lie group $G$ acts
transitively on $M$ in a way that preserves the Poisson bracket, then we may
look for functions which ``generate'' the action of $G$ in the same way that
the functions $x_{k}$ and $p_{k}$ generate the translational symmetries of
$\mathbb{R}^{2d}.$ The collection of such functions is called the ``moment
map'' for the action of $G.$ It is then reasonable to take this collection of
functions as our ``basic functions.'' The general theory guarantees that these
functions satisfy nice relations under the Poisson bracket, relations that are
closely related to the commutation relations for the Lie algebra of $G.$ So in
this case the Poisson bracket relations among our basic functions give us a
way of defining (generalized) canonical commutation relations. Even though we
will not typically have a result like the Stone-von~Neumann Theorem, there is
in many cases a preferred way of building a quantum Hilbert space which
satisfies the relevant commutation relations and the appropriate
irreducibility condition.

The next simplest example (after $\mathbb{R}^{2d}$) in which this scheme can
be carried out is the unit disk, acted on by the group $SU\left(  1,1\right)
$ of fractional linear transformations that map the disk onto itself. There is
a notion of Poisson bracket that is invariant under this action. The
quantization of the disk by the above approach leads to our friends the
weighted Bergman spaces, with the weight parameter $a$ being related to
$\hbar.$ The relevant commutation relations in this case are those of the Lie
algebra of $SU(1,1).$ It is possible to exponentiate the corresponding
operators to get the (projective) unitary representation of $SU(1,1)$ acting
in the weighted Bergman spaces, as described in Section \ref{su11.sect}. One
can similarly treat the unit ball in $\mathbb{C}^{d}$ and more generally
bounded symmetric domains. See \cite{KL1} and \cite{BLU} for an analysis of
the Toeplitz quantization of these spaces.

A larger class of examples is that of co-adjoint orbits of Lie groups. The
quantization of co-adjoint orbits gives an powerful method of constructing
unitary representations of Lie groups, a method pioneered by A. Kirillov and
B. Kostant, and since investigated in hundreds of papers. See the recent
survey article \cite{Ki}.

\subsection{The transform for compact Lie groups}

I will concentrate on the simplest non-commutative example of a compact Lie
group, even though the theory works in general. So let $K=\mathsf{SU}(2)$, the
group of $2\times2$ unitary matrices with determinant one. Explicitly,
\begin{equation}
\mathsf{SU}(2)=\left\{  \left.  \left(
\begin{array}
[c]{cr}%
\alpha & -\bar{\beta}\\
\beta & \bar{\alpha}%
\end{array}
\right)  \right|  \alpha,\beta\in\mathbb{C},\,\left|  \alpha\right|
^{2}+\left|  \beta\right|  ^{2}=1\right\}  .\label{su2}%
\end{equation}
(See Exercise \ref{su2.ex1}.) Note that $\mathsf{SU}(2)$ can be identified
with the unit sphere $S^{3}$ inside $\mathbb{C}^{2}=\mathbb{R}^{4}.$ In
particular $\mathsf{SU}(2)$ is a compact manifold of (real) dimension 3.

Now let $K_{\mathbb{C}}=\mathsf{SL}(2;\mathbb{C}),$ the group of $2\times2$
matrices with determinant one, that is,
\[
\mathsf{SL}(2;\mathbb{C})=\left\{  \left.  \left(
\begin{array}
[c]{cc}%
a & b\\
c & d
\end{array}
\right)  \right|  a,b,c,d\in\mathbb{C},\,ad-bc=1\right\}  .
\]
Then $\mathsf{SL}(2;\mathbb{C})$ is a 3-dimensional complex manifold, or a
6-dimensional real manifold. Furthermore, $\mathsf{SU}(2)$ sits
inside $\mathsf{SL}(2;\mathbb{C})$ in the same way that $\mathbb{R}^{3}$ sits
inside $\mathbb{C}^{3},$ namely, as a ``totally real submanifold of maximum dimension.''

There is a natural Laplacian operator on $K,$ namely, the spherical Laplacian,
thinking of $\mathsf{SU}(2)$ as $S^{3}\subset\mathbb{R}^{4}.$ This operator
will be denoted $\Delta_{K}.$ Geometrically, $\Delta_{K}$ is the
Laplace-Beltrami operator with respect to a bi-invariant Riemannian metric on
$\mathsf{SU}(2).$ I then want to consider the heat equation on $K,$ namely,
\begin{equation}
\frac{\partial u}{\partial t}=\frac{1}{2}\Delta_{K}u,\label{heat}%
\end{equation}
where $u(x,t)$ is a function on $K\times\left(  0,\infty\right)  .$ The
equation is subject to an initial condition of the form
\begin{equation}
\lim_{t\downarrow0}u\left(  x,t\right)  =f\left(  x\right)  .\label{initial}%
\end{equation}
We denote the (unique) solution to this equation schematically as
\begin{equation}
u(x,t)=e^{t\Delta_{K}/2}f,\label{heat.op}%
\end{equation}
where $e^{t\Delta_{K}/2}$ is the \textit{heat operator}. That is,
$e^{t\Delta_{K}/2}$ is short-hand for the operator that associates to a
function $f$ the solution at time $t$ of the heat equation with initial
condition $f.$ Note that formally the RHS of (\ref{heat.op}) satisfies the
heat equation, and that formally at $t=0$ the RHS equals $f.$ The expression
$e^{t\Delta_{K}/2}$ may be defined rigorously for example using the spectral
theorem. However, even if the initial function $f$ is smooth, $e^{t\Delta
_{K}/2}$ cannot necessarily be computed by means of the power series for the
exponential function.

The heat equation can be solved in the following way. We first find the
\textit{heat kernel} for $K,$ which is the fundamental solution at the
identity, denoted $\rho_{t}\left(  x\right)  .$ This means that
\begin{equation}
\frac{d\rho}{dt}=\frac{1}{2}\Delta_{K}\rho_{t}\label{fund.1}%
\end{equation}
and
\begin{equation}
\lim_{t\downarrow0}\rho_{t}\left(  x\right)  =\delta_{e}\left(  x\right)
.\label{fund.2}%
\end{equation}
Here $\delta_{e}\left(  x\right)  $ means a $\delta$-function at the identity.
It is known that the heat kernel $\rho_{t}$ exists and is unique. In this case
there is a fairly explicit formula for the heat kernel--see \cite{H3}.

Now let $dx$ denote the natural surface area measure on $\mathsf{SU}%
(2)=S^{3}.$ In group-theoretical language $dx$ is the Haar measure for
$\mathsf{SU}(2).$ Then (\ref{fund.2}) really means that for all continuous
functions $f,$%
\[
\lim_{t\downarrow0}\int_{K}\rho_{t}\left(  x\right)  f\left(  x\right)
\,dx=f\left(  e\right)  .
\]
Once we have the heat kernel $\rho_{t}\left(  x\right)  $ and the Haar measure
$dx$ we may express the heat operator $e^{t\Delta_{K}/2}$ as follows:
\begin{equation}
\left(  e^{t\Delta_{K}/2}f\right)  \left(  x\right)  =\int_{K}\rho_{t}\left(
xy^{-1}\right)  f\left(  y\right)  \,dy.\label{convolve}%
\end{equation}
Here $xy^{-1}$ refers to product and inverse in the group $\mathsf{SU}(2).$
The RHS of (\ref{convolve}) is a group-theoretical \textit{convolution} of the
function $\rho_{t}$ and the function $f.$

\begin{theorem}
For each fixed $t>0$ the heat kernel $\rho_{t}\left(  x\right)  $ has a unique
analytic continuation from $K=\mathsf{SU}(2)$ to $K_{\mathbb{C}}%
=\mathsf{SL}(2;\mathbb{C}).$
\end{theorem}

Note that here we are analytically continuing in the space variable $x,$ which
initially lived in $\mathsf{SU}(2)$ but is now extended by analytic
continuation to $\mathsf{SL}(2;\mathbb{C}).$ I will continue to call the
holomorphic function obtained by this analytic continuation $\rho_{t}.$ Let
$\mathcal{H}\left(  K_{\mathbb{C}}\right)  $ denote the space of (entire)
holomorphic functions on $K_{\mathbb{C}}=\mathsf{SL}(2;\mathbb{C}).$ Then we
are now ready to define the generalized Segal-Bargmann transform for $K.$ We
will now let Planck's constant $\hbar$ play the role of time in the heat equation.

\begin{definition}
\label{ch.def2}For each $\hbar>0,$ define a map
\[
C_{\hbar}:L^{2}\left(  K,dx\right)  \rightarrow\mathcal{H}\left(
K_{\mathbb{C}}\right)
\]
by
\[
\left(  C_{\hbar}f\right)  \left(  g\right)  =\int_{K}\rho_{\hbar}\left(
gx^{-1}\right)  f\left(  x\right)  \,dx,\quad g\in K_{\mathbb{C}}.
\]
\end{definition}

Here $\rho_{\hbar}$ refers to the analytically continued heat kernel, and
$gx^{-1}$ refers to the product of the element $g\in\mathsf{SL}(2;\mathbb{C})$
and the element $x^{-1}\in\mathsf{SU}(2)\subset\mathsf{SL}(2;\mathbb{C}).$
Because of the analytic continuation it makes sense to plug an element of
$\mathsf{SL}(2;\mathbb{C})$ into $\rho_{\hbar}.$ Since $\rho_{\hbar}\left(
g\right)  $ is holomorphic (by construction) as a function of $g,$ it is
easily seen that $\rho_{\hbar}\left(  gx^{-1}\right)  $ is holomorphic as a
function of $g$ for each fixed $x.$ It then follows that $C_{\hbar}f\left(
g\right)  $ is holomorphic as a function of $g$ for any $f\in L^{2}\left(
K,dx\right)  .$ If we restrict our attention to $g\in K,$ then we recognize
$C_{\hbar}f$ as $e^{\hbar\Delta_{K}/2}f.$ Thus we may also write
\[
C_{\hbar}f=\text{analytic continuation of }e^{\hbar\Delta_{K}/2}f.
\]
Again the analytic continuation is in the space variable, from $K=\mathsf{SU}%
(2)$ to $K_{\mathbb{C}}=\mathsf{SL}(2;\mathbb{C}).$

\begin{theorem}
\label{ch.thm2}For each $\hbar>0$ there exists a measure $\nu_{\hbar}$ on
$K_{\mathbb{C}}$ such that $C_{\hbar}$ is a unitary map from $L^{2}\left(
K,dx\right)  $ onto $\mathcal{H}L^{2}\left(  K_{\mathbb{C}},\nu_{\hbar
}\right)  . $
\end{theorem}

Let us see the analogy between this generalized Segal-Bargmann transform for
$\mathsf{SU}(2)$ and the $C_{\hbar}$ version of the Segal-Bargmann transform
for $\mathbb{R}^{d},$ as described in Theorem \ref{ch.thm1}. We may think of
$\mathbb{R}^{d}$ as a commutative group under addition. In that case we may
recognize the first expression in Theorem \ref{ch.thm1} as the convolution of
$\rho_{\hbar}$ with the function $f$ (with the group operation now written in
additive notation). Furthermore it may be verified by direct calculation (or
by consulting a standard text on partial differential equations) that the
function
\[
\rho_{t}\left(  x\right)  =\left(  2\pi t\right)  ^{-d/2}e^{-x^{2}/2t}
\]
is the heat kernel for $\mathbb{R}^{d}.$ (That is, it satisfies the heat
equation and concentrates to a $\delta$-function at the origin as
$t\downarrow0.$) Thus if we substitute the group $\mathbb{R}^{d}$ for the
group $\mathsf{SU}(2)$ in Definition \ref{ch.def2} we recover precisely the
$C_{\hbar}$ version of the Segal-Bargmann transform for $\mathbb{R}^{d}.$

The measure $\nu_{\hbar}$ should be the group-theoretical analog of the
measure $d\nu_{\hbar}\left(  z\right)  :=\left(  \pi\hbar\right)  ^{-d/2}%
\exp(-\left(  \operatorname{Im}z\right)  ^{2}/\hbar)\,dz$ on $\mathbb{C}^{d}.$
To see how to make this analogy we may observe that in the $\mathbb{C}^{d}$
case the density of the measure $\nu_{\hbar}$ satisfies the heat equation on
$\mathbb{C}^{d}=\mathbb{R}^{2d}$ (check!). Furthermore this density is
independent of $x=\operatorname{Re}z.$ Similarly, the measure $\nu_{\hbar}$ on
$K_{\mathbb{C}}$ has a density (with respect to the natural Haar measure on
$K_{\mathbb{C}}$) that satisfies a suitable heat equation on $K_{\mathbb{C}}$
and that is invariant under the action of $K.$ (The invariance means that the
density satisfies $\nu_{\hbar}\left(  gx\right)  =\nu_{\hbar}\left(  g\right)
$ for all $g\in K_{\mathbb{C}}$ and all $x\in K.$)

\textit{Remarks}. 1) There is also a version of the generalized Segal-Bargmann
transform for $K$ that is precisely analogous to the $B_{\hbar}$ form of the
Segal-Bargmann transform for $\mathbb{R}^{d}.$ (That is, precisely analogous
to the $B_{\hbar}$ transform as I have described it, which differs by some
factors of $\sqrt{2}$ from what Segal describes.) As in the $\mathbb{R}^{d}$
case, the formula for the transform $B_{\hbar}$ is the same as the formula for
$C_{\hbar},$ but the measures on $K$ and $K_{\mathbb{C}}$ are different in the
two cases. For $B_{\hbar}$ the measure on $K$ is the heat kernel measure
$\rho_{\hbar}\left(  x\right)  \,dx$, and the measure on $K_{\mathbb{C}}$ is
the full heat kernel measure $\mu_{t}\left(  g\right)  dg.$ The $B_{\hbar}$
transform is described in Theorem 1$^{\prime} $ of \cite{H1}. In contrast to
the $\mathbb{R}^{d}$ case, the two transforms for $K$ are not ``equivalent.''
That is, there is no change-of-variable on $K $ like the one on $\mathbb{R}%
^{d}$ that converts one transform into the other. In the $\mathbb{R}^{d}$ case
the two transforms are interchangeable, really just two different
normalizations of the same transform. (Cf. Section 3 of \cite{H5}.) In the
group case the two transforms are genuinely distinct, and the $C_{\hbar}$
version seems to be better-behaved, in part because it respects the symmetry
of left- and right-translations by $K.$

2) The generalized Segal-Bargmann transform can be constructed in a precisely
analogous way for an arbitrary connected compact Lie group $K$ (with a fixed
bi-invariant Riemannian metric). One defines the complexification
$K_{\mathbb{C}}$ of $K$ and the Laplacian $\Delta_{K},$ and everything goes
through exactly as above. I have restricted to the case $K=\mathsf{SU}(2)$
merely to keep the discussion as concrete and elementary as possible.

\subsection{What is ``right'' about this transform?}

We have already observed that when moving beyond the setting of $\mathbb{R}%
^{d}$ we cannot expect to have a nice analog of the canonical commutation
relations. Nevertheless, we want to have some way of deciding when we have the
``right'' definition of the Segal-Bargmann transform for a compact Lie group.
That is, why this transform and not some other? We have already seen two good
things about the transform $C_{\hbar}$: 1) it is unitary, and 2) when the
group $\mathsf{SU}(2)$ is replaced by the group $\mathbb{R}^{d}$ we get back
precisely the $C_{\hbar}$ form of the Segal-Bargmann transform for
$\mathbb{R}^{d}.$ I want to describe several additional aspects of this
transform that suggest it is in some sense ``right.'' Of course this does not
preclude the possibility of some other useful Segal-Bargmann transform for
$\mathsf{SU}(2)$ or some other group. (Indeed C. Villegas has introduced a
different transform for $S^{3}=\mathsf{SU}(2)$ which might be preferable in
connection with the study of the Kepler problem.)

\textit{The complex group as phase space}. If we are to think of the
Segal-Bargmann transform for $K=\mathsf{SU}(2)$ in the same way we think of
the Segal-Bargmann transform for $\mathbb{R}^{d},$ then we want to think of
the complex group $K_{\mathbb{C}}=\mathsf{SL}(2;\mathbb{C})$ as the phase
space corresponding to the configuration space $K=\mathsf{SU}(2).$ On the
other hand, in classical mechanics, if a given manifold $X$ is the
configuration space, then the phase space is usually taken to be the cotangent
bundle of $X,$ $T^{\ast}\!\!\left(  X\right)  .$ So we would like to be able
to identify the complex group $K_{\mathbb{C}}$ with $T^{\ast}\!\!\left(
K\right)  .$ This may be done as follows. We introduce the Lie algebra
$\mathsf{su}\left(  2\right)  $ of $\mathsf{SU}(2).$ By definition, the Lie
algebra is the set of all $2\times2$ matrices $Y$ such that $\exp tY$ lies in
$\mathsf{SU}(2)$ for all real $t,$ where $\exp$ is the matrix exponential and
is computed as a (convergent) power series. It is not to hard to show
(Exercise \ref{su2.ex2}) that $\mathsf{su}(2)$ is given explicitly as
\begin{equation}
\mathsf{su}(2)=\left\{  \left.  2\times2\text{ matrices }Y\right|  Y^{\ast
}=-Y\text{ and }\mathrm{trace}\left(  Y\right)  =0\right\}  .\label{su2.lie}%
\end{equation}
This is a 3-dimensional real vector space. Geometrically, $\mathsf{su}(2)$ can
be identified with the tangent space at the identity to $\mathsf{SU}(2).$ We
may similarly define the Lie algebra $\mathsf{sl}(2;\mathbb{C})$ of
$\mathsf{SL}(2;\mathbb{C}).$ Explicitly $\mathsf{sl}(2;\mathbb{C})$ may be
computed as the space of all $2\times2$ matrices with trace zero--a
3-dimensional complex vector space. Note that $\mathsf{sl}(2;\mathbb{C}%
)=\mathsf{su}(2)\oplus i\,\mathsf{su}(2).$

Now using the left action of $\mathsf{SU}(2)$ on itself, the cotangent bundle
of $\mathsf{SU}(2)$ can be trivialized. Thus $T^{\ast}\!\!\left(  K\right)  $
is diffeomorphic to $\mathsf{SU}(2)\times\mathsf{su}(2)^{\ast}.$ Using the
natural inner product on $\mathsf{su}(2),$ we may identify $\mathsf{su}(2)$
with $\mathsf{su}(2)^{\ast},$ so that $T^{\ast}\left(  K\right)  $ is
identified with $\mathsf{SU}(2)\times\mathsf{su}(2).$ We then make use of the
map
\[
\Phi:\mathsf{SU}(2)\times\mathsf{su}(2)\rightarrow\mathsf{SL}(2;\mathbb{C})
\]
given by
\[
\Phi\left(  x,Y\right)  =x\exp iY,\quad x\in\mathsf{SU}(2),Y\in\mathsf{su}%
(2).
\]
It turns out that $\Phi$ is a diffeomorphism of $T^{\ast}\!\left(
\mathsf{SU}(2)\right)  $ onto $\mathsf{SL}(2;\mathbb{C}).$ Note here that $Y$
is skew-adjoint with trace zero, so that $iY$ is self-adjoint with trace zero.
It follows that $\exp iY$ is self-adjoint and positive with determinant one.
So in order to express an arbitrary matrix $g$ in $\mathsf{SL}(2;\mathbb{C})$
as $\Phi\left(  x,Y\right)  ,$ we use the polar decomposition to express $g$
as $g=xp,$ with $x$ unitary with determinant one and $p$ self-adjoint and
positive with determinant one. Then $iY$ is the unique self-adjoint logarithm
of $p.$

The diffeomorphism $\Phi$ between $T^{\ast}\!\left(  \mathsf{SU}(2)\right)  $
and $\mathsf{SL}(2;\mathbb{C})$ is in a certain sense canonical--see \cite{H3}
or \cite{H4}. In particular the complex structure of $\mathsf{SL}%
(2;\mathbb{C})$ and the symplectic structure of $T^{\ast}\!\left(
\mathsf{SU}(2)\right)  $ fit together so as to form a K\"{a}hler manifold. So
indeed there is a natural way of identifying $K_{\mathbb{C}}=\mathsf{SL}%
(2;\mathbb{C})$ with the phase space over $K=\mathsf{SU}(2).$ This shows the
Segal-Bargmann transform for $K$ described above is reasonable. This
identification of $T^{\ast}\!\!\left(  K\right)  $ with $K_{\mathbb{C}}$ works
in a similar way for any compact Lie group $K.$

\textit{Additional results.} Some of what is known about the transform for
$K,$ beyond unitarity, seems to suggest that it behaves the way a
Segal-Bargmann transform ought to behave. For example, \cite{H2} gives a very
natural inversion formula for $C_{\hbar},$ which says roughly that the
``position wave function'' $f(x)$ can be recovered from the ``phase space wave
function'' $C_{\hbar}f$ by integrating out the momentum variables. This is the
group analog of the first part of Theorem \ref{inv.thm}. Further, \cite{H3}
gives physically natural (and non-obvious) phase space bounds on the transform
of an arbitrary function $f,$ namely, a group version of Exercise
\ref{husimi.ex} of the previous section.

\textit{Alternative constructions of the generalized Segal-Bargmann
transform.} I want to describe briefly two additional constructions that turn
out to produce precisely the same generalized Segal-Bargmann transform for
$\mathsf{SU}(2)$ (or for any compact Lie group $K$). So altogether there are
three very different constructions that all produce exactly the same
transform, which suggests that there is something right about this transform.

The first alternative approach was proposed by L. Gross and P. Malliavin
\cite{GM}, who derived the $B_{\hbar}$ form of the \textit{generalized}
Segal-Bargmann transform for a compact Lie group $K$ from the
infinite-dimensional \textit{ordinary} Segal-Bargmann transform. By modifying
the approach of Gross and Malliavin, Bruce Driver and I \cite{DH} derived the
$C_{\hbar}$ form of the transform for $K$ from the infinite-dimensional
classical transform. The idea is to start with a certain infinite-dimensional
linear space $\mathcal{A}$ and then to ``reduce'' by a certain action of the
loop group over $K.$ (See also \cite{H6,HS}.) This reduction turns
$\mathcal{A}$ into a single copy of the compact Lie group $K,$ and it turns
the ordinary Segal-Bargmann transform for $\mathcal{A}$ into the generalized
Segal-Bargmann transform for $K.$ Of course it was not obvious ahead of time
that doing the Segal-Bargmann transform for $\mathcal{A}$ and then reducing
down to $K$ would give the same result as doing the generalized Segal-Bargmann
transform for $K.$

The second alternative approach to the Segal-Bargmann transform for $K$ is
that of geometric quantization. Geometric quantization \cite{W} aims to
associate in as canonical a way as possible to a symplectic manifold $M$ (the
classical phase space) a Hilbert space $\mathcal{H}$ and to functions on $M$
operators in the Hilbert space $\mathcal{H}.$ It is generally accepted that
quantization cannot be done without some additional structure on $M$; in
geometric quantization this additional structure is taken to be a
``polarization'' on $M.$ Roughly speaking, a polarization means a choice of
$d$ variables out of the $2d$ variables on $M$ on which the functions in the
quantum Hilbert space should depend. So in the case of a system with
configuration space $\mathbb{R}^{d}$ and phase space $\mathbb{R}^{2d}$ we have
seen two possibilities for the quantum Hilbert space, $L^{2}\left(
\mathbb{R}^{d}\right)  $ and the Segal-Bargmann space. In $L^{2}\left(
\mathbb{R}^{d}\right)  $ we have functions that depend on the position
variables $x_{1},\cdots,x_{d}$ but are independent of the momentum variables
$p_{1},\cdots,p_{d}.$ In the Segal-Bargmann space we have functions that
depend on $z_{1},\cdots,z_{d}$ but are independent of $\bar{z}_{1},\cdots
,\bar{z}_{d}$ (in the sense that $\partial F/\partial\bar{z}_{k}=0$).

On the cotangent bundle of a compact Lie group $K$ we have two natural
polarizations. The first is the ``vertical polarization,'' which makes sense
for any cotangent bundle. Here we take the coordinates on $K$ itself as the
ones on which our functions depend, and we take the coordinates in the
cotangent spaces as the ones on which our functions will not depend. In terms
of the decomposition $T^{\ast}\!\left(  \mathsf{SU}(2)\right)  =\mathsf{SU}%
(2)\times\mathsf{su}(2),$ we want functions that depend on the ``position''
variable $x\in\mathsf{SU}(2)$ but not on the ``momentum'' variable
$Y\in\mathsf{su}(2).$ The second polarization is a complex polarization (or
``K\"{a}hler polarization'') which comes from the identification of $T^{\ast
}\!\left(  \mathsf{SU}(2)\right)  $ with $\mathsf{SL}(2;\mathbb{C}).$ This
identification makes $T^{\ast}\!\left(  \mathsf{SU}(2)\right)  $ into a
complex manifold, and so it makes sense to speak of functions that in
holomorphic local coordinates depend on $z_{1},z_{2},z_{3}$ but not on
$\bar{z}_{1},\bar{z}_{2},\bar{z}_{3}$.

Now in geometric quantization the Hilbert space is not actually a space of
functions, but rather a space of sections of a complex line bundle over the
phase space $M.$ These sections are required to be ``covariantly constant'' in
the direction of the polarization. It takes some time to unravel what all this
really means, but when the shouting and tumult have died down we have the
following result. Using the vertical polarization on $T^{\ast}\!\!\left(
K\right)  $ the quantum Hilbert space may be identified with $L^{2}\left(
K\right)  .$ (The space of sections of the line bundle gets identified with a
space of functions by trivializing the line bundle.) Using the complex
polarization the quantum Hilbert space may be identified with an $L^{2}$ space
of holomorphic functions on $K_{\mathbb{C}}$, with respect to a certain
measure. It turns out that the measure coming from geometric quantization
coincides exactly (up to an irrelevant overall constant) with the measure
$\nu_{\hbar}$ that appears in the generalized Segal-Bargmann space. (More
precisely, this is true provided that one includes the ``half-form
correction'' in the geometric quantization. Cf. Sect. 7 of \cite{H4} in which
I consider the geometric quantization without the half-form correction.) See
\cite{H7}.

Thus the process of geometric quantization reproduces the generalized
Segal-Bargmann space over $K_{\mathbb{C}}.$ Not only so, but geometric
quantization also reproduces the Segal-Bargmann transform for $K.$ That is,
there is in geometric quantization something called the ``pairing map.'' The
pairing map is a map between the quantum Hilbert spaces constructed using two
different polarizations. In general the pairing map need not be unitary.
However, in the case of the pairing map between the vertically polarized
Hilbert space over $K$ and the complex-polarized Hilbert space, the pairing
map is unitary and coincides precisely with the generalized Segal-Bargmann
transform. (All of this holds for an arbitrary compact Lie group $K$ and its
complexification $K_{\mathbb{C}}.$)

It is a seeming miracle that geometric quantization should reproduce the
generalized Segal-Bargmann space and transform. After all, the Segal-Bargmann
space and transform were defined in terms of heat kernels, and geometric
quantization seems to have nothing to do with heat kernels or the heat
equation. Clearly something very special is going on in this example that
deserves to be understood better.

We have, then, three completely different constructions of the Segal-Bargmann
transform for a compact Lie group $K.$ The first construction is in terms of
heat kernels, the second is by reduction from an infinite-dimensional linear
space, and the third is by geometric quantization. That all three
constructions yield the same transform suggests that we are doing something right.

In a more general setting, say in which the compact Lie group is replaced by a
more general Riemannian manifold $X,$ it is unlikely that all three of these
constructions will give the same answer. I hope that having these three
different approaches will give sufficient insight that one can see how to
construct a well-behaved Segal-Bargmann transform for some more general class
of manifolds $X.$ Time will tell!

\subsection{Exercises}

\begin{enumerate}
\item * Verify that the expression
\[
\left\{  f_{1},f_{2}\right\}  =-4i\left(  1-\left|  z\right|  ^{2}\right)
^{2}\left(  \frac{\partial f_{1}}{\partial z}\frac{\partial f_{2}}%
{\partial\bar{z}}-\frac{\partial f_{1}}{\partial\bar{z}}\frac{\partial f_{2}%
}{\partial z}\right)
\]
defines a Poisson bracket on the unit disk $\mathbb{D}$ that is invariant
under the action of $\mathsf{SU}(1,1).$

\item \label{su2.ex1}Verify that every element of the form (\ref{su2}) is
really unitary and has determinant one, and that every $2\times2$ unitary
matrix with determinant one can be expressed in this form.

\item  Verify that the function
\[
\rho_{t}\left(  x\right)  =\left(  2\pi t\right)  ^{-d/2}e^{-x^{2}/2t}
\]
on $\mathbb{R}^{d}$ satisfies the heat equation and that for every continuous
compactly supported function $f$ on $\mathbb{R}^{d},$%
\[
\lim_{t\downarrow0}\int_{\mathbb{R}^{d}}\rho_{t}\left(  x\right)  f\left(
x\right)  \,dx=f\left(  0\right)  .
\]

\item * a) Show that every element $g$ of $\mathsf{SL}(2;\mathbb{C})$ can be
written in the form
\begin{equation}
g=x_{1}e^{aH}x_{2}\label{kak}%
\end{equation}
with $a\in\mathbb{R}$, and $x_{1},x_{2}\in\mathsf{SU}(2).$ Here
\[
H=\left(
\begin{array}
[c]{cc}%
1 & 0\\
0 & -1
\end{array}
\right)  .
\]
\textit{Hint}: first write $g=x\exp p$ with $x\in\mathsf{SU}(2)$ and $p$
self-adjoint with trace zero. Then diagonalize $p.$

b) Consider the series expansion for the heat kernel on $\mathsf{SU}(2)$ (cf.
Eq. (11) of \cite{H1})
\begin{equation}
\rho_{t}\left(  x\right)  =\sum_{l}\left(  2l+1\right)  e^{-tl(l+1)/2}%
\,\mathrm{trace}\left(  \pi_{l}\left(  x\right)  \right)  ,\label{peter.weyl}%
\end{equation}
where $\pi_{l}$ is the irreducible representation of $\mathsf{SU}(2)$ of
dimension $2l+1,$ and where $l=0,1/2,1,3/2,2,\cdots.$ We want to analytically
continue $\rho_{t}$ from $\mathsf{SU}(2)$ to $\mathsf{SL}(2;\mathbb{C})$ by
analytically continuing (\ref{peter.weyl}) term-by-term. Show using
(\ref{kak}) that the analytically continued series converges uniformly on
compact subsets of $\mathsf{SL}(2;\mathbb{C}).$ (Cf. Sect. 4 of \cite{H1}.)
This shows that $\rho_{t}$ admits an analytic continuation from $\mathsf{SU}%
(2)$ to $\mathsf{SL}(2;\mathbb{C}).$ \textit{Hint}: what are the eigenvalues
of $H$ in the representation $\pi_{l}?$

\item \label{su2.ex2}Verify the description (\ref{su2.lie}) of the Lie algebra
$\mathsf{su}\left(  2\right)  $ of $\mathsf{SU}(2).$
\end{enumerate}

\section{To infinity and beyond}

In this section I will touch briefly on a few additional topics, to give the
flavor of them and to suggest directions for further reading.

\subsection{The infinite-dimensional theory}

I have already mentioned that Segal wished to consider systems with infinitely
many degrees of freedom, describing quantum \textit{field theory} instead of
ordinary quantum mechanics. This means that Segal wanted to let the dimension
$d$ tend to infinity. This limit raises several interesting technical issues.
Most important, there is no such thing as Lebesgue measure on an
infinite-dimensional space. Thus the ground state transformation, leading to
the $B_{\hbar}$ form of the Segal-Bargmann transform, is essential when
$d=\infty.$

So consider a real Hilbert space $X_{\mathbb{R}}$, which I assume is
infinite-dimensional and separable. We think of this as the $d\rightarrow
\infty$ limit of $\mathbb{R}^{d}.$ We need to try to construct the appropriate
measure on $\rho_{\hbar}$ on $X_{\mathbb{R}},$ which should be the
infinite-dimensional limit of the measures appearing in the $B_{\hbar}$ form
of the Segal-Bargmann transform for $\mathbb{R}^{d}.$ So we might imagine
picking an increasing sequence of finite-dimensional subspaces $V_{d}$ of
$X_{\mathbb{R}},$ with $\dim V_{d}=d$ and chosen so that the union of the
$V_{d}$'s is dense in $X_{\mathbb{R}}.$ Then we may consider a sequence of
measures $\rho_{\hbar}^{(d)}$ on $X_{\mathbb{R}}$ such that $\rho_{\hbar
}^{(d)}$ is concentrated on $V_{d}$ and given by
\[
d\rho_{\hbar}^{(d)}\left(  x\right)  =\left(  2\pi\hbar\right)  ^{-d/2}%
e^{-\left\|  x\right\|  ^{2}/2\hbar}\,dx,
\]
where $dx$ is the Lebesgue measure on $V_{d}$ and the constant in front
normalizes $\rho_{\hbar}^{(d)}$ to be a probability measure. We then need to
let $d$ tend to infinity.

Unfortunately, the limit $\lim_{d\rightarrow\infty}\rho_{\hbar}^{(d)}$ does
not exist as a measure on $X_{\mathbb{R}}.$ To see intuitively why this is so,
first consider the two-dimensional case. The measure $\rho_{\hbar}$ on
$\mathbb{R}^{2}$ is given explicitly as
\begin{align*}
d\rho_{\hbar}\left(  x,y\right)   & =\left(  2\pi\hbar\right)  ^{-1}%
e^{-\left(  x^{2}+y^{2}\right)  /2\hbar}\,dx\,dy\\
& =\left[  \left(  2\pi\hbar\right)  ^{-1/2}e^{-x^{2}/2\hbar}\,dx\right]
\left[  \left(  2\pi\hbar\right)  ^{-1/2}e^{-y^{2}/2\hbar}\,dy\right]  .
\end{align*}
Note that the measure factors as a measure in the $x$ variable times a measure
in the $y$ variable, both of which are probability measures, in fact, the
\textit{same} probability measure. We may say the same thing in probabilistic
language by saying that (with respect to $\rho_{\hbar}$) $x$ and $y$ are
independent and identically distributed. We may recognize the distribution of
$x$ or $y$ as normal with mean zero and variance $\hbar$. The same sort of
product decomposition will hold for the measures $\rho_{\hbar}^{(d)}$ in every
dimension $d.$ So now suppose that the $\rho_{\hbar}^{(d)}$'s did converge to
a probability measure $\rho_{\hbar},$ and let $\left\{  e_{i}\right\}  $ be an
orthonormal basis for $X_{\mathbb{R}}$ and $\left\{  x_{i}\right\}  $ the
coordinates with respect to this basis. Then the coordinates $x_{i}$ would
presumably be independent, with each $x_{i}$ normal with mean zero and
variance $\hbar.$ But if a vector $v$ is in $X_{\mathbb{R}}$ then the
coordinates $x_{i}$ of that vector satisfy $\Sigma x_{i}^{2}=\left\|
v\right\|  ^{2}<\infty.$ On the other hand, if $\left\{  x_{i}\right\}  $ are
independent normal random variables with mean zero and all having the same
variance $\hbar,$ then it is intuitively obvious that $\Sigma x_{i}^{2}%
=\infty$ with probability one. So roughly speaking the points of finite norm
in $X_{\mathbb{R}}$ (that is, all of $X_{\mathbb{R}}$!) constitute a set of
measure zero. So $\rho_{\hbar}$ cannot be a probability measure on
$X_{\mathbb{R}}.$

Even though $\rho_{\hbar}$ does not exist as a measure on $X_{\mathbb{R}},$ it
should exist as a measure on something. After all, it is possible to have an
infinite sequence of independent random variables with mean zero and variance
$\hbar.$ Following the approach of \cite{Gr} we consider a certain
``extension'' of $X_{\mathbb{R}},$ denoted $\overline{X}_{\mathbb{R}}.$ By
this I\ mean that $\overline{X}_{\mathbb{R}}$ is a Banach space and that there
is a continuous embedding of the Hilbert space $X_{\mathbb{R}}$ into
$\overline{X}_{\mathbb{R}}.$ If $\overline{X}_{\mathbb{R}}$ is sufficiently
much larger than $X_{\mathbb{R}},$ then in a natural way $\rho_{\hbar}$ may be
regarded as a measure on $\overline{X}_{\mathbb{R}}.$ The resulting measure
$\rho_{\hbar}$ is called a \textbf{Gaussian measure} on $\overline
{X}_{\mathbb{R}},$ and the subspace $X_{\mathbb{R}}\subset\overline
{X}_{\mathbb{R}}$ is called the \textbf{Cameron-Martin subspace}. The
Cameron-Martin subspace is a set of measure zero with respect to $\rho_{\hbar
}.$ (See also \cite{Ku}.)

The prototypical example is the following. We take $X_{\mathbb{R}}$ to be the
space of absolutely continuous functions $B:\left[  0,1\right]  \rightarrow
\mathbb{R}$ such that 1) $B\left(  0\right)  =0$ and 2) $\int_{0}^{1}\left|
dB/dt\right|  ^{2}dt<\infty,$ with inner product given by
\begin{equation}
\left\langle B_{1},B_{2}\right\rangle =\int_{0}^{1}\frac{dB_{1}}{dt}%
\frac{dB_{2}}{dt}\,dt.\label{inner.h1}%
\end{equation}
We take $\overline{X}_{\mathbb{R}}$ to be $\mathcal{C}_{0}\left(  \left[
0,1\right]  \right)  ,$ that is, the space of continuous functions $B:\left[
0,1\right]  \rightarrow\mathbb{R}$ such that $B\left(  0\right)  =0$. Then for
each $\hbar>0$ there exists a well-defined measure $\rho_{\hbar}$ on
$\mathcal{C}_{0}\left(  \left[  0,1\right]  \right)  $ that may be thought of
as the infinite-dimensional limit of the measures $\rho_{\hbar}^{(d)}$ on
$X_{\mathbb{R}}.$ The measure $\rho_{\hbar}$ has the highly non-rigorous
formal expression
\begin{equation}
d\rho_{\hbar}\left(  B\right)  =const.\,\exp\left[  -\frac{1}{2\hbar}\int
_{0}^{1}\left|  \frac{dB}{dt}\right|  ^{2}dt\right]  \,\mathcal{D}%
B,\label{formal}%
\end{equation}
where $\mathcal{D}B$ is the non-existent Lebesgue measure on $\mathcal{C}%
_{0}\left(  \left[  0,1\right]  \right)  $ and the constant is supposed to
normalize $\rho_{\hbar}$ to be a probability measure. Note that the expression
in the exponent is just $-\left\|  B\right\|  ^{2}/2\hbar,$ where the norm is
computed using the inner product (\ref{inner.h1}). Even though the measure
lives on $\mathcal{C}_{0}\left(  \left[  0,1\right]  \right)  ,$ the
properties of the measure are determined by the norm on $X_{\mathbb{R}}.$
Expressions of the form (\ref{formal}) are common in the physics literature.

The measure described in the previous paragraph is the \textbf{Wiener
measure}. As a measure on the space of continuous paths, it describes the
behavior of \textbf{Brownian motion}. The typical path $B$ (with respect to
the measure $\rho_{\hbar}$) is very wiggly and non-differentiable. A general
triple $\left(  X_{\mathbb{R}},\overline{X}_{\mathbb{R}},\rho_{\hbar}\right)
$ of the sort considered above is called an \textbf{abstract Wiener space}, in
honor of the motivating example of the Wiener measure. (The terminology is due
to Gross \cite{Gr}.)

We have then a good candidate for the domain Hilbert space of our
Segal-Bargmann transform in the infinite-dimensional case, namely,
$L^{2}(\overline{X}_{\mathbb{R}},\rho_{\hbar}).$ We now need to find the right
range Hilbert space. So consider the complexified Hilbert space $X_{\mathbb{C}%
}=X_{\mathbb{R}}+iX_{\mathbb{R}},$ the family $V_{d}^{\mathbb{C}}=V_{d}%
+iV_{d}$ of finite-dimensional subspaces, and the family $\mu_{\hbar}^{(d)}$
of measures given by
\[
d\mu_{\hbar}^{(d)}\left(  z\right)  =\left(  \pi\hbar\right)  ^{-d}%
e^{-\left\|  z\right\|  ^{2}/\hbar}\,dz,
\]
where $dz$ is Lebesgue measure on $V_{d}^{\mathbb{C}}$. One can consider the
limit $\mu_{\hbar}$ of these measures, which exists as a measure on a certain
extension $\overline{X}_{\mathbb{C}}$ of $X_{\mathbb{C}},$ where $\overline
{X}_{\mathbb{C}}$ is a complex Banach space. (As on the domain side, we have
$\mu_{\hbar}\left(  X_{\mathbb{C}}\right)  =0.$) In the case of the Wiener
measure, $\overline{X}_{\mathbb{C}}$ may be taken to be the space of
continuous functions $Z:\left[  0,1\right]  \rightarrow\mathbb{C}$ with
$Z\left(  0\right)  =0.$

Now there exists a perfectly suitable notion of what it means for a function
on a complex Banach space such as $\overline{X}_{\mathbb{C}}$ to be
holomorphic, and so it seems plausible that we should take the Segal-Bargmann
space to be the space of holomorphic functions on $\overline{X}_{\mathbb{C}}$
that are square-integrable with respect to $\mu_{\hbar}.$ Unfortunately, this
definition does not work, because in the infinite-dimensional case the space
of square-integrable holomorphic functions is \textit{not} a closed subspace
of $L^{2}\left(  \overline{X}_{\mathbb{C}},\mu_{\hbar}\right)  $ and therefore
not a Hilbert space. There are then two approaches to defining the
Segal-Bargmann space. The first approach is essentially to define the
Segal-Bargmann space to be the closure in $L^{2}\left(  \overline
{X}_{\mathbb{C}},\mu_{\hbar}\right)  $ of the space of holomorphic functions.
This is the approach used in \cite{HS,DH}. (See also \cite{Sh,Su}.)

Another approach to the Segal-Bargmann space is to consider holomorphic
functions on $X_{\mathbb{C}}$ itself. In that case the $L^{2}$ norm is
meaningless (since the measure $\mu_{\hbar}$ is not defined on $X_{\mathbb{C}}
$), but we can define a norm as follows. Suppose $F$ is a holomorphic function
on $X_{\mathbb{C}}.$ Then define $\left\|  F\right\|  _{\hbar}$ by
\[
\left\|  F\right\|  _{\hbar}^{2}=\sup_{d}\int_{V_{d}^{\mathbb{C}}}\left|
F\left(  z\right)  \right|  ^{2}\,d\mu_{\hbar}^{(d)}\left(  z\right)  ,
\]
where the $V_{d}^{\mathbb{C}}$'s are the finite-dimensional subspaces
introduced above. The Segal-Bargmann space is then defined to be
\[
\mathcal{H}^{\hbar}\left(  X_{\mathbb{C}}\right)  =\left\{  F:\mathcal{H}%
_{\mathbb{C}}\rightarrow\mathbb{C}\left|  F\text{ is holomorphic and }\right.
\left\|  F\right\|  _{\hbar}<\infty\right\}  .
\]
It turns out that $\left\|  F\right\|  _{\hbar}$ is a norm on $\mathcal{H}%
^{\hbar}\left(  X_{\mathbb{C}}\right)  $ and that $\mathcal{H}^{\hbar}\left(
X_{\mathbb{C}}\right)  $ becomes a Hilbert space if we take the inner product
to be
\[
\left\langle F_{1},F_{2}\right\rangle =\sup_{d}\int_{V_{d}^{\mathbb{C}}%
}\overline{F_{1}\left(  z\right)  }F_{2}\left(  z\right)  \,d\mu_{\hbar}%
^{(d)}(z).
\]
This is the form of the Segal-Bargmann space used in \cite{S3,BSZ} (except
that Segal always uses anti-holomorphic rather than holomorphic functions).

We now state a form of the Segal-Bargmann theorem for the infinite-dimensional case.

\begin{theorem}
\label{bt.inf}For all $f\in L^{2}(\overline{X}_{\mathbb{R}},\rho_{\hbar})$
there exists a unique holomorphic function $B_{\hbar}f$ on $X_{\mathbb{C}}$
whose restriction to $X_{\mathbb{R}}$ is given by
\begin{equation}
B_{\hbar}f\left(  y\right)  =\int_{\overline{X}_{\mathbb{R}}}f\left(
y-x\right)  \,d\rho_{\hbar}\left(  x\right)  ,\quad y\in X_{\mathbb{R}%
}.\label{bt.integral}%
\end{equation}
(The integral is well-defined and convergent for all $y\in X_{\mathbb{R}}.)$
Furthermore, $B_{\hbar}$ is a unitary map of $L^{2}(\overline{X}_{\mathbb{R}%
},\rho_{\hbar})$ onto $\mathcal{H}^{\hbar}\left(  X_{\mathbb{C}}\right)  .$
\end{theorem}

\textit{Remarks}. 1) Note that we compute the value of $B_{\hbar}f$ directly
on $X_{\mathbb{R}}$ by the integral (\ref{bt.integral}). To get the value on
$X_{\mathbb{C}}$ we analytically continue from $X_{\mathbb{R}}$ to
$X_{\mathbb{C}}.$

2) A simple change of variable shows that in the finite-dimensional case the
transform defined here agrees with that of Section 6. Once one knows a
reasonable amount about Gaussian measure spaces and about the space
$\mathcal{H}^{\hbar}\left(  X_{\mathbb{C}}\right)  $, the proof is a
straightforward reduction to the finite-dimensional case.

3) To verify that the integral in the theorem makes sense one needs to know
that the measure $\rho_{\hbar}$ is ``quasi-invariant'' under translations in
the direction of $X_{\mathbb{R}}.$ (This quasi-invariance is the content of
the Cameron-Martin Theorem.)

4) Theorem \ref{bt.inf} is similar to Proposition 4.7 and Theorem 4.8 of
\cite{GM}. (Cf. Corollary 11 of \cite{HS}.)

\subsection{Coherent states}

In a holomorphic $L^{2}$ space, the \textbf{coherent states} are the unique
elements $\phi_{z}\in\mathcal{H}L^{2}\left(  U,\alpha\right)  $ such that
\begin{equation}
F\left(  z\right)  =\left\langle \phi_{z},F\right\rangle \label{coherent1}%
\end{equation}
for all $F\in\mathcal{H}L^{2}\left(  U,\alpha\right)  .$ (In the standard
lingo of quantum physics a ``state'' means simply a non-zero element of the
relevant Hilbert space.) The states $\phi_{z}$ are the same as those in
Section 2. That is, the coherent states are given by
\begin{equation}
\phi_{z}\left(  w\right)  =\overline{K\left(  z,w\right)  },\label{coherent2}%
\end{equation}
where $K\left(  z,w\right)  $ is the reproducing kernel. Using the basic
property (\ref{coherent1}) of the $\phi_{z}$'s we see that the reproducing
kernel is just the inner product of the coherent states:
\begin{equation}
K\left(  z,w\right)  =\left\langle \phi_{z},\phi_{w}\right\rangle
.\label{coherent3}%
\end{equation}

Meanwhile, let's consider $L^{2}\left(  \mathbb{R}^{d},dx\right)  $ and the
``invariant'' form $C_{\hbar}$ of the Segal-Bargmann transform. (A similar
analysis can be done with the other forms.) We now want to define coherent
states $\psi_{z}$ in $L^{2}\left(  \mathbb{R}^{d},dx\right)  .$ These are the
unique states $\psi_{z}\in L^{2}\left(  \mathbb{R}^{d},dx\right)  $ such that
\begin{equation}
C_{\hbar}f\left(  z\right)  =\left\langle \psi_{z},f\right\rangle
_{L^{2}\left(  \mathbb{R}^{d},dx\right)  }.\label{coh.def2}%
\end{equation}
Since the Segal-Bargmann transform is unitary, we have
\[
C_{\hbar}f\left(  z\right)  =\left\langle \psi_{z},f\right\rangle
_{L^{2}\left(  \mathbb{R}^{d},dx\right)  }=\left\langle C_{\hbar}\psi
_{z},C_{\hbar}f\right\rangle _{\mathcal{H}L^{2}\left(  \mathbb{C}^{d}%
,\nu_{\hbar}\right)  }.
\]
Comparing this with (\ref{coherent1}) we see that
\[
C_{\hbar}\psi_{z}=\phi_{z}.
\]
So if you prefer you may define the coherent states in $L^{2}\left(
\mathbb{R}^{d},dx\right)  $ by
\begin{equation}
\psi_{z}=C_{\hbar}^{-1}\phi_{z}.\label{coh.def3}%
\end{equation}
From (\ref{coherent3}) and (\ref{coh.def3}) we see that $K\left(  z,w\right)
=\left\langle \psi_{z},\psi_{w}\right\rangle .$

Recalling the formula for the $C_{\hbar}$ form of the Segal-Bargmann transform
we see that the states satisfying (\ref{coh.def2}) are
\[
\psi_{z}\left(  x\right)  =\left(  2\pi\hbar\right)  ^{-d/2}e^{-\left(
\bar{z}-x\right)  ^{2}/2\hbar}.
\]
Doing some algebra we get that
\begin{equation}
\psi_{z}\left(  z\right)  =c_{z}e^{-i\left(  \operatorname{Im}z\right)  \cdot
x/\hbar}e^{-\left(  x-\operatorname{Re}z\right)  ^{2}/2\hbar}%
\label{coh.formula}%
\end{equation}
where the constant $c_{z}$ is given by
\[
c_{z}=\left(  2\pi\hbar\right)  ^{-d/2}e^{\left(  \operatorname{Im}z\right)
^{2}/2\hbar}e^{i\operatorname{Im}z\cdot\operatorname{Re}z/\hbar}.
\]
From (\ref{coh.formula}) we see that $\psi_{z}$ is a Gaussian centered at the
point $\operatorname{Re}z$ and multiplied by a constant and $e^{-i\left(
\operatorname{Im}z\right)  \cdot x/\hbar}.$ A function of this sort are called
\textbf{Gaussian wave packet}; it is the oscillating ``wave'' $e^{-i\left(
\operatorname{Im}z\right)  \cdot x/\hbar}$ multiplied by a Gaussian.

These states are very special. For example, they are ``minimum uncertainty''
states. This means that they give equality in the inequality of the Heisenberg
uncertainty principle. One should think of $\psi_{z}$ as being the closest
thing there is to a quantum state with position $\operatorname{Re}z$ and
momentum $\operatorname{Im}z.$ This is the idea that is intended to be
conveyed by the word ``coherent''--these states are as localized in phase
space as is consistent with the uncertainty principle. In addition, these
states behave in a very simple way with respect to the time-evolution of a
quantum harmonic oscillator.

Let us express the isometricity of the Segal-Bargmann transform in terms of
the coherent states. The isometricity of $C_{\hbar}$ tells us that for all
$f,g\in L^{2}\left(  \mathbb{R}^{d},dx\right)  $ we have
\[
\int_{\mathbb{R}^{d}}\overline{f\left(  x\right)  }g\left(  x\right)
\,dx=\int_{\mathbb{C}^{d}}\overline{C_{\hbar}f\left(  z\right)  }C_{\hbar
}g\left(  z\right)  \nu_{\hbar}\left(  z\right)  \,dz.
\]
But by (\ref{coh.def2}), $C_{\hbar}g\left(  z\right)  =\left\langle \psi
_{z},g\right\rangle $ and $\overline{C_{\hbar}f\left(  z\right)
}=\left\langle f,\psi_{z}\right\rangle ,$ so
\begin{equation}
\int_{\mathbb{R}^{d}}\overline{f\left(  x\right)  }g\left(  x\right)
\,dx=\int_{\mathbb{C}^{d}}\left\langle f,\psi_{z}\right\rangle \left\langle
\psi_{z},g\right\rangle \nu_{\hbar}\left(  z\right)  \,dz.\label{res1}%
\end{equation}

Now let $\left|  \psi_{z}\right\rangle \!\left\langle \psi_{z}\right|  $ be
the operator given by
\[
\left|  \psi_{z}\right\rangle \!\left\langle \psi_{z}\right|  f=\psi
_{z}\left\langle \psi_{z},f\right\rangle ,
\]
which is essentially just projection onto the state $\psi_{z}.$ (The
projection would have a factor of $\left\|  \psi_{z}\right\|  ^{2}$ in the
denominator.) This is part of the ``Dirac notation'' commonly used in physics.
To understand the logic behind this notation note that $\left\langle
f,\psi_{z}\right\rangle \left\langle \psi_{z},g\right\rangle $ is just the
inner product of $f$ with $\left|  \psi_{z}\right\rangle \!\left\langle
\psi_{z}\right|  g.$ The Dirac notion expresses the inner product with a
vertical line, so $\left\langle f\left|  \psi_{z}\right.  \right\rangle $
instead of $\left\langle f,\psi_{z}\right\rangle .$ So in Dirac notation
$\left\langle f,\psi_{z}\right\rangle \!\left\langle \psi_{z},g\right\rangle $
becomes $\left\langle f\left|  \psi_{z}\right.  \right\rangle \left\langle
\psi_{z}\left|  g\right.  \right\rangle $ which is supposed to be notationally
indistinguishable from the inner product of $f$ with $\left|  \psi
_{z}\right\rangle \!\left\langle \psi_{z}\right|  g.$ So then (\ref{res1}) can
be rewritten by formally bringing the integral inside the inner product to
give
\begin{equation}
\left\langle f,g\right\rangle =\left\langle f,\left(  \int_{\mathbb{C}^{d}%
}\left|  \psi_{z}\right\rangle \!\left\langle \psi_{z}\right|  \nu_{\hbar
}\left(  z\right)  \,dz\right)  g\right\rangle .\label{res2}%
\end{equation}
If this holds for all $f$ and $g$ then the operator inside the parentheses on
the right in (\ref{res2}) must be the identity operator:
\begin{equation}
\int_{\mathbb{C}^{d}}\left|  \psi_{z}\right\rangle \!\left\langle \psi
_{z}\right|  \nu_{\hbar}\left(  z\right)  \,dz=I.\label{res3}%
\end{equation}

Equation (\ref{res3}) is called a \textbf{resolution of the identity}. Note
that both sides are operators in $L^{2}\left(  \mathbb{R}^{d},dx\right)  ,$
even though the integral is over $\mathbb{C}^{d}$. This is because the
coherent states $\psi_{z}$ are elements of $L^{2}\left(  \mathbb{R}%
^{d},dx\right)  ,$ with parameter $z$ in $\mathbb{C}^{d}.$ Formally,
(\ref{res3}) is equivalent to the isometricity of the Segal-Bargmann
transform. This resolution of the identity first appears in the 1960 paper of
John Klauder \cite{K}. (Klauder uses a different normalization.) The
resolution of the identity is often a useful way to think about the
Segal-Bargmann transform (or its generalizations). The weakness of this point
of view is that there is no straightforward way to express the surjectivity of
the Segal-Bargmann transform (that it maps \textit{onto} the space of
square-integrable holomorphic functions) in terms of the coherent states.
Still, it is useful to be able to go back and forth between the transform
point of view and the coherent state point of view.

One can think about Toeplitz operators in terms of the coherent states
$\phi_{z}\in\mathcal{H}L^{2}\left(  \mathbb{C}^{d},\nu_{\hbar}\right)  .$ If
$f$ is a not-necessarily-holomorphic function on $\mathbb{C}^{d},$ then the
Toeplitz operator $T_{f}$ on $\mathcal{H}L^{2}\left(  \mathbb{C}^{d}%
,\nu_{\hbar}\right)  $ may be expressed as
\begin{equation}
T_{f}=\int_{\mathbb{C}^{d}}f\left(  z\right)  \left|  \phi_{z}\right\rangle
\!\left\langle \phi_{z}\right|  \nu_{\hbar}\left(  z\right)  \,dz.\label{res4}%
\end{equation}
I leave it as an (instructive) exercise to the reader to verify this
expression, using properties of Toeplitz operators and of the coherent states.
Note that taking $f\equiv1$ in (\ref{res4}) gives the analog of (\ref{res3})
in $\mathcal{H}L^{2}\left(  \mathbb{C}^{d},\nu_{\hbar}\right)  .$

Numerous other kinds of coherent states have been considered. See for example
the books \cite{KS} and \cite{P}.

\subsection{K\"{a}hler quantization}

A K\"{a}hler manifold is a complex manifold $M$ with a symplectic structure
(i.e., a nice Poisson bracket) in which the two structures satisfy a natural
compatibility condition. The simplest example is $\mathbb{C}^{d}$ itself. The
theory of geometric quantization \cite{W} gives you a way of associating
Hilbert space with certain K\"{a}hler manifolds. In the case of $\mathbb{C}%
^{d}, $ the resulting Hilbert space is (or can be identified with) the
Segal-Bargmann space. In general the Hilbert space is a space of $L^{2}$
holomorphic sections of a holomorphic line bundle over $M.$ In the case of
$\mathbb{C}^{d}$ this line bundle is holomorphically trivial, which means that
the Hilbert space can be identified with an $L^{2}$ space of holomorphic
functions--the Segal-Bargmann space. So these Hilbert spaces of holomorphic
sections of line bundles should be thought of as generalizations of the
Segal-Bargmann space, in which $\mathbb{C}^{d}$ is replaced by some other
K\"{a}hler manifold. Another example is the unit disk (with an $SU\left(
1,1\right)  $-invariant symplectic structure) in which case the space of
sections may be identified with one of the weighted Bergman spaces. Note that
Planck's constant is a parameter in the geometric quantization scheme;
different values of $\hbar$ give different values of $a$ in the weighted
Bergman spaces. I have already mentioned this example in connection with
generalized canonical commutation relations. However, the method of K\"{a}hler
quantization applies to arbitrary K\"{a}hler manifolds, not assumed to have
any symmetry condition. One interesting case is that of \textit{compact}
K\"{a}hler manifolds. In this case the quantum Hilbert space is
finite-dimensional, reflecting the finite size of the classical phase space.
There is much interesting topology in the line bundles in this case.

These $L^{2}$ spaces of holomorphic sections allow much of the same structure
as our spaces $\mathcal{H}L^{2}\left(  U,\alpha\right)  .$ In particular,
pointwise evaluation is continuous, so there is a reproducing kernel and the
holomorphic subspace is a closed subspace. So you have coherent states as
above and you can define Toeplitz operators in a similar fashion to what we consider.

For a sampling of papers on this subject, see the works of Klimek and
Lesniewski \cite{KL1,KL2}, Coburn \cite{C}, Bordemann, Meinrenken, and
Schlichenmaier \cite{BMS}, Borthwick, Lesniewski, and Upmeier \cite{BLU}, and
Borthwick, Paul, and Uribe \cite{BPU}. The expository paper \cite{Bo} gives a
(fairly) gentle introduction to some of the techniques. As always, the book of
Woodhouse \cite{W} gives valuable background material. (But there is a very
large amount of information in \cite{W} and it is not always easy to extract
what is relevant to a particular application.)

\end{document}